\documentclass[prb,preprint]{revtex4-1}
\usepackage[dvips]{graphicx}
\usepackage{slashed}
\usepackage{amssymb}
\usepackage{natbib}
\usepackage{amsmath}
\usepackage{url}
\usepackage{color}
\usepackage{hyperref}
\usepackage[normalem]{ulem}
\usepackage{xcolor}
\usepackage[utf8]{inputenc}
\usepackage[english]{babel}
\begin{document}
\newcommand{\dd}{{\rm d}}
\newcommand{\fracc}[2]{\frac{\textstyle{#1}}{\textstyle{#2}}}
%---------------------------------------------
%%%%%%%%%Eduardo macro (remove afterwards)%%%%%%%%%%%%%%%%%%%
\newcommand{\red}{\textcolor{red}}
\newcommand\redout{\bgroup\markoverwith
{\textcolor{red}{\rule[.5ex]{2pt}{0.6pt}}}\ULon}
\newcommand{\nocontentsline}[3]{}
\newcommand{\tocless}[2]{\bgroup\let\addcontentsline=\nocontentsline#1{#2}\egroup}
\newcommand{\beq}{\begin{equation}}
\newcommand{\eeq}{\end{equation}}
\newcommand{\vare}{\varepsilon}
%\tocless\section{hide}
%\tocless\subsection{subhide}
\newtheorem{theorem}{Lemma}

\title{Metric Relativity and the Dynamical Bridge: highlights of Riemannian geometry in physics}

\author{Mario Novello}
\email{novello@cbpf.br}
\affiliation{Instituto de Cosmologia Relatividade e Astrofisica (ICRA/CBPF) \\
Rua Dr. Xavier Sigaud 150, Urca 22290-180 Rio de Janeiro, RJ -- Brazil}

\author{Eduardo Bittencourt}
\email{eduardo.bittencourt@icranet.org}
\affiliation{CAPES Foundation, Ministry of Education of Brazil, Bras\'ilia, Brazil
and Physics Department, \lq\lq La Sapienza" University of Rome, I-00185 Rome -- Italy}

%\pacs{02.40.-k, 02.40.Ky, 03.50.-z, 03.65.-w}
\date{\today}

\begin{abstract}
We present an overview of recent developments concerning modifications of the geometry of space-time to describe various physical processes of interactions among classical and quantum configurations. We concentrate in two main lines of research: the Metric Relativity and the Dynamical Bridge. We describe the notion of equivalent (dragged) metric $ \widehat{g}_{\mu\nu}$ which is responsible to map the path of any accelerated body in Minkowski space-time onto a geodesic motion in such associated $\widehat{g}$ geometry.

Only recently the method introduced by Einstein in general relativity was used beyond the domain of gravitational forces to map arbitrary accelerated bodies submitted to non-Newtonian attractions onto geodesics of a modified geometry. This process has its roots in the very ancient idea to treat any dynamical problem in Classical Mechanics as nothing but a problem of static where all forces acting on a body annihilates themselves including the inertial ones. This general procedure, that concerns arbitrary forces -- beyond the uses of General Relativity that is limited only to gravitational processes -- is nothing but the relativistic version of the d'Alembert method in classical mechanics and consists in the principle
of Metric Relativity. The main difference between gravitational interaction and all other forces concerns the universality of gravity which added to the interpretation of the Equivalence Principle allows all associated geometries -- one for each different body in the case of non-gravitational forces -- to be unified into a unique Riemannian space-time structure. The same geometrical description appears for electromagnetic waves in the optical limit within the context of nonlinear theories or material medium. Once it is largely discussed in the literature, the so-called \textit{analogue models of gravity}, we will dedicate few sections on this emphasizing their relation with the new concepts introduced here.

Then we pass to the description of the Dynamical Bridge formalism which states the dynamic equivalence of non-linear theories (driven by arbitrary scalar, spinor or vector fields) that occur in Minkowski background to theories described in associated curved geometries generated by each one of these fields. We shall see that it is possible to map the dynamical properties of a theory, say Maxwell electrodynamics in Minkowski space-time, into Born-Infeld electrodynamics described in a curved space-time the metric of which is defined in terms of the electromagnetic field itself in such way that it yields the same dynamics. It is clear that when considered in whatever unique geometrical structure these two theories are not the same, they do not describe the same phenomenon. However we shall see that by a convenient modification of the metric of space-time an equivalence appears that establishes a bridge between these two theories making they represent the same phenomenon. This method was recently used to achieve a successful geometric scalar theory of gravity. At the end we briefly review the proposal of geometrization of quantum mechanics in the de Broglie-Bohm formulation using an enlarged non-Riemannian (Weyl) structure.
\end{abstract}

\maketitle
\tableofcontents

\section{Preliminary comments}

The success of general relativity (GR) led to the general acceptance that the introduction of a geometry to describe a physical process yields a unique structure identified to the space-time. However, this is a rather general mathematical procedure and one can applies the modification of the background geometry to substitute the acceleration not only by gravity but for any kind of force; introducing specific and limited geometries for different observers in many distinct situations seems to be a very useful tool.

One could accept that the achievements of a given theory may depend on the uniqueness of its representation. However, in more recent years we have learned that the possibility of presenting an alternative equivalent description of a given phenomenon may be an important theoretical tool. The presentation of a theory under distinct formulations has been of great help in many occasions, as we will verify with direct examples in these notes. We shall see how it is possible to modify the background geometry of space-time (intrinsic to the description of gravitational interactions according to the rules and proposals of GR) to describe the effect of forces of distinct and multiple natures. A well-known example concerns the propagation of light inside moving dielectrics.

We shall describe how very different modifications of the space-time geometry have been used to describe accelerated paths in Minkowski
space-time as geodesics in an associated curved Riemannian geometry and even to exhibit the equivalence dynamics of various fields in distinct associated geometries.

An important step in the construction of the theory of special relativity was the hypothesis that a specific proper time can be assigned for each body. Distinct observers establish distinct times. This proposal highlights the importance of an individual body in the description of its kinematics and stresses the importance of the global aspects of the physical description in the domain of mechanics. Such an individualization was new in the standard program of physics that usually is based, in a very broad sense, to develop formalisms that intended to be universal.

After the success of this procedure there has been a need to recover the global and unified character of the theoretical framework. This has indeed come about through the next big step made by the entrance of a dynamical scheme associated to the universal gravitational interaction, which becomes identified to an attribute of the structure of space-time. There, the geometry of space-time allows all test bodies to be considered as free particles in a curved world and to follow the geodesics on this curved space-time, as far as gravity is taken into account solely. If a body suffers a particular acceleration of non-gravitational character, this freedom disappears and the body follows a non-geodesic path.

The purpose of these notes is to describe some new results that intend, in a very particularly and precise way, to extend the idea that it is possible to deal with any kind of force through similar lines as suggested by GR. This means to introduce the idea that modifications of the geometry of space-time may describe a large variety of physical processes. In other words, to generalize the approach of GR in order to represent all kind of acceleration as nothing but a modification of the geometry. Along this line the path of any body on which an arbitrary force has acted upon, can be described as if this body is free, without any interaction, in a particular associated metric structure that depends on both, the characteristics of the force and the kinematics of the body. Let us point out that there are obvious distinctions between the case of GR and all other cases. This is due to the universality of gravitational processes. Indeed, one of the postulates of GR states that there exists one and only one geometric structure of space-time as perceived for all that exists. Besides that, the uniqueness of the scenario where all events occurs lies on the hypothesis that its geometry is controlled by the gravitational phenomena. This unification requires the neglect of all other kind of interaction and  must be interpreted as some sort of approximation related to the degree of amplitude of its corresponding description. This beautiful requirement of GR appears in each observation as specific examples of geometries obtained as a particular solutions of its dynamics. In these notes we will analyze another possibility.

The original GR proposal was concerned with the possibility of eliminating the acceleration induced on an arbitrary body $\mathbb{A}$ by a gravitational field by a convenient modification of the metric associated to the space where $\mathbb{A}$ is propagating. GR shows that all gravitational effects can be equivalently described in terms of such an universal modification of the space-time geometry.

However, from the technical point of view this is only one possibility. Indeed, as we shall review in these notes, this should not be considered the only possibility and one could well make other conventions to ascribe specific metrics to different events. According to this point of view, to each interaction a particular modification of the metric environment in which a body moves in such a way that any kind of force can be eliminated by this interpretation is likely to be associated. After this characterization of its own metric structure, the body is interpreted as realizing a free motion. In other words, the metric becomes just a convention to eliminate the force that drives the motion of a body. The main direct consequence of this is that, contrary to the principles of GR, these metrics do not need any additional constraint associated to its dynamics.

The very fact that GR assumes that the universality of gravity is the origin to accept its geometric interpretation and once this modification of the geometry is universal, completely independent of any particular process,  makes obligatory the existence of a specific dynamics of the space-time metric. In the case of other forces, once each process has its own origin that produces a metric in order to compensate the force, there is no room to impose an extra dynamics for the geometry: its characterization is specified by the interaction.

It is clear that this strategy has a drawback: we lose the uniqueness of space-time. However such uniqueness may be understood as nothing but a suitable and conventional way to describe the universality of gravitational interaction. Let us remark that in the original paper that generated the modern point of view of geometrical spaces, the great mathematician B. Riemann \cite{riemann} supported such a proposal, which was not the point of view emphasized and developed in the theory of general relativity. On the eve of his 150th death anniversary, we intend to rescue the seminal ideas concerning the role of the Riemannian geometries in Physics developed by him, which were not investigated before.

There is no better way to start such analysis than considering the analogy with the photon propagation inside a moving dielectric. In general, we know that the light path acquires an acceleration inside a medium. In the early twenties Gordon showed that it is possible to describe this path as a geodesics in a modified metric. This means that a change of the geometry can eliminate the acceleration in a very similar way as it is done in GR. It is clear that such geometric description is not indispensable. The choice of such point of view yields a new path of investigation which was named analogue models of gravity. This means to mimic gravitational configurations by means of non-gravitational interactions. In the case of Gordon, he used electromagnetic forces to represent gravitational interactions. We shall see that the interest on Gordon\rq s approach goes beyond its original proposal once it allows a generalization to include all kinds of accelerated paths, independently of the origin of the force that produced it and for any kind of massive or massless body. This result seemed to be hidden and its importance diminished during more than half a century.

The main reason for this concerns the way gravity was related to the geometry of space-time and the various unsuccessful proposals of
unification of electrodynamics into an unique geometric framework. We know that such unfruitful works were related to the non universality of the non-gravitational forces. However there is another component to be taken into account as soon as we realize that this method is nothing but the relativistic version of an ancient idea of d\rq Alembert\rq s to transform a dynamical problem into a static one. We shall see that this reduction is made by a modification of the evaluation of distances in the space-time through a change in its geometry. We call such procedure metric relativity (MR).

We shall finally enter the microphysics and the domain of quantum properties. We shall see that an unexpected novelty on the structure of 3-D space may be hidden in the quantum world. This will be described by using the de Broglie-Bohm causal formulation of quantum mechanics. We shall be prepared to make another drastic change to our description of the motion of bodies in space. According to Riemann in his Habilitation Dissertation, one should wounder about the extension of our ideas of geometry beyond the observation limits of the infinitely great and the infinitely small. In these notes we have made constant references to GR and the modification on the structure of geometry that it proposes in our neighborhood and in the immense regions of the universe. We will now look to the other direction, to the one that consists the micro-world. How should we treat the geometry in these extremely small regions?

Since ancient times, Euclidean geometry was considered the most adequate mathematical formulation to describe the physical space.
However, its validity can only be established a posteriori for its construction yields useful notions to connect physical
quantities such as the Euclidean distance between two given points. Special relativity modified the notion of 3-dimensional Euclidean space to incorporate time in a four-dimensional continuum (Minkowski space-time). Later on, GR generalized the absolute Minkowski space-time to describe gravitational phenomena by considering the space-time manifold as a dynamical structure constituting a Riemannian manifold.

According to Riemann, the matter of which geometry is actually realized in Nature has to be determined by physical experiments. Instead of imposing a priori that non-relativistic quantum mechanics has to be constructed over a Euclidean 3-dimensional background as it is traditionally done, we shall see that quantum effects can also be interpreted as a manifestation of a non-Euclidean structure derived from a variational principle. We then arrive at a special subclass of a more general structure than Riemannian geometries, known as Weyl space which, in the de Broglie-Bohm vision of quantum phenomenon, should be used to ``determine the measure-relations of space".

\newpage

\section{Brief mathematical compendium}\label{definition}
In this section we display some formula and definitions that will be used throughout this paper. The lector interested in more details on the origin and developments that led to these expressions can check the references, mainly the textbooks, in the bibliography at the end \cite{adler-bazin,anderson,rindler,synge,tolman}.

The Riemannian manifold is characterized by a second order symmetric tensor  $g_{\mu\nu}$ and a connection $\Gamma^{\mu}_{\alpha\beta}$ related by the formula
$$\Gamma^{\mu}_{\alpha\beta} = \frac{1}{2} \, g^{\mu\nu} \, \left( g_{\nu\alpha , \beta} + g_{\nu\beta , \alpha} - g_{\alpha \beta , \nu}  \right). $$
The Minkowski metric $\gamma_{\mu\nu}$ takes the following form when expressed in a Cartesian coordinate system ${\rm diag}( 1, -1, -1, -1).$ In this coordinate system it is denoted by $\eta_{\mu\nu}$. Greek indices run from $0$ to $3$. The covariant derivative ($;$) is written as
\begin{equation}
\label{sec_der}
v_{\mu \, ; \, \nu} = v_{\mu, \nu} - \Gamma^{\alpha}_{\mu\nu} v_{\alpha},
\end{equation}
where $ v_{\mu , \nu} \equiv \partial_{\nu} v_{\mu} $ denotes partial derivatives. The metricity Riemannian condition is expressed as
$$ g_{\mu\nu \, ;\lambda} = 0.$$
Then, it follows
$$ v^{\alpha}{}_{;\mu ;\nu} - v^{\alpha}{}_{;\nu ;\mu} = R^{\alpha}{}_{\beta\mu\nu} \, v^{\beta}, $$
where $R^{\alpha}{}_{\beta\mu\nu}$ is the Riemann curvature tensor. In terms of the connection we can write
$$ R^{\mu}_{~ \epsilon\alpha\beta} = \Gamma^{\mu}_{~\epsilon \alpha , \beta} - \Gamma^{\mu}_{~\epsilon \beta , \alpha} +
\Gamma^{\mu}_{\beta\sigma} \,  \Gamma^{\sigma}_{\epsilon\alpha} - \Gamma^{\mu}_{~\alpha\sigma} \,  \Gamma^{\sigma}_{\beta\epsilon}. $$
The curvature tensor satisfies the algebraic identities $R_{\mu\nu\alpha\beta} = - \, R_{\mu\nu\beta\alpha} = - \,
R_{\nu\mu\alpha\beta} = R_{\alpha\beta\mu\nu},$
and the Bianchi identities
$$R^{\mu\nu}_{~~~\alpha\beta ; \lambda} + R^{\mu\nu}_{~~~ \lambda\alpha ; \beta} +  R^{\mu\nu}_{~~~\beta \lambda ; \alpha} = 0.$$
Contacting indices, it then follows
$$R^{\mu\nu}_{~~~; \nu } - \frac{1}{2} \,  R_{, \nu} \, g^{\mu\nu} = 0, $$
which implies, through the GR equations, the conservation of the energy-momentum tensor. We note that this equation is obtained from Hilbert Lagrangian
$$\delta S=\delta \, \int \sqrt{- \, g} \, R = 0,$$
where we use $\delta\sqrt{-g} = - \frac{1}{2} \, \sqrt{-g}\, g_{\mu\nu} \, \delta g^{\mu\nu}$.

\tocless\subsection{Non metricity}
We shall deal with a generalization of Riemannian geometry that was introduced by H. Weyl \cite{Weyl2} defined by the non-metricity condition:
\[
\label{no_met}
g_{\mu\nu \, ; \, \lambda} = f_{\lambda} \, g_{\mu\nu},
\]
where $f_{\lambda}$ is an arbitrary vector. It then follows that the affine connection is given by
$$\Gamma^{\lambda}_{\mu \nu} = \left\{^{\lambda}_{\mu \nu}\right\}  - \, \frac{1}{2} \, \left( \delta^{\lambda}_{\mu} \, f_{\nu} + \delta^{\lambda}_{\nu} \, f_{\mu} - g_{\mu \nu} \,  f^{\lambda} \right),$$
where $\{^{\lambda}_{\mu \nu}\}$ is the Christoffel symbol constructed with the metric tensor.

\tocless\subsection{Duality}
The Levi-Civita completely anti-symmetric object $\epsilon_{\alpha\beta\mu\nu}$ takes the value $1$ when indices are $(0123)$ or any even permutation, $-1$ for odd permutations and vanishes for repeated indices. We can then construct the tensor
$$ \eta_{\alpha\beta\mu\nu} =  \sqrt{-g} \, \varepsilon_{\alpha\beta\mu\nu} $$
where $g$ is the determinant of $g_{\mu\nu}.$ Using this object we define a dual, that is, for any anti-symmetric tensor  $F_{\mu\nu} = - \, F_{\nu\mu}$ we construct its dual by the relation:
$$ F^{*}_{\mu\nu} \equiv \frac{1}{2} \, \eta_{\mu\nu\alpha\beta} \, F^{\alpha\beta}. $$
Thus, $F^{**}_{\mu\nu} = - \, F_{\mu\nu}.$

It is useful to construct the quantity $g_{\alpha\beta\mu\nu} \equiv g_{\alpha\mu} g_{\beta\nu} - g_{\alpha\nu} g_{\beta\mu} $
that satisfy the symmetries
$$ g_{\alpha\beta\mu\nu} = - g_{\alpha\beta\nu\mu} = - g_{\beta\alpha\mu\nu} = g_{\mu\nu\alpha\beta}.$$

It is easy to see that $g_{\alpha\beta\mu\nu}$ is the dual of $\eta^{*}_{\alpha\beta\mu\nu}$, i.e.,
$$ \eta^{*}_{\alpha\beta\mu\nu} = - \, \,g_{\alpha\beta\mu\nu} $$
and inversely
$$ g^{*}_{\alpha\beta\mu\nu} =  \, \eta_{\alpha\beta\mu\nu}. $$
Note that $\varepsilon_{\alpha \beta \mu \nu}$ is a pseudo-tensor, although $\eta_{\alpha \beta \mu \nu}$ is a true tensor, that is
$$ \eta^{\mu \nu \rho \sigma}=g^{\mu \alpha}g^{\nu \beta}g^{\rho \varepsilon} g^{\sigma \lambda}\eta_{\alpha \beta \varepsilon
\lambda}. $$
Thus, we obtain
$$ \eta^{\alpha \beta \mu \nu}=-\frac{1}{\sqrt{-g}} \varepsilon^{\alpha \beta \mu \nu}. $$
All the contractions of the Levi-Civita tensor with itself are collected below:
\begin{eqnarray}
\eta^{\alpha \beta \mu \nu}\eta_{\rho \sigma \varepsilon \lambda} &=&- \delta^{\alpha \beta \mu \nu}_{\rho \sigma \varepsilon \lambda}, \nonumber \\
\eta^{\sigma \nu \rho \varepsilon}\eta_{\lambda \alpha \beta \varepsilon} &=& - \delta^{\sigma \nu \rho}_{\lambda \alpha \beta}, \nonumber \\
\eta_{\alpha \beta \varepsilon \lambda}\eta^{\sigma \nu \varepsilon \lambda} &=& -2\delta^{\sigma \nu}_{\alpha \beta},\\
\eta^{\sigma \nu \varepsilon \lambda}\eta_{\beta \nu \varepsilon \lambda} &=& -6 \delta^{\sigma}_{\beta}, \nonumber \\
\eta^{\alpha \beta \mu \nu}\eta_{\alpha \beta \mu \nu} &=& -24,\nonumber
\end{eqnarray}
where
\[
\delta^{\mu \nu \beta}_{\lambda \alpha \rho}= \det \left|
\begin{array}{l}
\delta^{\mu}_{\lambda} \quad \delta^{\mu}_{\alpha} \quad
\delta^{\mu}_{\rho}\\
\delta^{\nu}_{\lambda} \quad \delta^{\nu}_{\alpha} \quad
\delta^{\nu}_{\rho}\\
\delta^{\beta}_{\lambda} \quad \delta^{\beta}_{\alpha} \quad
\delta^{\beta}_{\rho}
\end{array}
\right|.
\]

\tocless\subsection{Decomposition of an anti-symmetric tensor: the Faraday tensor}
An arbitrary observer endowed with a normalized 4-velocity $ v^{\mu}$ can decompose any second order anti-symmetric tensor  $F_{\mu\nu}$ into its electric and magnetic parts under the form:
\[
F_{\mu \nu}=-v_{\mu}E_{\nu}+v_{\nu}E_{\mu}+ \eta_{\mu \nu}{}^{\rho \sigma}v_{\rho}H_{\sigma},
\]
where the electric $(E_{\mu})$ and magnetic $(H_{\mu})$ vectors are defined by
\[
\nonumber
E_{\mu} = F_{\mu \alpha}v^{\alpha}, \quad {\rm and} \quad H_{\mu} = F^{*}_{\mu \alpha}v^{\alpha}= \frac{1}{2} \eta_{\mu
\alpha}^{~~\rho \sigma}F_{\rho \sigma}V^{\alpha}
\]
It then follows that these vectors are defined in the 3-space orthogonal to the observer with velocity $v^{\mu}$, that is
$$E_{\mu} v^{\mu} = 0\quad {\rm and} \quad H_{\mu} v^{\mu} = 0$$

The six degrees of freedom of $F_{\mu \nu}$ are represented by the $3 + 3$ quantities $E_{\mu}$ e $H_{\mu}.$

The gauge invariant scalars constructed with the electromagnetic field represented by the Faraday tensor $ F_{\mu\nu} $ are given by
$$ F \equiv F_{\mu\nu} \, F ^{\mu\nu} \quad {\rm and} \quad G \equiv F_{\mu\nu}^{*}\, F ^{\mu\nu} $$

The following algebraic identities hold
\begin{eqnarray}
&&^{\ast}F^{\mu\alpha}\,{}^{\ast}F_{\alpha\nu}-F^{\mu\alpha}F_{\alpha\nu}=\frac{1}{2} F \delta^{\mu}_{\ ~\nu},\label{algrel1}\\
&&\stackrel{\ast}{F^{\mu\alpha}}F_{\alpha\nu}=-\frac{1}{4} G \delta^{\mu}_{\ ~\nu},\label{algrel2}\\
&&F^\mu{}_\alpha F^\alpha{}_\beta F^\beta{}_\nu=-\frac{G}{4}\,{}^{*}F^{\mu}{}_{\nu}-\frac{F}{2}F^\mu{}_{\nu},\label{algrel3}\\
&&F^\mu{}_\alpha F^\alpha{}_\beta F^\beta{}_\lambda F^\lambda{}_\nu=\frac{G^2}{16}\delta^\mu{}_{\nu}-\frac{F}{2}F^\mu{}_{\alpha}F^\alpha{}_{\nu}. \quad \;\label{algrel4}
\end{eqnarray}

\tocless\subsection{Weyl tensor}

It is possible to decompose the Riemann curvature tensor $R_{\alpha\beta\mu\nu}$ in its irreducible parts: the conformal Weyl tensor $W_{\alpha\beta\mu\nu}$ and its traces
$$R_{\alpha\beta\mu\nu} = W_{\alpha\beta\mu\nu} + M_{\alpha\beta\mu\nu} - \frac{1}{6} R g_{\alpha\beta\mu\nu},$$
where
$$2 M_{\alpha\beta\mu\nu} = R_{\alpha\mu} g_{\beta\nu} + R_{\beta\nu} g_{\alpha\mu} - R_{\alpha\nu} g_{\beta\mu} - R_{\beta\mu} g_{\alpha\nu}.$$

The Weyl tensor has only ten independent components. The remaining ten components of Riemann tensor are provided by the Ricci tensor $R_{\mu\nu}=R^{\alpha}{}_{\mu\alpha\nu}$ and the scalar curvature $R=R^{\alpha}{}_{\alpha}$.

The ten independent components of Weyl tensor are separated into its electric and magnetic parts according to any  observer endowed
with velocity $v^{\mu}.$ Indeed we write
$$E_{\alpha\beta} =  - W_{\alpha\mu\beta\nu} v^{\mu} v^{\nu}, \quad {\rm and} \quad  H_{\alpha\beta} =  - W^{\ast}_{\alpha\mu\beta\nu} v^{\mu} v^{\nu}.$$

Thus electric and magnetic tensors are symmetric, traceless and orthogonal to the observer:

\begin{equation}
\nonumber
E_{\mu\nu} = E_{\nu\mu}, \quad E_{\mu\nu} v^{\mu} = 0\quad {\rm and} \quad E_{\mu\nu} g^{\mu\nu} = 0
\end{equation}
and
\begin{equation}
\nonumber
H_{\mu\nu} = H_{\nu\mu}, \quad H_{\mu\nu} v^{\mu} = 0 \quad {\rm and} \quad H_{\mu\nu} g^{\mu\nu} = 0.
\end{equation}

From these symmetries it follows that the dual operation is independent of the indices pair
of Weyl tensor in which it is applied. Note that this is not the case for the Riemann tensor, for which the condition of independence, that is,
$$R_{\alpha\beta\mu\nu}^{*} =  R_{\alpha\beta}{}^{*}_{\mu\nu}$$
occurs only in the Einstein spaces where
$$R_{\mu\nu} = \frac{R}{4} g_{\mu\nu}.$$

\tocless\subsection{Conformal transformation}

A conformal transformation is the map from the metric
$g_{\mu\nu}(x)$ into $\tilde{g}_{\mu\nu}(x)$ defined by
$$\tilde{g}_{\mu\nu}\,(x^\alpha) = \Omega^{2}\,(x^\alpha)\, g_{\mu\nu}\,(x^\alpha),$$
where $\Omega^2(x^{\alpha})$ is an arbitrary function. Then,
$$\tilde{g}^{\mu\nu}\,(x^\alpha) = \Omega^{-\,2} \,(x^\alpha)\, g^{\mu\nu}\,(x^\alpha),$$
which yields for the affine connection
$$ \tilde{\Gamma}^{\alpha}_{\mu \nu} = \Gamma^{\alpha}_{\mu \nu}+ \frac{1}{\Omega} \left( \Omega_{, \mu} \delta^{\alpha}_{~\nu}+
\Omega_{, \nu} \delta^{\alpha}_{~\mu}- \Omega_{, \lambda} g^{\alpha \lambda}g_{\mu \nu} \right) $$
and the curvature tensor
$$ {\tilde{R}^{\alpha\beta}}_{~~\mu\nu} = \Omega^{-\,2} \,{R^{\alpha\beta}}_{\mu\nu} - \frac{1}{4} {\delta^{[\,\alpha}}_{[\,\mu} \, {M^{\beta\,]}}_{\nu\,]} $$
where $ {M^{\alpha}}_{\beta} \equiv 4 \,\Omega^{-\,1} \, (\Omega^{-\,1})_{,\,\beta;\,\lambda} g^{\alpha\lambda} - 2 \, (\Omega^{-\,1})_{,\,\mu} \,(\Omega^{-\,1})_{,\,\nu} \,g^{\mu\nu} \delta^{\alpha}_{\beta}$. The squared brackets mean anti-symmetrization. Contracting the indices of the above expression we obtain the transformations of the Ricci tensor and the scalar curvature $R$, respectively
$$ {\tilde{R}^{\alpha}}_{~\mu} = \Omega^{-\,2} \,{R^{\alpha}}_{\mu} - \frac{1}{2}\,   {M^{\alpha}}_{\mu} - \frac{1}{4} M
\delta^{\alpha}_{\mu} $$
and
$$ \tilde{R} =  \Omega^{-\,2} [\,R + 6\,\Omega^{-\,1} \, \Box\,\Omega \,].$$
Finally, collecting these transformations we find the invariance of the Weyl conformal tensor
$$ {\tilde{W}^{\alpha}}_{~~\beta\mu\nu} = {W^{\alpha}}_{\beta\mu\nu}.$$

\tocless\subsection{Projection tensor}

Let $v^{\mu}$ represent a normalized congruence in a given space-time endowed with a metric $ g_{\mu\nu}$. The corresponding projector  $h_{\mu \nu}$ is defined by
\begin{equation}
\label{2agosto2014}
h_{\mu \nu}  \equiv g_{\mu \nu}-v_{\mu} \, v_{\nu}.
\end{equation}
It projects quantities defined in space-time in the rest-frame of $v^{\mu}.$ Indeed, it satisfies:

\[
h_{\alpha \beta} h^{\beta}_{~\nu} = h_{\alpha \nu}, \quad \mbox{and} \quad h_{\alpha \beta}v^{\beta}=0.
\]
Note that $h_{\mu \nu}$ is a symmetric tensor $ h_{\mu \nu}= h_{\nu \mu}.$ We can then write the distance from two arbitrary points $ P $ and $ Q$ as
$$ds^2(P,Q)=g_{\mu \nu}dx^{\mu}dx^{\nu}=h_{\mu \nu}dx^{\mu}dx^{\nu}+(v_{\mu}dx^{\mu})^2.$$

\tocless\subsection{Kinematical parameters}
Let us consider a congruence of curves $\Gamma$ that can be assigned by parameters $s_i$ on each curve that will be called its proper time. We denote the displacement vector $\vec{z}$, the vector connecting two curves of $\Gamma$ assigned by the same value of $s$.

Defining the tensor $Q_{\alpha \beta} \equiv h_{\alpha}{}^{\mu} h_{\beta}{}^{\nu}\, v_{\mu ; \nu}$ and decomposing it into its irreducible components, we set
$$Q_{\alpha \beta}= \frac{\theta}{3}\ h_{\alpha \beta}+ \sigma_{\alpha \beta}+\omega_{\alpha \beta},$$
where
$$\theta \equiv h^{\alpha \lambda}v_{\alpha ; \lambda}= v^{\alpha}_{~\ ; \alpha}$$
is the expansion factor, the traceless symmetric part
$$\sigma_{\alpha \beta} \equiv \frac{1}{2}\ h^{\mu}_{~(\alpha}\ h_{\beta )}^{~\lambda} v_{\mu ; \lambda}- \frac{1}{3}\ \theta h_{\alpha \beta}$$
is the shear tensor and the anti-symmetric part
$$\omega_{\alpha \beta} \equiv \frac{1}{2}\ h_{[\alpha}^{~\mu}\ h_{\beta ]}^{~\lambda} v_{\mu ; \lambda}$$
is the vorticity tensor.

The parentheses denote symmetrization. It follows that $\sigma_{\mu \nu}v^{\mu} = 0$ and $\omega_{\mu \nu}v^{\mu} = 0$. We can then write
$$v_{\mu \, ;\, \nu} = Q_{\mu\nu} + a_{\mu} \, v_{\nu}= \frac{\theta}{3}\ h_{\mu\nu}+ \sigma_{\mu\nu}+\omega_{\mu\nu} + a_{\mu} \, v_{\nu},$$
where the acceleration is defined by $ a_{\mu} \equiv \dot{v}_{\mu} = v_{\mu;\nu}v^{\nu}$. From independent projections of Eq.\ (\ref{sec_der}), we can obtain the evolution equations for the kinematical quantities\cite{nov_bit_sal}. In particular, the evolution of the expansion coefficient $\theta$ is given by the Raychaudhuri equation
$$\dot\theta + \frac{\theta^2}{3} + 2(\sigma^2-\omega^2) - a^{\alpha}{}_{;\alpha}=R_{\mu\nu}v^{\mu}v^{\nu}.$$
Particular situation for this equation will be used afterwards when we discuss analogue models of gravity.

\tocless\subsection{The energy-momentum tensor}
The energy-momentum tensor can be written in terms of a fluid by the choice of a particular frame represented by an observer endowed with a four-velocity field $v^{\mu}$, yielding the decomposition
\[
T_{\mu\nu}=\rho \, v_{\mu} \,v_{\nu}-p \,h_{\mu\nu}+q_{(\mu} \,v_{\nu)}+\pi_{\mu\nu}\,,
\]
where the ten independent quantities $\rho,p,q^{\alpha}$ and $\pi_{\alpha \beta}$ are obtained through the projections of $T_{\mu\nu}$ onto $v^{\alpha}$ and the space orthogonal to it.

Explicitly, the scalars $\rho$ and $p$ (energy density and pressure) are defined by $\rho=T^{\alpha\beta} \, v_{\alpha} \,v_{\beta}$ and $p=-\frac{1}{3} \,h_{\alpha\beta} \,T^{\alpha\beta}$, the heat flux is $q^{\alpha }=h^{\alpha\beta} \, v^{\gamma}\, T_{\beta\gamma}$, and the traceless symmetric anisotropic pressure is $\pi^{\alpha\beta} = h^{\alpha\mu} \, h^{\beta\nu} \, T_{\mu\nu}+p \,h^{\alpha\beta}$.

\tocless\subsection{The formula of the determinant}

The determinant of the matrix representation of a mixed tensor $\textbf{T}=T^{\alpha}{}_{\beta}$ may be calculated through its characteristic polynomial due to the Cayley-Hamilton theorem:

\begin{equation}
\label{determinant}
\det\textbf{T} = -\frac{1}{4}\left[{\rm Tr} (\textbf{T}^{4}) - \frac{4}{3} \,{\rm Tr} (\textbf{T}) \, {\rm Tr} (\textbf{T}^{3}) - \frac{1}{2} \, \left({\rm Tr} (\textbf{T}^{2})\right)^{2} + \left({\rm Tr} (\textbf{T})\right)^{2} \, {\rm Tr} (\textbf{T}^{2}) - \frac{1}{6} \, \left( {\rm Tr} (\textbf{T}) \right)^{4}\right].
\end{equation}

\tocless\subsection{Dirac spinors and the Clifford algebra}

We will deal here with fields $\Psi$ that are four-components Dirac spinors\cite{copenhagen2}. The vector and axial currents are constructed with $\Psi$  namely,
\[
J^{\mu}\equiv \overline{\Psi} \gamma^{\mu}  \Psi, \quad {\rm and} \quad  I^{\mu}\equiv \overline{\Psi} \gamma^{\mu} \gamma_{5} \Psi.
\]

For completeness we recall $\overline{\Psi} \equiv  \Psi^{+} \gamma^{0}$, where $\Psi^{+}$ is the complex conjugate of $\Psi$. The Clifford algebra is the algebra of the Dirac matrix $\gamma_{\mu}$ defined by its basic property
\begin{equation}
\gamma_{\mu} \, \gamma_{\nu} + \gamma_{\nu} \, \gamma_{\mu} = 2 \eta_{\mu\nu} \, {\bf 1}
\label{11agosto20}
\end{equation}
where ${\bf 1}$ is the identity of the Clifford algebra. We use the following notation
\[
\nonumber
\gamma^{0} =  \left(
\begin{array}{cc}
I_{2}&0\\
0&-\,I_{2}
\end{array}
\right), \quad
\gamma_{k} =  \left(
\begin{array}{cc}
0&\sigma_{k}\\
-\sigma_{k}&0
\end{array}
\right), \quad
\gamma_{5} =  \left(
\begin{array}{cc}
0&I_{2}\\
I_{2}&0
\end{array}
\right),
\]
where $I_2$ represents the $2\times2$ identity matrix and $\sigma_k$ are the Pauli matrices, which satisfy $\sigma_{i} \, \sigma_{j} = i \, \epsilon_{ijk} \, \sigma_{k} + \delta_{ij}$, and we set

\[
\sigma_{1} =  \left(
\begin{array}{cc}
0&1\\
1&0
\end{array}
\right), \quad
\sigma_{2} =  \left(
\begin{array}{cc}
0&-i\\
i&0
\end{array}
\right), \quad
\sigma_{3} =  \left(
\begin{array}{cc}
1&0\\
0&-1
\end{array}
\right).
\]

The $\gamma_{5}$-matrix anti-commutes with all $\gamma_{\mu}$ and is defined in terms of them by
$$\gamma_{5} = \frac{i}{4 !} \, \eta^{\alpha\beta\mu\nu} \, \gamma_{\alpha} \gamma_{\beta} \gamma_{\mu} \gamma_{\nu} = i
\gamma_{0} \gamma_{1} \gamma_{2} \gamma_{3},$$
where the second equality is valid in a Cartesian coordinate system in the Minkowski background. The $\gamma_{5}$ is Hermitian and the others $\gamma_{\mu}$ obey the self-adjoint relation
\[
\gamma_{\mu}^{+} = \gamma^{0} \gamma_{\mu} \gamma^{0}.
\]

Any spinor can be decomposed into its left- and right-handed parts through the identity

\begin{equation}
\Psi =  \Psi_{L} + \Psi_{R} = \frac{1}{2} ({\bf 1} + \gamma_{5})\Psi + \frac{1}{2} ({\bf 1} - \gamma_{5})\Psi
\label{domingo5}
\end{equation}

Then
$$\overline{\Psi}_{L} \, \Psi_{L} = 0, \quad {\rm and} \quad \overline{\Psi}_{R} \, \Psi_{R} = 0.$$

\tocless\subsection{Pauli-Kofink identity}

The properties needed to analyze nonlinear spinors are contained in the Pauli-Kofink (PK) relation. These are identities that establish a set of relations concerning elements of the four-dimensional Clifford algebra. The main property states that, for any element $Q$ of this algebra, the PK relation ensures the validity of the identity:
\begin{equation}
(\overline{\Psi} Q \gamma_{\lambda} \Psi) \gamma^{\lambda} \Psi = (\overline{\Psi} Q \Psi)  \Psi  - (\overline{\Psi} Q \gamma_{5} \Psi) \gamma_{5} \Psi.
\protect\label{H5}
\end{equation}
for $Q$ equal to ${\bf 1}$, $\gamma^{\mu}$, $\gamma_{5}$, $\gamma^{\mu} \gamma_{5}$ and $\sigma^{\mu\nu}\equiv(\gamma^{\mu} \gamma^{\nu}-\gamma^{\nu} \gamma^{\mu})/2$. As a consequence of this relation we obtain two extremely important facts: (i) the norm of the currents $J_{\mu}$ and $I_{\mu}$ have the same value and opposite sign; (ii) vectors  $J_{\mu}$ and $I_{\mu}$ are orthogonal. Thus, $J_{\mu}$ is a time-like vector and $I_{\mu}$ is space-like.

Pauli-Kofink formula also implies some identities which will be used later on:
\begin{subequations}
\label{27dez915}
\begin{eqnarray}
J_{\mu} \, \gamma^{\mu} \, \Psi &\equiv&  ( A  + i B \gamma_{5}) \, \Psi,\\
I_{\mu} \, \gamma^{\mu} \, \gamma_{5} \,  \Psi &\equiv& -  ( A  + i B \gamma_{5}) \, \Psi, \\
I_{\mu} \, \gamma^{\mu} \, \Psi &\equiv&  ( A  + i B \gamma_{5}) \, \gamma_{5}  \Psi, \\
J_{\mu} \, \gamma^{\mu} \, \gamma_{5} \, \Psi &\equiv& - ( A  + i B \gamma_{5}) \, \gamma_{5} \Psi,
\end{eqnarray}
\end{subequations}
where $A \equiv  \overline{\Psi} \, \Psi$ and   $B \equiv i \overline{\Psi} \, \gamma_{5} \Psi.$ Note that both quantities $A$
and $B$ are real.

\tocless\subsection{Dirac dynamics}

We shall deal with two dynamics for the spinor fields: a linear and a non-linear. For the linear case we take Dirac theory
\begin{equation}
i\gamma^{\mu} \partial_{\mu} \, \Psi - \mu \, \Psi = 0,
\label{domingo1}
\end{equation}
where $\mu$ is the mass. We use the conventional units where $\hbar = c = 1$. The corresponding Lagrangian is
\begin{equation}
L = L_{D} - \mu \,\bar{\Psi} \, \Psi  = \left( \frac{i}{2} \bar{\Psi} \gamma^{\mu} \partial_{\mu} \Psi - \frac{i}{2} \partial_{\mu} \bar{\Psi} \gamma^{\mu} \Psi\right) - \mu \,\bar{\Psi} \, \Psi
\label{20mar17}
\end{equation}
Note that on-mass-shell Dirac Lagrangian vanishes $ L(\mbox{oms}) = 0$. From the decomposition in a right $\Psi_{R}$ and left-handed $\Psi_{L}$ helicity it follows that the mass-term mix both
helicities:
\begin{equation} i\gamma^{\mu} \partial_{\mu} \, \Psi_{L} - \mu \, \Psi_{R} = 0,
\label{20mar1750}
\end{equation}
\begin{equation} i\gamma^{\mu} \partial_{\mu} \, \Psi_{R} - \mu \, \Psi_{L} = 0.
\label{20mar1751}
\end{equation}

\tocless\subsection{Heisenberg dynamics}

The Heisenberg self-interaction Lagrangian:
\begin{equation}
L = L_{D} - V(\Psi). \protect\label{H1}
\end{equation}
The potential $V$ is constructed with the two scalars $A$ and $B:$
\begin{equation}
V = s \left( A^{2} +  B^{2} \right),
\protect\label{H3}
\end{equation}
where $s$ is a real parameter of dimension $[{\rm length}]^{2}.$ The corresponding equation of
motion is
\begin{equation}
i\gamma^{\mu} \partial_{\mu} \, \Psi^{H} - 2 s \, (A + i  \, B \gamma_{5}) \, \Psi^{H} = 0.
\label{domingo2}
\end{equation}
Correspondingly, we have
\begin{equation}
i \partial_{\mu} \,\overline{\Psi}^{H} \, \gamma^{\mu} + 2 s \, \overline{\Psi}^{H} \, (A + iB \gamma_{5}) = 0
\label{21mar712}
\end{equation}

The Heisenberg potential $V_{H}$ can be written in an equivalent and more suggestive form in terms of the associated currents $J_{\mu}$ and $I_{\mu}.$ As a direct consequence of Pauli-Kofinki identities, Heisenberg potential $V$ is nothing but the norm of the four-vector current $J^{\mu},$ that is $A^{2} + B^{2} = J^{\mu} \, J_{\mu}$. Note that on-mass-shell, Heisenberg Lagrangian takes the value of its potential $L(\mbox{oms}) = V_{H}$.

\tocless\subsection{Gauge invariance}

The dynamics displayed by both Dirac and Heisenberg equations of motion are invariant under the map
\begin{equation}
\widetilde{\Psi} = S \, \Psi,
\label{1junho715}
\end{equation}
where $S$ is a unitary matrix satisfying $S^{-1} S_{, \mu} = c_{\mu} \, I$. From Noether theorem this imply that the current $J_{\mu}$ is conserved. When the transformation $S$ is space-time dependent one has to introduce a modification on the derivative as much the same as it occurs for tensors in arbitrary coordinate transformation when a covariant derivative is defined. We shall deal with this spinor covariant derivative latter on.

\tocless\subsection{Chiral invariance}

Chiral transformation is defined by the map
$$ \Psi^{'} = \gamma_{5} \, \Psi.$$
Dirac equation is invariant under this map only for massless neutrino equation. On the other hand, Heisenberg equation is invariant under chirality. Indeed, we have, for the conjugate spinor:
$$ \overline{\Psi}^{'} = -  \overline{\Psi} \gamma_{5},$$
which implies
$$ A^{'} = - \, A, \quad {\rm and} \quad B^{'} = - \, B $$
consequently the Lagrangian remains the same. Although the constant $s$ is not a ``mass", it provides the similar mixing of Heisenberg spinors $\Psi_{L}$ and  $ \Psi_{R}.$  Indeed, we have
\begin{equation}
i\gamma^{\mu} \partial_{\mu} \, \Psi_{L} - 2 s \, (A - iB\gamma_5)  \,
\Psi_{R} = 0 \label{21mar7}
\end{equation}
\begin{equation}
i\gamma^{\mu} \partial_{\mu} \, \Psi_{R} - 2 s \, (A + iB\gamma_5 ) \,
\Psi_{L} = 0 \label{21mar71}
\end{equation}

\tocless\subsection{Current Conservation}

Let us introduce an arbitrary parameter $\epsilon$ in the Heisenberg equation for $\Psi$
\begin{equation}
i\gamma^{\mu} \partial_{\mu} \, \Psi - 2 s \, (A + i \, \epsilon \, B \gamma_{5}) \, \Psi = 0
\label{domingo2e}
\end{equation}
and for $\overline\Psi$
\begin{equation}
i \partial_{\mu} \,\overline{\Psi} \, \gamma^{\mu} + 2 s \, \overline{\Psi} \, (A + i \,\epsilon \,B \gamma_{5}) = 0
\label{21mar7121}
\end{equation}

We will now show that the vector current $ J^{\mu}$ is conserved for any value of $\epsilon$, but the axial vector current $ I^{\mu}$ is conserved only for the Heisenberg equation when $ \epsilon = 1.$ Thus, multiplying by $\overline{\Psi}$ Eq.\ (\ref{domingo2e}) and by $\overline\Psi$ Eq.\ (\ref{21mar7121}) and adding these results, it follows that indeed
$$ \partial_{\mu} \, J^{\mu} = 0. $$
Let us now do the similar procedure by multiplying further by a $\gamma^{5}. $ We have, respectively:
\begin{equation}
i \,\overline{\Psi} \,  \gamma^{\mu} \gamma_{5} \partial_{\mu} \Psi= 2s \,i \, ( 1 - \epsilon ) A B. \label{12agosto1}
\end{equation}
and
\begin{equation}
i \,\partial_{\mu} \overline{\Psi} \,  \gamma^{\mu} \gamma_{5} \Psi = 2s \,i \, ( 1 - \epsilon ) A B. \label{12agosto2}
\end{equation}
It then follows the result that the axial current satisfies the equation

\begin{equation}
\partial_{\mu} I^{\mu} = 4 s \, ( 1 - \epsilon ) \, AB.
\label{12agosto3}
\end{equation}
which shows that only in the case of Heisenberg choice $ ( \epsilon = 1 )$ the axial current is conserved.

\tocless\subsection{Weyl Integrable Space (WIS)}

We shall see that it is possible to provide a geometric formulation of the non-relativistic quantum mechanics in de Broglie-Bohm approach. For this we need some mathematical properties of the 3-D WIS (which is called the quantum WIS or Q-WIS). Contrary to the Riemannian geometry, which is completely specified by a metric tensor, the Weyl space defines an affine geometry. This means that the covariant derivative which is defined in terms of a connection $ \Gamma^{m}_{i k}$ depends not only on the metric coefficients but also on a vector field $f_a(x)$. latin indeces run from $1$ to $3$.

For instance, the covariant derivative of a given vector $X_{a}$ is
\begin{equation}
X_{a \, ; b} = X_{a \, , \, b} - \Gamma^{m}_{a b} \, X_{m}.
\end{equation}

The non-metricity of the Weyl geometry implies that rulers, which are standards of length measurement, changes while we transport it
by a small displacement $\dd x^{i}.$ This means that a ruler of length $l$ will change by an amount
\begin{equation}
\delta \, l = l \, f_{a} \, \dd x^{a}.
\end{equation}

As a consequence, the covariant derivative of the metric tensor does not vanishes as in a Riemannian geometry but instead it is given by
\begin{equation}
g_{a b  \,; \, k} = f_{k} \, g_{a b },
\end{equation}
which is the analogous of the 4-D equation (\ref{no_met}). Using Cartesian coordinates, it follows that the expression for the connection in terms of the vector $f_{k}$ takes the form
\begin{equation}
\label{conn}
\Gamma^{k}_{a b} = - \, \frac{1}{2} \, \left( \delta^{k}_{a} \,f_{b} +  \delta^{k}_{b} \, f_{a} - g_{a b} \,  f^{k} \right).
\end{equation}

The WIS case is provided by the condition that the vector $f_{i}$ is a gradient of a function, i.e. $f_i\equiv f_{,\, i}$. This property ensures that the length does not change its value along a closed path
\begin{equation}
\oint \, \dd l = 0.
\end{equation}

As a matter of convenience, we define
\begin{equation}
f = -4 \, \ln \, \Omega.
\end{equation}

Then the 3-D Ricci tensor is constructed with the affine connection (\ref{conn}) and is explicitly given by
\[
R_{ij}=2\frac{\Omega_{,ij}}{\Omega}-6\frac{\Omega_{,i}\Omega_{,j}}{\Omega^2}+2g_{ij}\left[\frac{\nabla^2 \Omega}{\Omega} +\frac{\vec{\nabla} \Omega .\vec{\nabla} \Omega}{\Omega^2}\right].
\]
The scalar of curvature $R$ becomes
\begin{equation}
R =  \,8 \, \frac{\nabla^{2} \Omega}{\Omega} \quad.
\end{equation}

\newpage

\section{Introduction}
The first attempts to interpret the motion of the bodies in classical mechanics geometrically, i.e. in terms of geodesics, refer to the works of M. de Maupertuis \cite{Maupertuis}, in which the principle of least action was formulated for the first time. The appearance of an effective geometry describing the path of bodies becomes even more evident with Jacobi's formulation of Maupertuis' principle \cite{lanczos}.

For the electromagnetism, such geometrical description appeared only in 1923, when Gordon \cite{gordon} made a seminal suggestion to treat the propagation of electromagnetic waves in a moving dielectric by a modification of the metric structure of the background. He showed that the acceleration of the photons inside a dielectric medium can be eliminated by a procedure similar to the one made in GR, that is, by changing the rules of distances in space and time in the interior of the dielectric, at least in what concerns its propagation. He proposed to interpret the propagation of electromagnetic waves inside a moving dielectric as geodesics not in the standard background geometry $ \gamma_{\mu\nu} $, but instead in the effective metric
\begin{equation}
g^{\mu\nu} = \gamma^{\mu\nu} + ( \epsilon \mu - 1 ) \,v^{\mu} \,v^{\nu},
\label{251211}
\end{equation}
where $ \epsilon$ and $ \mu$ are constant parameters that characterize the dielectric and $v^{\mu}$ is the four-velocity of the material under consideration (which is not necessarily constant). Later, it was recognized that this interpretation could be used to describe nonlinear structures even when $\epsilon$ and $\mu $ depend on the intensity of the electromagnetic field \cite{nov01} or more complicated functions of the field strengths as we shall see later on. In all these cases, the causal cone,
describing the propagation of photons is associated to an effective metric, that does not coincide with the null-cone of the background metric. The origin of this modification is due to the presence of the moving dielectric, which changes the paths of the electromagnetic waves inside this medium and shall be described in next sections. Before this, let us introduce a more general question: could such particular description of the electromagnetic waves in moving dielectrics be generalized for other cases, in which accelerated paths of arbitrary bodies submitted to other kind of forces would be described as geodesic motions in an associated metric? We shall see that the answer is affirmative and such kinematical map depends only upon the acceleration of the body.

This method allows us to undertake the geometrization of any force in the sense that arbitrary accelerated body in a given metric substratum $\gamma_{\mu\nu}$ is equivalently described as geodesic motion in an effective geometry $\widehat{g}_{\mu\nu}$. We shall call this proposal as the principle of Metric Relativity (MR). When the acceleration is produced by gravity this procedure led to the geometrization of the gravitational field as it was done by GR. According to the MR, the effects of the particle acceleration caused by any kind of force can be described as geodesics on the associated metric $\widehat{g}_{\mu\nu}.$  In the case of gravity this should be identified with Einstein's approach.

\section{The principle of metric relativity}
The principle of metric relativity (MR) states that the motion of an arbitrarily accelerated body in a flat Minkowski space-time can be equivalently described as free of any force and following a geodesic in an associated geometry \cite{nov_bit_drag}, which we call dragged metric (DM). Although the analysis we present in this section can be dealt with any underlying metric structure, that is, flat or curved space-time, we limit this section to the case of process occurring in the flat Minkowski background. Its generalization for arbitrary curved background is a direct task and we will present some examples of it in a later section.

Let us start with the case in which the acceleration vector $ a_{\mu}$ is the gradient of a function, that is, the force acting on the body under observation comes from a potential. We set
\begin{equation}
a_{\mu} = \partial_{\mu}\Psi.
\label{31dez152}
\end{equation}

Using the freedom in the definition of the four-vector $v^{\mu}$ of the body, we set $\eta_{\mu\nu}v^{\mu}v^{\nu} = 1$ and the acceleration is orthogonal to the velocity $ a_{\mu} \, v^{\mu} =0.$

Motivated by Gordon's approach, where the effective metric depends on the external vector field associated to the dielectric motion, here we consider that the associated DM, for any congruence of curves $\Gamma(v)$, takes the form
\begin{equation}
\widehat{g}^{\mu\nu} = \eta^{\mu\nu} + \beta \, v^{\mu} \,v^{\nu}.
\label{13junho1}
\end{equation}
The covariant expression is
\begin{equation}
\widehat{g}_{\mu\nu} = \eta_{\mu\nu} - \frac{\beta}{ ( 1 + \beta)} \, v_{\mu} \,v_{\nu},
\label{13junho12}
\end{equation}
in order to have $\delta^{\mu}_{\nu}=\widehat g^{\mu\alpha}\widehat g_{\alpha\nu}$. The covariant derivative of an arbitrary vector $ S^{\mu}$ in this metric is defined in the standard way
$$ S^{\alpha}{}_{; \, \mu} = S^{\alpha}{}_{, \, \mu} + \widehat{\Gamma}^{\alpha}_{\mu\nu} \, S^{\nu}, $$
where the Christoffel symbol is constructed with metric (\ref{13junho1}). We set $ \widehat{v}^{\mu} \equiv \widehat g^{\mu\nu}v_\nu=\sqrt{1+\beta}v^\mu$.
In order to identify this congruence generated by $\widehat{v}_{\mu}$ to the one associated with $v_{\mu},$ we require that $\sqrt{1+\beta}$ be constant along the motion, that is $ v^{\mu} \, \partial_{\mu} \beta=0$. The congruence $\Gamma(\widehat{v}) $ will be geodesic if
$$\widehat{v}_{\mu ,\nu} \, \widehat{v}^{\nu} - \widehat{\Gamma}^{\epsilon}_{\mu\nu} \, \widehat{v}_{\epsilon} \, \widehat{v}^{\nu} = 0. $$
Note that the hat symbol denotes objects defined in the metric $\widehat{g}^{\mu\nu}$.

The description of an accelerated curve in a flat space-time as a congruence of geodesics in the DM can be rewritten as
\begin{equation}
\left( v_{\mu , \nu} - \widehat{\Gamma}^{\epsilon}_{\mu\nu} \, v_{\epsilon} \right) \, v^{\nu} = 0.
\label{31dez1}
\end{equation}
Once the acceleration in the background is defined by $a_{\mu} = v_{\mu , \nu} \, v^{\nu} $ and using equation (\ref{31dez152}), the condition of geodesics in the DM takes the form
$$ \partial_{\mu} \Psi = \widehat{\Gamma}^{\epsilon}_{\mu\nu} \, v_{\epsilon} \, v^{\nu}.$$
The rhs can be written as
$$\widehat{\Gamma}^{\epsilon}_{\mu\nu} \, v_{\epsilon} \, v^{\nu} =\frac{1 + \beta}{2} \, v^{\alpha} \, v^{\nu} \,
\widehat{g}_{\alpha\nu , \mu}.$$

Using Eqs.\ (\ref{13junho1}) and (\ref{13junho12}), we obtain that the coefficients of the DM are given in terms of the acceleration
$$a_{\mu} + \frac{1}{2} \, \partial_{\mu} \ln (1 + \beta) = 0.$$
Thus, we have proved the following

\begin{theorem}
\label{drag_lemma}
Given a congruence of accelerated curves $ \Gamma(v) $ in Minkowski space-time driven by a potential,  $a_{\mu} = \partial_{\mu} \Psi,$  it is always possible to construct an associated DM of the form
\begin{equation}
\label{drag_metric}
\widehat{g}^{\mu\nu} =  \eta^{\mu\nu} + \beta \, v^{\mu} \,v^{\nu},
\end{equation}
such that  the paths of the curves become geodesics in this DM and
$$1 + \beta = e^{- 2 \Psi}.$$
\end{theorem}
Let us compare this method to the original proposal of the GR for the description of the motion of a body around the weak gravitational field of the sun: they provide the same metric. In other words, as far as we limit to weak fields, the motion of test particles in the framework of GR is the same as in the DM formulation: the accelerated paths are mapped into geodesics of the DM.

\vspace{0.50cm}
\textbf{The GR description}
\vspace{0.50cm}
\begin{itemize}
\item{Assume that only the $v^{0}$ component matters and the other terms $v^i$ are very small in comparison $ v^{0};$}
\item{The acceleration suffered by this velocity field is small and does not depend on time;}
\item{Define a metric $ g_{\mu\nu} \approx \eta_{\mu\nu} + \gamma_{\mu\nu};$ then GR states that the body follows a geodesic in this metric. At this regime, the only component of $\gamma_{\mu\nu}$ that enters in the geodesic equation is $ \gamma_{00} \equiv \Psi ;$}
\item{$\Psi$ satisfies the Laplace equation $\nabla^{2}\Psi = 0.$}
\end{itemize}

\vspace{0.50cm}
\textbf{The DM description}
\vspace{0.50cm}
\begin{itemize}
\item{Assume that only the $v^{0}$ component matters and the other terms $v^i$ are very small in comparison $v^{0};$}
\item{The acceleration suffered by this velocity field is small and does not depend on time;}
\item{The above Lemma states that the body follows a geodesic in the metric $ g^{\mu\nu} \approx \eta^{\mu\nu} + \beta \, v^{\mu} \, v^{\nu},$ which yields $\beta=\exp{(-2\,\Psi)}-1$ and $g_{00}=\Psi$ in the weak field limit;}
\item{$\Psi$ satisfies the Laplace equation $\nabla^{2}\Psi = 0.$}
\end{itemize}

In respect to the motion of individual bodies, we see that the distinction from DM to GR concerns the independence of the geometry on the actual motion of the body once in GR the geodesics are constructed in a universal metric.

\subsection{Kinematical parameters in the DM}
Let us evaluate the modification that the passage from $ v_{\mu}$ to $ \widehat{v}_{\mu}$ implies for the kinematical parameters. First, we set for the projector the formula (see Eq. \ref{2agosto2014})
\begin{equation}
\widehat{h}_{\mu\nu} = \widehat{g}_{\mu\nu} - \widehat{v}_{\mu} \, \widehat{v}_{\nu}.
\label{29jul20141}
\end{equation}
Thus $\widehat{h}_{\mu\nu} =  \, h_{\mu\nu}.$
For the expansion factor
$$ \widehat{\theta} = \frac{1}{\sqrt{- \widehat{g}}} \, \partial_{\mu} ( \, \sqrt{- \widehat{g}} \, \widehat{v}^{\mu}) $$
using
$$\sqrt{- \widehat{g}} = \frac{\sqrt{- \eta}}{ (1 + \beta)^{1/2}} $$
we obtain the value $\widehat{\theta} = \sqrt{ 1 + \beta} \, \theta.$

For the shear tensor, we get
$$ \widehat{\sigma}_{\mu\nu} = \sqrt{ 1 + \beta} \,\, \sigma_{\mu\nu} $$
and for the vorticity
$$ \widehat{\omega}_{\mu\nu} = \frac{\omega_{\mu\nu}}{\sqrt{ 1 + \beta}}.$$
This procedure of eliminating the acceleration, modify the remaining kinematical parameters only by a multiplicative factor.

Using the formula for the determinant of $\widehat{g}_{\mu\nu}$ and a convenient coordinate system, we can rewrite the DM as
$$ \widehat{g}_{\mu\nu} = \eta_{\mu\nu} - (1 + \widehat{g}) \, v_{\mu} \, v_{\nu}.$$
We can equivalently state that the potential of the external force measures the logarithm of the determinant of the DM, that is $\Psi = \ln\sqrt{-\widehat g}.$

\subsection{The curvature of the DM}
In the case the background metric is not flat or if we use an arbitrary coordinate system the connection is given by the sum of the corresponding background one and a tensor, that is
\begin{equation}
\widehat{\Gamma}^{\epsilon}_{\mu\nu} = \Gamma^{\epsilon}_{\mu\nu} + K^{\epsilon}_{\mu\nu}.
\end{equation}
In the case of the Minkowski background a direct calculation gives for the connection the form
\begin{equation}
\widehat{\Gamma}^{\epsilon}_{\mu\nu}  = K^{\epsilon}_{\mu\nu} = v^{\epsilon} \, ( a_{\mu}\, v_\nu + a_{\nu} \, v_{\mu} ) - a^{\epsilon} \, v_{\mu} \, v_{\nu},
\label{31dez17}
\end{equation}
where we have considered Lorentzian coordinates and the particular case where the congruence has no expansion, no shear and no vorticity, that is $ v_{\mu , \nu} = a_{\mu} \, v_{\nu}$ just to simplify the calculations. Then, $K^{\epsilon}_{\mu\epsilon} = a_{\mu}$. The contracted Ricci curvature has the expression
\begin{equation}
\widehat{R}_{\mu\nu} = a_{\mu , \nu} - a_{\mu} \, a_{\nu}  + (a^{\lambda} \, a_{\lambda} + a^{\alpha}{}_{, \alpha} ) \, v_{\mu} \, v_{\nu},
\end{equation}

Noting that $ \widehat{a}^{\mu} = a_{\nu} \,\widehat{q}^{\mu\nu} = a^{\mu} $ it follows that $ a^{\mu} \, a_{\mu} = a_{\mu} \, a_{\nu} \,\eta^{\mu\nu}= a_{\mu} \, a_{\nu} \,\widehat{q}^{\mu\nu} .$ The scalar of curvature $ \widehat{R} = \widehat{R}_{\mu\nu} \, \widehat{q}^{\mu\nu} $ is
\begin{equation}
\widehat{R} = ( 2 + b ) \, a^{\alpha}{}_{, \alpha}.
\end{equation}

These expressions can be rewritten in a covariant way by noting that
$$a_{\mu , \nu} \equiv a_{\mu ; \nu} + \widehat{\Gamma}^{\epsilon}_{\mu\nu} \, a_{\epsilon},$$
which yields
\begin{equation}
\widehat{R}_{\mu\nu} = a_{\mu \,;\, \nu} - a_{\mu} \, a_{\nu}  - (a^{\lambda} \, a_{\lambda} - a^{\alpha}{}_{;\,  \alpha} ) \, v_{\mu} \, v_{\nu},
\label{8jan2}
\end{equation}
and for the scalar curvature $\widehat{R}$ the form
\begin{equation}
\widehat{R} = ( 2 + b ) \, (a^{\alpha}{}_{; \alpha}- a^{\lambda} \, a_{\lambda} ).
\end{equation}

\subsection{Classical dynamics}
A curious consequence of the above calculations can guide us in the possible equations of motion for the classical long-range fields. Indeed, let us consider a congruence of curves $ x^{\mu} =  x^{\mu}(s)$ for an arbitrary parameter $s.$ Restricted to the case in which the congruence has no expansion $(\theta = 0),$  no shear $(\sigma_{\mu\nu} = 0)$ and no vorticity $( \omega = 0)$, the remaining component of the derivative of the velocity field, the acceleration, is given by the gradient of a function as $ a_{\mu} = \partial_{\mu} \Psi.$ Thus, we have $v^{\mu}{}_{, \nu} = \partial^{\mu} \Psi \, v_{\nu}.$ Using the equation of the propagation of the expansion
$$ \dot{\theta} + \frac{\theta^{2}}{3} + 2 \, \sigma^{2} - 2 \, \omega^{2} - a^{\alpha}{}_{; \alpha} = R_{\mu\nu} \, v^{\mu} \, v^{\nu} $$
in the Minkowski background it follows that $\nabla^{2} \Psi = 0$, which is the Laplace equation for the Newtonian potential.

In the next sections we present some applications of this procedure. We will apply, for instance, the approach of the construction of the DM to obtain in a new perspective the propagation of electromagnetic rays inside a moving dielectric and in the case of a scalar field. Before this, however, let us consider a general case of accelerated paths.

\section{Geometrizing accelerated paths}\label{geo_acc_path}
We turn now to the case in which the norm of the vector is not a constant of motion, that is $ N \equiv v_{\mu} \, v_{\nu} \, \eta^{\mu\nu} \neq 1.$ We restrict our analysis in this section to the case in which the congruence $ \Gamma(v) $ has no vorticity and set $v_{\mu} =\partial_{\mu} \chi.$ Then for the acceleration it follows
$$ a_{\mu} = v_{\mu , \nu} \, v^{\nu} = \frac{1}{2} \, \partial_{\mu}  N.$$

The associated DM is given again by the formula
\begin{equation}
\widehat{g}^{\mu\nu} =  \eta^{\mu\nu} + \beta \, v^{\mu}\,v^{\nu},
\label{8junho1}
\end{equation}
where we impose that the vector $ \widehat{v}_{\mu} = v_{\mu} $ is normalized: $\widehat{v}_{\mu} \, \widehat{v}^{\mu}
=\widehat{v}_{\mu} \, \widehat{v}_{\nu} \, \widehat{g}^{\mu\nu}= 1$. Then, it follows that
$$ 1 + \beta \, N = \frac{1}{N}.$$
The equation of the geodesics in the DM takes the form

\begin{equation}
\left( \widehat{v}_{\mu , \nu} - \widehat{\Gamma}^{\epsilon}_{\mu\nu} \, \widehat{v}_{\epsilon}
\right) \, \widehat{v}^{\nu} = 0,
\label{8junho3}
\end{equation}
which can be rewritten as
$$ \frac{1}{N} \, v_{\mu , \nu} \, v^{\nu} - \frac{1}{2} \, \widehat{g}_{\alpha\beta ,\mu} \,\widehat{v}^{\alpha} \, \widehat{v}^{\beta} = 0 $$
and, after a direct calculation, yields nothing but the previous condition
$$ a_{\mu} =  \frac{1}{2} \, \partial_{\mu} N.$$
This ends the proof of the following lemma:

\begin{theorem}
For any congruence $\Gamma(v)$ of accelerated curves in Minkowski space-time such that
$$ (v_{\mu , \nu} -  v_{\nu , \mu} ) \, v^{\nu} = 0 $$
and $v_{\mu} \, v_{\nu} \, \eta^{\mu\nu} = N$, it is always possible to construct an associated DM of the form
$$ \widehat{g}^{\mu\nu} =  \eta^{\mu\nu} + \beta \, v^{\mu} \,v^{\nu}, $$
such that  the paths of the curves become geodesics in this geometry. The parameters of the metric satisfy the condition $
N(1 + \beta \, N) = 1$.
\end{theorem}

We can then rewrite the metric under the form
\begin{equation}
\widehat{g}^{\mu\nu} =  \eta^{\mu\nu} + \frac{( 1 - N )}{N^{2}} \, v^{\mu}
\,v^{\nu}.
\label{29jul201417}
\end{equation}
Note that the origin of the acceleration is not taken into account. For a unified treatment of the body and the source of its acceleration we should look into the description of the field responsible for this acceleration. Could this field of force to be described in this space endowed with a DM too? We shall see that this is indeed possible.

\subsection{General case}
In the precedent sections we limited our analysis to the case in which the acceleration is given by a unique function. Let us now pass to more general situation. In order to geometrize any kind of force we must deal with a larger class of geometries that are constructed not only with the normalized velocity $ v^{\mu}$ but also in terms of its acceleration $ a^{\mu}.$ The most general form of DM that allows the description of accelerated bodies as true geodesics in a modified geometry has the form
\begin{equation}
\widehat{g}^{\mu\nu} =  \eta^{\mu\nu} + b \, v^{\mu} \, v^{\nu} + m \, a^{\mu} \, a^{\nu} + n a^{(\mu} \, v^{\nu)},
\label{301216}
\end{equation}
where we denoted $a^{(\mu} \, v^{\nu)}\equiv a^{\mu} \, v^{\nu}+ a^{\nu} \, v^{\mu}$. The three arbitrary parameters $ b, m, n$ are related to the three degrees of freedom of the acceleration vector. The corresponding covariant form of the metric is given by

\begin{equation}
\widehat{g}_{\mu\nu} =  \eta_{\mu\nu} + B \, v_{\mu} \, v_{\nu} + M \,
a_{\mu} \, a_{\nu} + N a_{(\mu} \, v_{\nu)},
\label{30121620}
\end{equation}
in which $ B, M, N$ are given in terms of $ b, m,n$ by
$$ B = - \, \frac{b \, ( 1 - m \, a^2 ) + n^{2} \,a^2}{(1 + b) \, ( 1 - m \, a^2) + n^{2} \, a^2}, $$
$$ M = \frac{n^{2}-m(1+b)}{(1 + b) \, (1 - m \,a^2) + n^{2} \, a^2},$$
$$ N = - \, \frac{n}{ (1 + b) \, (1 - m \,a^2) + n^{2} \, a^2},$$
where $a^2=-a_{\mu}a^{\mu}$.
In this case the equation that generalizes the geodesic condition\ (\ref{31dez1}) takes the form
\begin{equation}
a_{\mu} = \frac{1}{2} \, \left[ (1 + b) \, v^{\lambda} \, v^{\nu} + n \, a^{\lambda} \, a^{\nu}\right]  \,
(\widehat{g}_{\lambda\mu \, , \, \nu} + \widehat{g}_{\lambda\nu \, , \, \mu} - \widehat{g}_{\mu \nu \, , \, \lambda}).
\end{equation}
This equation can be cast in the following formal expression
\begin{equation}
a_{\mu} = -\fracc{b_{,\mu}}{2(1+b)} - n \, \omega_{\mu\nu} \, a^{\nu},
\end{equation}
where $\omega_{\mu\nu}\equiv v_{[\mu,\nu]}-a_{[\mu} \, v_{\nu]}$ is the vorticity tensor. Solving this equation for these functions provides the most general expression for any acceleration.

With these results, we can transform the path of any particle submitted to any kind of force as a geodetic motion in the dragged metric. We can understand this method as a relativistic extension of the d'Alembert principle, corresponding to all types of motion, i.e., the acceleration is geometrized through the DM approach.

\subsection{DM's unchange orthogonal directions}
\begin{theorem}
Let $Z^{\mu}$ be tangent to a geodesic curve in space-time embodied with metric $g^{\mu\nu}$ and consider $\widehat{g}^{\mu\nu}$ as the DM for a congruence of curves $v^{\mu}$. If vectors  $ Z^{\mu}$ and $ v^{\mu}$ are orthogonal in the sense of $g^{\mu\nu} $, then $ Z^{\mu}$  will also be geodesic in the DM $\widehat{g}^{\mu\nu}$.
\end{theorem}

The proof of this lemma is straightforward. Set $ Z^{\mu} Z^{\nu} g_{\mu\nu} = 1 $ and define $\Phi=Z_{\mu} v_{\nu} g^{\mu\nu}$. Consider $\widehat{Z}_{\mu} = \lambda \, Z_{\mu}$, then
$$ \widehat{Z}^{\mu} = \left( \eta^{\mu\nu} + b \, v^{\mu} \, v^{\nu} \right) \, \lambda \, Z_{\nu} = \lambda \, ( Z^{\mu} + b \, \Phi \, v^{\mu}).$$ Then, for $ \Phi = 0$ it follows that $\lambda = 1.$  $ \widehat{Z} ^{\mu} $ will be a geodesic in the DM under the condition $Z_{\mu \, , \nu} \, Z^{\nu} - \widehat{\Gamma}^{\varepsilon}_{\mu\nu} \, Z_{\varepsilon} \, Z^{\nu} = 0. $

Now, using the expression of the Christoffel symbol in terms of $\widehat g^{\mu\nu}$, we obtain
$$ \widehat{\Gamma}^{\varepsilon}_{\mu\nu} \, Z_{\varepsilon} \, Z^{\nu} = -  Z^{\alpha} \, Z_{\nu} \, \left[\frac{b \, v_{\alpha} \, v^{\nu}}{ 2 \, ( b + 1) } \right]_{, \, \mu} = 0.$$

Thus if $Z_{\mu}$ is geodesic in the metric $g_{\mu\nu}$ it will also be in the DM, which proves the lemma.

\subsection{Distinct accelerated particles in the same dragged metric}
In this section we face the problem to encounter other accelerated vector fields that satisfy the same requirements to follow a geodesic motion in the same DM that a previously given vector field. For it, consider an accelerated congruence $v^{\mu}$ in Minkowski spacetime following a geodetic motion in the metric (\ref{301216}). We have shown that $a_{\mu}=\partial_{\mu}\Psi-n\omega_{\mu\nu}a^{\nu},$ where $1+b=e^{-2\Psi}$. Note that the intrinsic properties (mass, charge etc.) of the particle represented by $v^{\mu}$ are contained in $\Psi$ and $n$.

Now, consider another four-vector $\tilde v^{\mu}$ such that it follows a geodetic motion in a DM

\begin{equation}
\label{drag_v_l}
\widehat q_{(\tilde v)}^{\mu\nu}=\eta^{\mu\nu} + \tilde b\,\tilde v^{\mu}\tilde v^{\nu}.
\end{equation}
For convenience, suppose that $\tilde v^{\mu}\tilde v_{\mu}=1$ and $\tilde v_{\mu,\nu}=\tilde a_{\mu}\tilde v_{\nu}$. This last assumption implies that $\tilde a_{\mu}=\partial_{\mu}\tilde\Psi$, where $\tilde\Psi$ contains all information about kinematical features of $\tilde v^{\mu}$. We then ask if it is possible that these congruences $v^{\mu}$ and $\tilde v^{\mu}$ could agree with the same metric structure. In principle, we suppose that $\tilde v^{\mu}$ lies on the plane generated by $v^{\mu}$ and $a^{\mu}$, i.e.,

\begin{equation}
\label{vlva}
\tilde v^{\mu}=p\,v^{\mu}+q\,a^{\mu},
\end{equation}
where $p$ and $q$ are arbitrary functions. As we set $\tilde v^{\mu}\tilde v_{\mu}=1$, this restricts the coefficients $p$ and $q$ to $p^2-q^2a^2=1$. Therefore, from the equation of motion for $\tilde v^{\mu}$, i.e., $\tilde v_{\mu,\nu}\tilde v^{\nu}=\tilde a_{\mu}$, we obtain

\begin{equation}
\label{alva}
\tilde a_{\mu}=(p\,\dot p +q\,p'+p\,q\,a^2)v_{\mu}+\left[2-p^2+\fracc{p}{q\,a^2}(p\,\dot p +q\,p')\right]a_{\mu}.
\end{equation}
where $\dot p\equiv p_{,\mu}v^{\mu}$ and $p'\equiv p_{,\mu}a^{\mu}$. Now we look for the case $\tilde a_{\mu}\propto a_{\mu}$ which implies that $p\dot p +qp'+pqa^2=0$. This assumption is made in order to answer if it is possible that different particles subjected to the same force on the background can follow a geodetic motion in the same DM. Under these considerations, Eq.\ (\ref{alva}) yields

\begin{equation}
\label{alva2}
\tilde a_{\mu}=2(1-p^2)a_{\mu}.
\end{equation}
Eq.\ (\ref{alva2}) implies that

\begin{equation}
\label{alvapsi}
\partial_{\mu}\tilde\Psi=2(1-p^2)(\partial_{\mu}\Psi- n\omega_{\mu\nu}a^{\nu}).
\end{equation}
Note that the potential $\tilde\Psi$ must satisfy this equation in order that the acceleration of $v^{\mu}$ be proportional to the acceleration of $\tilde v^{\mu}$. If they follow geodesics in the same geometry, then an extra proposition must be satisfied

\begin{equation}
\label{drag_vleqv}
\widehat q_{(\tilde v)}^{\mu\nu}=\widehat q^{\mu\nu}\Longrightarrow \tilde b\,\tilde v^{\mu}\tilde v^{\nu}= b\,v^{\mu}v^{\nu} + m\,a^{\mu}a^{\nu} + n\,a^{(\mu}v^{\nu)}.
\end{equation}
Substituting (\ref{vlva}) into (\ref{drag_vleqv}), we obtain

\begin{equation}
\label{eqcoef_ql_q}
\begin{array}{lcl}
p&=&\sqrt{\fracc{e^{-2\Psi}-1}{e^{-2\tilde\Psi}-1}},\\
q&=&n\left[(e^{-2\Psi}-1)(e^{-2\tilde\Psi}-1)\right]^{-1/2},\\
m&=&\fracc{n^2}{e^{-2\Psi}-1}.
\end{array}
\end{equation}
Remark that we still have an arbitrariness in the choice of $n$. Once it is fixed, then we can uniquely determine the DM in which $v^{\mu}$ and $\tilde v^{\mu}$ follow geodesics. This could be the case of particles in a electromagnetic field with the same charge-mass ratio.

All these analysis concerns the map that relates accelerated paths of a given background metric, Minkowski or not, into
geodesics of the modified DM. It is worth to note that this map is independent on the other kinematical parameters of the
congruence. Indeed, an arbitrary congruence of curves are characterized by 10 parameters: the acceleration $ a_{\mu},$ the expansion $\theta,$
 the shear $ \sigma_{\mu\nu}$ and the vorticity $ \omega_{\mu\nu}.$ The most general expression for the derivative of $ v_{\mu}$ is
provided by (see the definition of these quantities in Sec. [I])

$$ v_{\mu \, , \nu} = \frac{\theta}{3} \, h_{\mu\nu} + \sigma_{\mu\nu} + \omega_{\mu\nu} + a_{\mu} \, v_{\nu}. $$
in which we are considering the background metric as Minkowski just for simplicity. It can be easily generalized for
arbitrary Riemannian geometry. The condition to map an arbitrary accelerated curve to a geodesic in a DM is given by Eq. (\ref{31dez1}), where the important term to be analyzed is $\widehat{\Gamma}^{\varepsilon}_{\mu\nu} \, v_{\varepsilon} \, \widehat{v}^{\nu},$ that is $$\widehat{\Gamma}^{\varepsilon}_{\mu\nu} \, v_{\varepsilon} \, \widehat{v}^{\nu} = \left( \frac{1 + b}{2} \right) \, v^{\lambda} \, v^{\nu} \left(\eta_{\lambda\nu} - \frac{b}{(1 + b)} \, v_{\lambda} \, v_{\nu} \right)_{, \, \mu}. $$
Using the properties of the kinematical quantities entering in the expression of the derivative of $v_{\mu}$ it follows that indeed
only the acceleration survives in this formula.

\subsection{Accelerated particles in rotating frames in Minkowski space-time}
Let us consider a simple example concerning the acceleration of a body in flat Minkowski spacetime written in the cylindrical coordinate system $(t, r, \phi, z)$ to express the following line element
\begin{equation}
\label{ds2_gir}
ds^2=a^2[dt^2 - dr^2 - dz^2 + g(r)d\phi^2 + 2h(r)d\phi dt],
\end{equation}
where $a$ is a constant. We choose the following local tetrad frame given implicitly by the $1$-forms

\begin{equation}
\begin{array}{lcl}
\theta^0&=&a(dt + h d\phi),\\[1ex]
\theta^1&=&a dr,\\[1ex]
\theta^2&=&a \Delta d\phi,\\[1ex]
\theta^3&=&a dz,
\end{array}
\end{equation}
where we define $\Delta = \sqrt{h^2 - g}$. The only non-identically zero components of the Riemann tensor $R^{A}{}_{BCD}$ in the tetrad frame are

\begin{eqnarray}
R^{0}{}_{101}&=&\frac{1}{4a^2}\left(\frac{h'}{\Delta}\right)^2=R^{0}{}_{202}, \nonumber \\[2ex]
R^{0}{}_{112}&=&-\frac{1}{2a^2}\left(\frac{h''}{\Delta}-\frac{h'\Delta'}{\Delta^2}\right), \nonumber \\[2ex]
R^{1}{}_{212}&=&\frac{1}{a^2}\left[\frac{\Delta''}{\Delta}-\frac{3}{4}\left(\frac{h'}{\Delta}\right)^2\right],
\label{no_sta_ricci}
\end{eqnarray}
where $(\,'\,)$ means derivative with respect to the radial coordinate $r$. The
equations of general relativity for this geometry have two simple solutions that we shall analyze below.

In the case of $R^{A}{}_{BCD}$=0, we get $h'=0;\hspace{1cm} \Delta''=0.$ Solving these equations, we find $h\equiv \mbox{const.}$ and $\Delta\equiv\omega r,$ where $\omega$ is a constant. Therefore, Eq. (\ref{ds2_gir}) takes the form
$$ ds^2=a^2[dt^2 - dr^2 - dz^2 + (h^2-\omega^2r^2)d\phi^2 + 2h\,d\phi dt]. $$
This metric corresponds to the Minkowski one with a topological defect in the angular coordinate.

If we consider the observer field
$$ v^{\mu} =\frac{1}{a\sqrt{h^2-\omega^2r^2}} \delta^{\mu}_2,$$
with $h^2-\omega^2r^2>0$, describing a closed time-like curve, this path corresponds to an acceleration vector given by
$$ a_{\mu} = \left(0,  \frac{\omega^2r}{(h^2-\omega^2r^2)}, 0, 0 \right).$$
This means that $ a_{\mu} = \partial_{\mu} \Psi,$ where
$$ 2\Psi = - \,\ln (h^2-\omega^2r^2).$$

We are in a  situation similar to the previous lemma since the acceleration is a gradient. The parameter $b$ of the DM, given by the expression (\ref{ds2_gir}), can be written down as
$$ 1 + b = h^2-\omega^2r^2, $$
and for the DM, we get

$$\fracc{ds^2}{a^{2}} = \fracc{\omega^4r^4-\omega^2r^2h^2+1}{(h^2- \omega^2r^2)^2} dt^2 + d\phi^2 + \fracc{2h}{h^2-\omega^2r^2}d\phi \ dt -dr^2-dz^2.$$
Note that the accelerated closed time-like curves (CTC) in the original background are mapped into closed time-like geodesics (CTG) in the DM. If one calculates the scalar curvature of this metric it is possible to see that there exists a real singularity in $r=h/\omega$ and that the $0-0$ component of the metric changes sign at
$$ \omega^{2} \, r_{\pm}^{2} = \frac{h^{2} \,  \pm \, \sqrt{h^{4} - 4}}{2}.$$

\subsection{Accelerated particles in curved space-times}

In this section we will show that the above method of eliminating the external force by a change on the underlying geometry is a very general procedure and can be realized in arbitrary background. We will consider some specific examples in certain curved space-times that are solutions of the equation of motion of general relativity. We choose three well-known geometries: Schwarzschild, G\"odel universe and the Kerr solution in which accelerated particles have very peculiar properties. In these Riemannian manifolds we analyze some examples of accelerated paths that will be described as geodesics in the associated DM.

\subsubsection{Schwarzschild geometry}
We start with the Schwarzschild metric in the $ (t, r, \theta, \varphi)$ coordinate system
$$ds^2=\left(1-\frac{r_h}{r} \right) dt^{2} - \frac{1}{\left( 1 - \frac{r_h}{r}\right)} \, dr^{2} - r^{2} (d\theta^{2} + \sin^{2} \theta \, d\varphi^{2}),$$
and choose the trajectory described by the four-velocity
$$ v_{\mu} = \sqrt{1 - \frac{r_h}{r}} \, \delta_{\mu}^{0}.$$
The corresponding acceleration is
$$ a_{\mu} = \left( 0, - \, \frac{r_h}{2 (r^{2} - r \, r_h)}, 0, 0 \right). $$
In this case the acceleration is gradient of the function $\Psi$ given by
$$ \Psi = - \, \frac{1}{2} \,  \ln\left(1 - \frac{r_h}{r}\right),$$
where the DM coefficient is $b=-r_h/r,$ and, therefore, the DM takes the form
$$\widehat{ds}^2 =  dt^{2} - \frac{1}{( 1 - \fracc{r_h}{r})} \, dr^{2} - r^{2} (d\theta^{2} + \sin^{2} \theta \, d\varphi^{2}).$$
Straightforwardly, the only non-vanishing Ricci tensor components of this metric are
$$R^{1}_{1} = -2R^{2}_{2} = -2R^{3}_{3} =\, \frac{r_h}{r^{3}}.$$

According to GR the energy-momentum tensor corresponding to this DM has only an anisotropic pressure term (which is proportional to the components $C_{0\mu0\nu}$ of the conformal tensor of this metric). Such fluid has no physical meaning because the energy density is identically zero. In spite of the apparent singularity at $r=r_h$, all Debever invariants are finite everywhere except at the origin $r=0$.

\subsubsection{G\"odel\rq s geometry}
Let us now turn our analysis to the G\"odel geometry \cite{godel}. In the cylindrical coordinate system this metric is given by Eq.\ (\ref{ds2_gir}), where $a$ is a constant related to the vorticity $ a = 2/\omega^{2}$ and
$$ h(r) = \sqrt{2} \sinh^2 r,\quad \mbox{and}\quad g(r) = \sinh^2 r (\sinh^2 r-1).$$
For the sake of completeness we note the nontrivial contravariant terms of this
metric are
\begin{eqnarray}
g^{00} &=& \frac{1 - \sinh^{2} r}{a^{2} \,\cosh^{2} r}, \nonumber \\
g^{02} &=& \frac{\sqrt{2}}{a^{2} \, \cosh^{2} r}, \\
g^{22} &=& \frac{- \, 1}{a^{2} \, \sinh^{2} r \, \cosh^{2} r}.\nonumber
\end{eqnarray}

Previously, it was pointed out the a-causal properties of a particle moving into a circular orbit around the $z-$axis with four-velocity \cite{novellonamiemilia}
$$ v^{\mu} = \left(0, 0, \frac{1}{a \, \sinh r \, \sqrt{\sinh^{2} r
- 1}}, 0 \right). $$
This path corresponds to an acceleration given by
$$ a^{\mu} = \left(0,  \frac{\cosh r \, [2 \, \sinh^{2} r - 1]}{a^{2} \,  \sinh r \,
[\sinh^{2} r - 1]}, 0, 0 \right). $$
This means that $ a_{\mu} = \partial_{\mu} \Psi,$ where
$$ \Psi = - \,\ln (\sinh r \, \sqrt{\sinh^{2}r - 1}).$$

We are again in the situation where the acceleration is a gradient. Therefore, the parameter of the DM is given by
$$1 + b = \sinh^{2}r ( \sinh^{2}r - 1 ). $$
The DM has the following expression
$$ \frac{\widehat{ds}^2}{a^{2}} = \fracc{3-\sinh^4r}{(\sinh^2r-1)^2}dt^2 + d\phi^2 + 2\fracc{\sqrt{2}}{\sinh^2r-1}d\phi \ dt - dr^2 - dz^2.$$

From the analysis of geodesics in G\"odel's geometry the domain $ r < r_{c} $ where $
\sinh^{2}r_{c} = 1 $ separates the causal from the non-causal regions of the spacetime. This is related to the fact that a geodesic that reaches the value $r=0$ will be confined within the domain $ \Omega_{i}$ defined by the region $0<r<r_{c}$ \cite{novellosoarestiomno}. However, the gravitational field is finite in the region $ r = r_{c}.$ Nothing similar happens in the DM, since at $ \sinh^{2}r=1 $ there exists
a real singularity in the DM. Only the exterior domain is allowed. This means that for this kind of accelerated path in G\"odel geometry the allowed domain for the DM is precisely the whole non-causal region.

\subsubsection{Kerr metric}
Let us turn now to the DM approach in the case the background is the Kerr metric. In the Boyer-Lindquist coordinate system this metric is given by

$$ds^2 = \left(1-\fracc{2Mr}{\rho^2}\right) dt^2 - \fracc{\rho^2}{\Sigma}dr^2 - \rho^2d\theta^2 + \fracc{4Mra\sin^2\theta}{\rho^2} dt d\phi - \left[(r^2+a^2) \sin^2\theta + \fracc{2Mra^2\sin^4\theta}{\rho^2} \right] d\phi^2,$$
where $\Sigma=r^2+a^2-2Mr$ and $\rho^2 = r^2+a^2\cos^2\theta$. On the equatorial plane ($\theta=\pi/2$) consider the following vector field
$$ v^{\mu} = \left(0, 0, 0, \fracc{r}{\sqrt{-(r^2+a^2)^2+a^2\Sigma}} \right).$$
This path corresponds to an acceleration given by
$$ a_{\mu} = \left(0, -\fracc{r^3-Ma^2}{r^4+r^2a^2+2Mra^2}, 0, 0 \right).$$
This means that $ a_{\mu} = \partial_{\mu} \Psi,$ where
$$ 2\Psi = -\ln\left[-\left(r^2+a^2+\fracc{2Ma^2}{r}\right)\right].$$
Over again, we choose an accelerated path that can be represented by a gradient. The parameter $b$ is given by
$$ 1 + b = -\left(r^2+a^2+\fracc{2Ma^2}{r}\right),$$
and the DM, on the equatorial plane, takes the form
$$ \fracc{\widehat{ds}^2}{a^{2}}= \fracc{1}{(1+b)^2} \left(1 - \fracc{2M}{r} - b\fracc{4M^2a^2}{r^2} \right) dt^2 + d\phi^2 + \fracc{4Ma}{(1+b)r} dtd\phi - \fracc{r^2}{\Delta}dr^2.$$

These last cases (G\"odel and Kerr metrics) show a very curious and intriguing property: the accelerated CTC's at their respective metrics are transformed into geodesic curves, that is CTG's. Moreover, the DMs display a real singularity excluding the causal domain in both cases.

\newpage

\section{Accelerated photons in moving dielectrics}\label{accel_mov_diel}
The suggestion to transform an accelerated path in Minkowski background into a geodesics in an associated curved space comes from the analysis of the seminal work of Gordon \cite{gordon}. The description of this procedure becomes simpler and transparent using the Hadamard's method \cite{hadamard}. This will allow us to have a deep idea of Gordon\rq s approach and will enlighten more complex cases that appear when we treat nonlinear dielectrics. This method will be used also in the next sections for the analysis of photon propagation in nonlinear electrodynamics. A worthwhile application of the analysis of the modification of the geometry to deal with photon propagation inside the medium has been called ``mimic gravity". This means to find specific electromagnetic configurations such that the photon propagation is described by geodesics in geometries that are identical or at least similar to specific solutions of GR. We shall deal in particular with two cases that concern spatially homogeneous and isotropic cosmology and the rotating G\"odel universe.

\subsection{A short overview on Hadamard's method}\label{had_meth}
In the classical theory of hyperbolic partial differential equations, the analysis of propagation phenomena is done using a method that was developed by the French mathematician Jacques Hadamard and concerns the study of discontinuities across some given surfaces \cite{hadamard}. According to his method, wave propagation can be studied by following the evolution of wave fronts, through which the field is continuous, but some of its derivatives, in general, are not. The method is sufficiently general to be applied to in arbitrary dimensions $D$. Nevertheless, we will concentrate here only in $D=4$.

To be specific, let $\Sigma$ be a discontinuity hyper-surface, i.e. a definite boundary that clearly characterizes the field disturbance and its surrounding field configurations in space-time. Let $\Sigma$ be defined by the equation
\begin{equation}
\Sigma(x^{\mu})=0.
\end{equation}
This surface locally outlines two regions 1 and 2 in space-time. The discontinuity $[f]\Bigl\vert_{\Sigma}$ of a given space-time function $f(x^{\alpha})$ across $\Sigma$ is defined as the limit
\begin{equation}\label{DiscDef}
\left[ f(x) \right]\Bigl\vert_\Sigma = \lim_{\epsilon \to 0^+}
\bigl( f^{(1)}(x + \epsilon) - f^{(2)}(x - \epsilon)\bigr),
\end{equation}
where $f^{(1)}$ and $f^{(2)}$ are to be understood as the values of the function $f$ in the domains 1 and 2 respectively. The basic idea of the method is to investigate the compatibility between some assumed field discontinuities and the structure of a given partial differential equation. We start by assuming that the field itself is continuous, but this is not true for the highest order derivatives. Assuming, for instance that the first derivative is discontinuous across $\Sigma$ i.e.
\begin{equation}
\left[ f_{,\alpha} \right]\Bigl\vert_\Sigma\neq0,
\end{equation}
Hadamard showed that the differentials of the function $f$ in both domains $df^{(1)}=\partial_{\alpha}f^{(1)}dx^{\alpha}$ and $ df^{(2)} = \partial_{\alpha}f^{(2)}dx^{\alpha}$ have to be continuous. He then obtained the following crucial result
\begin{equation}
\left[ df \right]\Bigl\vert_\Sigma=\left[ \partial_{\alpha}f \right]\Bigl\vert_\Sigma dx^{\alpha}=0
\end{equation}
which means that $\left[ \partial_{\alpha}f \right]\Bigl\vert_\Sigma$ is orthogonal to the hypersurface. In other words, there exist a non-null scalar $\chi(x)$ such that
\begin{equation}
\left[ \partial_{\alpha}f \right]\Bigl\vert_\Sigma=\chi(x)k_{\alpha}
\end{equation}
where $k_{\alpha}\equiv\partial_{\alpha} \Sigma$ is the gradient of the surface $\Sigma = constant$ of discontinuity.

The type of discontinuity assumed depends on the order of the differential equation under investigation. In the context of second order equations, derivatives of order $D\geq 2$ will in general be discontinuous. When we have a discontinuity of this type we say that there exist a \textit{shock wave} on the surface $\Sigma$. These shocks are always present when the fields are described by hyperbolic PDE's. It is important to stress that the discontinuities are propagated along the surface $\Sigma$ in a specific way that is determined by the equations of motion.

At this point it is instructive to show how the method works in the context of Maxwell's linear electrodynamics. We assume that the electromagnetic field tensor $F_{\mu\nu}$ is such that
\begin{equation}
[F_{\mu\nu}]\Bigl\vert_{\Sigma} = 0,\quad \mbox{and} \quad [F_{\mu\nu,\lambda}]\Bigl\vert_{\Sigma} = f_{\mu\nu}(x)  k_{\lambda},
\protect\label{N8}
\end{equation}
where $f_{\mu\nu}$ is a non-null antisymmetric tensor. Evaluating the discontinuities of Maxwell's equations in vacuum
\begin{equation}
F^{\mu\nu}_{\phantom a\phantom a;\nu}=0\quad \mbox{and} \quad \stackrel{\ast}{F^{\mu\nu}}_{;\nu}=0
\end{equation}
across $\Sigma$ we obtain respectively
\begin{equation}
f^{\mu\nu}  k_{\nu} = 0.
\protect\label{N10}
\end{equation}
\begin{equation}
f_{\mu\nu}  k_{\lambda} + f_{\nu\lambda}  k_{\mu} + f_{\lambda\mu}
k_{\nu} = 0.
\protect\label{N13}
\end{equation}
Contracting Eq.\ (\ref{N13}) with $k^{\lambda}$ and using in Eq.\ (\ref{N10}), it results
\begin{equation}
g^{\mu\nu}k_{\mu} k_{\nu}   = 0
\protect\label{N16}
\end{equation}
that is
\begin{equation}
g^{\mu\nu}\frac{\partial \Sigma}{\partial x^{\mu}}\frac{\partial \Sigma}{\partial x^{\nu}}= 0
\protect\label{N161}
\end{equation}
Taking the derivative of this expression and noting that $ k_{\mu}$ is a gradient it follows
\begin{equation}
k_{\mu,\lambda} k^{\lambda} = 0. \protect\label{N17}
\end{equation}
This result shows that the waves propagate through null geodesics in the metric background.

\subsection{Light paths on moving dielectric: generalized Gordon method}
Apart from the Faraday tensor $F_{\mu\nu}$, let us define the skew-symmetric object $P_{\mu\nu}$ representing the electromagnetic field inside the material medium also. These tensors are expressed in terms of the field strengths $E^{\mu}$ and $H^{\nu}$ and field excitations $D^{\mu}$ and $B^{\mu}$ as follows

\begin{equation}
\nonumber
\begin{array}{lcl}
F_{\mu\nu} & \equiv & E_{\mu}  \, v_{\nu} - E_{\nu} \, v_{\mu} + \eta_{\mu\nu}{}^{\alpha\beta} \, v_{\alpha} \, B_{\beta},\\[2ex]
P_{\mu\nu} & \equiv & D_{\mu}v_{\nu} - D_{\nu} \, v_{\mu} + \eta_{\mu\nu}{}^{\alpha\beta} \, v_{\alpha} \, H_{\beta},
\end{array}
\end{equation}
where $v^{\mu}$ is a given four-vector comoving with the dielectric and $\eta_{\mu\nu}{}^{\alpha\beta}$ is the Levi-Civita tensor. We assume that the electromagnetic properties of the medium are characterized by the constitutive relations where
$$D_{\alpha}=\epsilon_{\alpha}{}^{\nu} \, E_{\nu}, \hspace{.5cm} B_{\alpha}=\mu_{\alpha}{}^{\nu} \, H_{\nu},$$
where $\epsilon_{\alpha}{}^{\nu} $ and $\mu_{\alpha}{}^{\nu}$ are arbitrary tensors that depend on $ ( E, H ).$ Consider Maxwell equations on dielectric media \cite{landau1971} with permittivity $\epsilon$ and permeability $\mu$ that characterize the dielectric:

\begin{equation}
\label{max_diel}
P^{\mu\nu}{}_{;\nu}=0,\quad \mbox{and} \quad ^*F^{\mu\nu}{}_{;\nu}=0.
\end{equation}
From now on, we take the background metric as flat Minkowski space-time and assume that $\mu\equiv\mu_0$ is a constant and $\epsilon=\epsilon(E)$, where $E \, \equiv \, \sqrt{-E_{\alpha} \, E^{\alpha}}$ and $E^{\alpha}$ is the electric field. It is straightforward to generalize these equations to arbitrary curved space-time. Indeed, suppose an observer with velocity $v^{\mu}$ co-moving with the dielectric and such that $v^{\mu}{}_{;\nu}=0$. Then, Eqs.\ (\ref{max_diel}) written in terms of the displacement vectors $D^{\mu}$ and $B^{\mu}$ become

\begin{equation}
\begin{array}{l}
D^{\mu}{}_{;\nu} \, v^{\nu} - D^{\nu}{}_{;\nu} \, v^{\mu}+\eta^{\mu\nu\alpha\beta} \, v_{\alpha} \, H_{\beta;\nu} \, = \, 0,\\[2ex]
B^{\mu}{}_{;\nu} \, v^{\nu}-B^{\nu}{}_{;\nu} \, v^{\mu}-\eta^{\mu\nu\alpha\beta} \, v_{\alpha} \, E_{\beta;\nu}\, =\,0.
\end{array}
\end{equation}

The projection with respect to $v^{\mu}$ yields the four independent non-linear equations of motion describing the electromagnetic field inside the dielectric medium:

\begin{equation}
\begin{array}{l}
\label{max_diel_proj}
\epsilon \, E^{\alpha}{}_{;\alpha}-\fracc{\epsilon' \, E^{\alpha} \, E^{\beta}}{E}E_{\alpha;\beta} \, = \,0,\\[2ex]
\mu_0 \, H^{\alpha}{}_{;\alpha}\, =\, 0,\\[2ex]
\epsilon \, \dot E^{\lambda} - \fracc{\epsilon'E^{\lambda} \, v^{\alpha} \, E^{\mu}}{E}E_{\mu;\alpha} + \eta^{\lambda\beta\rho\sigma} \, v_{\rho} \, H_{\sigma;\beta} \, = \,0,\\[2ex]
\mu_0 \,  \dot H^{\lambda}-\eta^{\lambda\beta\rho\sigma} \, v_{\rho} \, E_{\sigma;\beta}=0.
\end{array}
\end{equation}
We define the unitary vector $l^{\mu}$ by setting $E^{\mu}\equiv E \, l^{\mu}$, where $l^{\mu}$ satisfies $l_{\alpha} \,l^{\alpha}=-1$ and
use Hadamard conditions to obtain the propagation waves. Then, the discontinuities of Eqs.\ (\ref{max_diel_proj}) become ${[E_{\mu,\lambda}]}_{\Sigma} = e_{\mu} \, k_{\lambda}$ and ${[H_{\mu,\lambda}]}_{\Sigma}\, = \,h_{\mu} \, k_{\lambda},$ where $e_{\mu}(x)$ and $h_{\mu}(x)$ are the amplitudes of the discontinuities and $k_{\mu}$ is the wave vector. Thus, it follows that

\begin{equation}
\begin{array}{l}
\epsilon \, k^{\alpha} \, e_{\alpha}-\fracc{\epsilon'}{E}\, E^{\alpha} \, e_{\alpha} \, E^{\beta} \,k_{\beta}\, =\, 0,\\[2ex]
\mu_0 \, h^{\alpha} \, k_{\alpha} \, = \,0,\\[2ex]
\epsilon \, k^{\alpha} \,v_{\alpha} \, e^{\mu} - \fracc{\epsilon'}{E} \, E^{\lambda} \, e_{\lambda} \, v^{\alpha} \, k_{\alpha} \,E^{\mu} + \eta^{\mu\nu\alpha\beta} \, k_{\nu} \, v_{\alpha} \,h_{\beta}\, =\,0,\\[2ex]
\mu_0 \, k_{\alpha} \, v^{\alpha} \, h^{\lambda}-\eta^{\lambda\beta\rho\sigma} \, k_{\beta} \, v_{\rho} \,e_{\sigma} \,= \,0,
\end{array}
\end{equation}
where $\epsilon'$ is the derivative of $\epsilon$ with respect to $E$. Combining these equations and multiplying by $E_{\mu}$, we obtain the dispersion relation
\begin{equation}
\left(\eta^{\mu\nu}+(\mu_0 \, \epsilon-1+\mu_0 \, \epsilon'E) \,v^{\mu} \,v^{\nu}-\fracc{\epsilon'}{\epsilon\, E}E^{\mu} \,E^{\nu}\right) \,k_{\mu}\,k_{\nu}\,=\,0.
\end{equation}
We see that the envelop of discontinuity propagates differently from Minkowski light-cone of the linear Maxwell theory. In this case, the causal structure is given by an effective Riemannian geometry. Mathematically, the metric tensor is a covariant tensor of rank 2 but sometimes we shall call ``metric" a contravariant tensor of rank 2. In particular, this is the way the Gordon metric appears naturally as $\widehat g^{\mu\nu}$. From this point of view, $k_{\mu}$ is null-like in $\widehat g^{\mu\nu}$, namely,
\begin{equation}
\widehat g^{\mu\nu} \, k_{\mu} \, k_{\nu}=0.
\end{equation}
The expression of the effective geometry is given by

\begin{equation}
\widehat g^{\mu\nu} = \eta^{\mu\nu} + (\mu_0 \,\epsilon-1+\mu_0 \,\epsilon' \,E) \,v^{\mu} \,v^{\nu}-\fracc{\epsilon' \,E}{\epsilon} \,l^{\mu} \,l^{\nu}.
\end{equation}
A simple calculation show that its inverse is

\begin{equation}
\widehat g_{\mu\nu} = \eta_{\mu\nu} - \left( 1-\fracc{1}{ \mu_0 \,\epsilon(1+\xi)}\right) \,v_{\mu} \,v_{\nu}+\fracc{\xi}{1+\xi} \, l_{\mu} \, l_{\nu},
\end{equation}
where
\begin{equation}
\nonumber
\xi\equiv\fracc{\epsilon'E}{\epsilon}.
\end{equation}

In particular, when $\epsilon$ is a constant, this formula reduces to Gordon metric
\begin{equation}
\widehat g^{\mu\nu} = \eta^{\mu\nu} + (\epsilon\mu_0 - 1) \, v^{\mu} \,v^{\nu},
\label{22set2014}
\end{equation}
which depends only on the dielectric properties $\mu_0$, $\epsilon$ and its velocity $v^{\mu}$. The magnitude $N\equiv\eta^{\mu\nu}v_{\mu}v_{\nu}$ of the wave vector in Minkowski space-time (written in terms of dielectric properties) is determined by Gordon dispersion relation

\begin{equation}
\label{mod_gord_k}
\begin{array}{l}
\widehat g^{\mu\nu}k_{\mu}k_{\nu} = \left(\eta^{\mu\nu} + (\epsilon\mu_0 - 1) \,
v^{\mu}v^{\nu}\right)k_{\mu}k_{\nu}=0,\Longrightarrow\\[2ex]
\Longrightarrow N=(1-\mu_0\epsilon)(k^{\alpha}v_{\alpha})^2,
\end{array}
\end{equation}
where $k^{\alpha}v_{\alpha}\equiv k_{\alpha}v_{\beta}\eta^{\alpha\beta}$ is evaluated in the Minkowski metric.

The analysis of the wave propagation in material media and the study of effective geometry are particularly interesting in the investigation of analogue model \cite{nov_viss_volo} for the understanding of kinematical properties at very small scale of astrophysical objects \cite{novello4}.

\subsection{Effective gravity}
The property of the propagation of electromagnetic waves in a moving dielectric has been used to analyze certain particularities of gravity (and vice-versa \cite{felice1971}). We then turn our attention to the curved geometry that controls the photon propagation in a moving dielectric. We start by noting that there is no restriction whatsoever on the particularities of this motion. We can then choose particular displacements of the dielectrics to reproduce specific geometries that could help in the investigation of specific metrics that are particular solutions of GR. Let us show a simple example of this.

Select a given congruence of curves that has no shear, no vorticity and no acceleration. Thus we can set
$$ v_{\mu\, , \nu} = \frac{\theta}{3} \, h_{\mu\nu} $$
where $  h_{\mu\nu} = \eta_{\mu\nu} -  v_{\mu} \, v_{\nu}.$ Define $\widehat{v}_{\mu} = v_{\mu}$, thus
$$\widehat{v}^{\mu} = \epsilon \, \mu \, v^{\mu}$$
where $ v^{\mu} = \eta^{\mu\nu} \, v_{\nu}$ and $\widehat v^{\mu}=\widehat g^{\mu\nu}v_{\nu}$.
The Christoffel symbol for the effective metric are given by
$$ \widehat{\Gamma}^{\lambda}_{\mu\nu} = \frac{\theta}{3} \, ( 1 - \epsilon \, \mu) \,  v^{\lambda} \, h_{\mu\nu}.$$
For the contracted curvature this form yields
$$ R_{\mu\nu} = \frac{(\epsilon \, \mu - 1)}{3} \, ( \dot{\theta} + \theta^{2} )\, h_{\mu\nu}.$$
From the Raychaudhuri equation (see Sec.\ [\ref{definition}]), it follows
$$ \dot{\theta} + \frac{\theta^{2}}{3} = 0.$$

We can use these properties to combine such geometry with the equations of GR and  define the Einstein tensor $ \widehat{G}_{\mu\nu} \equiv \widehat{R}_{\mu\nu} - \frac{1}{2} \, \widehat{R} \, \widehat{g}_{\mu\nu}$ that allows us to define an equivalent source for this geometry defined by
$$ \widehat{T}_{\mu\nu} = \frac{(\epsilon \mu - 1)}{9} \, \theta^{2} \, \left( \eta_{\mu\nu} + \frac{(3 - \epsilon \mu)}{\epsilon \mu} \, v_{\mu} \, v_{\nu} \right)$$
That is, the motion of the photon inside the dielectric endowed with such effective metric mimics situations that are obtained through solutions of
Einstein's equations.

\subsection{The uses of effective metric}
Let us now provide another simple example of a physical system where the exceptional dynamics may be relevant to real practical situations and may be used to investigate dynamical aspects of fields in curved space-times. The basic system consists of Maxwell's electrodynamics inside (linear) dielectric medium in motion. The quantities we will deal with consists in tensors $F_{\mu\nu}$ and $P_{\mu\nu}$ and a normalized time-like vector field $v^{\mu}(x)$ that is $\eta_{\mu\nu}v^{\mu}v^{\nu}=1$ which represents the velocity field of the material. For pedagogical reasons we restrict our analysis to the case of isotropic constitutive laws provided by
\begin{equation}
D_{\alpha}=\epsilon(x)E_{\alpha}\quad \mbox{and}\quad B_{\alpha}=\mu(x)H_{\alpha}.
\end{equation}
The equations of motion are (\ref{max_diel}) and the study of field discontinuities inside such material leads to Gordon's metric (\ref{22set2014}). Thus, the wave fronts inside an isotropic, heterogeneous and linear material is described by a null vector $k_{\mu}$ with respect to this
effective metric, i.e., $\widehat{g}^{\mu\nu}k_{\mu}k_{\nu}=0$. Furthermore, we showed above that $k_{\mu}$ is a geodesic with respect to $\widehat{g}_{\mu\nu}$. This result was obtained several times in the literature of analogue models, enabling the study of kinematical aspects of fields that mimic the presence of gravitation. On the other hand, there exist various situations, typical of certain class of materials, where it is possible to go beyond this kinematical analogy. In these cases the field dynamics itself is described by its dependence on the effective metric, as in the case of exceptional dynamics.

In fact, using the Cayley-Hamilton formula it follows that
\begin{equation}
\label{D}
\widehat{g}\equiv \det(\widehat{g}_{\mu\nu})=-\frac{1}{\mu\epsilon}.
\end{equation}
Also, from the definition of the effective metric we obtain the identity
\begin{equation}
\label{P}
\widehat{F}^{\mu\nu}\equiv\widehat{g}^{\mu\alpha}\widehat{g}^{\nu\beta}F_{\alpha\beta}=\mu P^{\mu\nu}.
\end{equation}
Using (\ref{D}) and (\ref{P}) simultaneously, it is possible to write the first of Maxwell equations (\ref{max_diel}) in the form
\begin{equation}
\left(\frac{1}{\mu} \, \widehat{F}^{\mu\nu}\right)_{,\nu}=0.
\end{equation}
In the case of impedance matched materials where the ratio $\epsilon/\mu$ is constant, we can rewrite this equation in the very suggestive form
\begin{equation}
(\sqrt{-\widehat{g}} \, \widehat{F}^{\mu\nu})_{,\nu}=0,
\end{equation}\\
Finally the complete set of Maxwell equations (\ref{max_diel}) in this medium can be written as
\begin{equation}
\widehat{\nabla}_{\nu}\widehat{F}^{\mu\nu}=0\quad\quad \widehat{\nabla}_{[\alpha}\widehat{F}_{\mu\nu]}=0,
\end{equation}
where $\widehat{\nabla}_{\alpha}$ is the covariant derivative written in terms of $\widehat{g}_{\mu\nu}$. This means that the effective metric, that describes the characteristics, has an active part in the very description of the field dynamics. This situation can be used as a tool to investigate simultaneously kinematical and dynamical aspects of fields interacting with gravity in laboratories. We shall see afterwards that this dynamic equivalence can be useful to describe nonlinear field theories in the context of equivalent metric.

\subsection{Moving dielectrics: alternative expressions}\label{mov_diel}
Let us go back to Gordon's paper and its generalization that is characterized by the associated metric
\begin{equation}
\widehat{g}^{\mu\nu} = \eta^{\mu\nu} + ( \epsilon \mu - 1 ) \, v^{\mu} \,v^{\nu},
\label{2512112}
\end{equation}
where $\epsilon$ and $\mu$ are parameters that characterize the dielectric and $v^{\mu}$ is the four-velocity of the dielectric. Note that this metric can also be used to describe non-linear structures when $\epsilon$ and $\mu $ depends on the intensity of the field \cite{nov01}.

We shall analyze this propagation under the new perspective presented in the previous section. We start by noting that the net effect of the motion of a dielectric is to produce an acceleration for the light ray that propagate inside it. Indeed, we note that the quantity $ N = k_{\mu} \, k_{\nu} \, \eta^{\mu\nu} $ is not null and varies through the motion of the dielectric. Once $ k_{\mu}$ is a gradient
we set $k_{\mu}=\partial_{\mu}\Sigma$. Let us construct, following the expression of the DM defined previously, the metric
\begin{equation}
\label{bin_met}
\widehat{q}^{\mu\nu} =  \eta^{\mu\nu} + \beta \, k^{\mu} \,k^{\nu},
\end{equation}
where we set that $\widehat{k}_{\mu} = k_{\mu}$. Correspondingly, the contravariant expression is
$$ \widehat{k}^{\mu} = (1 + \beta \, N) \, k^{\mu}.$$
Imposing that this vector has unit norm in the DM it follows that $N( 1 + \beta \, N) = 1$. Thus, the optic ray $ \widehat{k}_{\mu} $ becomes a geodesic in this DM. This is a trivial consequence of the fact that this vector is both the gradient of a function and has constant norm. We then rewrite the Gordon result in the form
\begin{theorem}
In a moving dielectric the optic rays propagate through geodesics in the associated DM of the form
$$ \widehat{q}^{\mu\nu} =  \eta^{\mu\nu} + \frac{1 -  N}{N^{2}} \, k^{\mu}\, k^{\nu},$$
where $ N= (1 - \mu_{0} \, \epsilon) \, (k_{\mu} \, v_{\nu} \, \eta^{\mu\nu})^{2}.$
\end{theorem}

There are two equivalent forms to define the DM for the path of photons in a moving dielectric, that is:
\begin{itemize}
\item{The Gordon metric
$$ \widehat{g}^{\mu\nu} = \eta_{\mu\nu} + (\epsilon \, \mu - 1)\, v^{\mu} \, v^{\nu};$$}
\item{The dragged form
\begin{equation}
\widehat{q}^{\mu\nu} =  \eta^{\mu\nu} + \frac{1 -  N}{N^{2}} \, k^{\mu} \, k^{\nu}.
\label{7outubro2014}
\end{equation}}
\end{itemize}

Note that we have used the fact that $ N $ is constant along the ray, that is $ N_{, \, \mu} \, k^{\mu} = 0$. Thus the above Lemma is valid only for rays $ k_{\mu}$ such that its angle with the motion of the dielectric is preserved along the ray.

The interesting remark concerns the fact that, although in the Gordon analysis the associated metric depends explicitly on the velocity of the dielectric, in the present formulation this dependence is hidden in the form of the scalar product $ (k_{\mu} \, v_{\nu} \, \eta^{\mu\nu}) $ and the direction of the ray $ k_{\mu} $ appears explicitly in the driven metric. This difference plays a very important role in the cases of the search of the properties that mimics GR using electromagnetic effective geometry, as for instance in the case of non-gravitational black-hole as we shall see later.

\subsection{Polynomial Metrics}
Gordon approach depends explicitly on the velocity $v^{\alpha}$ of the dielectric. Nevertheless such form of introducing an effective metric is not unique as we have shown in Eq. (\ref{7outubro2014}). Indeed, it is possible to construct another metric $\widehat q_{\mu\nu}$ that allows to arrive at the same results obtained by Gordon and besides reduces the dependence on the four-velocity $v^{\alpha}$. From practical reasons, it might be useful to weaken this constraint of Gordon approach, once it is easier to determine the shape of the electromagnetic wave packet in the laboratory than constructing nonlinear dielectric media with arbitrary tensorial parameters $\epsilon_{\alpha\beta}$ and $\mu_{\alpha\beta}$---despite of the great advances in this research area recently \cite{smith,schuring}.

Let us now show that exists a class of geometries which play the same role as Gordon metric depending only on the angle $k_{\alpha}v^{\alpha}$ between the wave vector $k_{\alpha}$ and the dielectric four-vector $v^{\alpha}$. This is achieved by generalizations of results of previous sections. Let us list some examples:\\

\underline{Case $A$}: this approach corresponds to Sec.\ [\ref{mov_diel}] where the metric is given by Eq.\ (\ref{bin_met}). We transform there the non-normalized wave vector $k_{\mu}$ in Minkowski background in a normalized time-like vector in $\widehat q_{\mu\nu}$, namely, $\widehat{N}_{(q)}=(1+\beta \, N) \, N\, \equiv\, 1$. It does not violate Lorentz invariance, because everything happens inside the dielectric. We note that it is not possible to fix $\widehat N$ equal to zero, otherwise the metric is ill-defined. Therefore, $k_{\mu}$ is not a null-like vector in the $\widehat Q$-metric. Another feature is that the magnitude of the dielectric four-vector
$$\widehat q^{\mu\nu} \, v_{\mu} \, v_{\nu}= 1+\beta \, (k_{\mu} \, v_{\nu} \, \eta^{\mu\nu})^2,$$
is not necessarily positive definite allowing observers with velocity great than speed of light inside the medium In the laboratory, the angle between these two vectors is easier to manipulate than the dielectric velocity field only. We expect that this fact could be of reasonable utility in the research of analog models. For instance, if we set
$$ 1 + \, \beta(k_{\mu} \, v_{\nu} \, \eta^{\mu\nu})^2=0,$$
then, using Eq.\ (\ref{mod_gord_k}), we obtain
$$\beta=\fracc{(\mu_0 \, \epsilon-1)}{N}.$$
Substituting this result into the equation for the norm of $k_{\mu}$ in $\widehat q_{\mu\nu}$, yields
$$ \mu_0 \, \epsilon N  = 1.$$
Therefore, the metric $\widehat q^{\mu\nu}$ with this particular value of $\beta$ produces the following outcome: the wave vector $k_{\mu}$ becomes a normalized and time-like vector, while the dielectric velocity $v^{\mu}$, which was a time-like vector in the Minkowski background, becomes a null geodesic in $\widehat q_{\mu\nu}$. Thus, the causal structure is no more determined by $k_{\mu}$.

Remark that the metric $\widehat q^{\mu\nu}$ presented in the precedent sections is not unique. We can enlarge the set of metrics that have the same properties showed above adding other terms to $\widehat q^{\mu\nu}$ provided the condition\ (\ref{2}) is valid. To exemplify these cases we consider:\\

\underline{Case $B$}: the polynomial metric is given by

$$\widehat m^{\mu\nu} = \eta^{\mu\nu} + \beta \, k^{\mu} \,k^{\nu}+ \delta a^{\mu}a^{\nu}.$$
It is straightforward to show that its inverse has an extra term
$$\widehat m_{\mu\nu} = \eta_{\mu\nu} + B \, k_{\mu} \,k_{\nu}+ \Delta a_{\mu}a_{\nu}+ \Lambda  a_{(\mu}k_{\nu)},$$
where $(\,)$ means symmetrization. The coefficients of the inverse metric are
$$B = - \fracc{\beta( 1 - \delta a^2 )}{X},\quad \Delta = -\fracc{\delta(1+\beta N)}{X} \quad \mbox{and} \quad \Lambda = \fracc{\beta\delta \dot N}{2X}.$$
Here, we defined $a^2\equiv-a^{\alpha}a_{\alpha}$, $\dot N\equiv N_{,\mu}k^{\mu}$ and
$$X=1-\delta a^2+\beta N-\beta\delta\left(\fracc{\dot N^2}{4}+Na^2\right).$$

The appearance of an extra term also happens with the inverse metric when we consider instead of $a^{\mu}a^{\nu}$ a term of the form $a^{(\mu}k^{\nu)}$. In both cases an extra term is necessary breaking the polynomial symmetry between the metric and its inverse. Nevertheless we will present the calculations for this case focusing only on the metric containing the term $a^{\mu}a^{\nu}$ and indicating that the results are very similar when the other term is considered separately.

The geodetic motion condition for the wave vector leads to
$$\widehat{N}_{(m)}=(1+\beta N)N+\fracc{\delta}{4}\dot N^2=0.$$
Note that this approach permits a null geodesic motion for the wave vector $k_{\mu}$. This is the simplest case in which we regain the main Gordon result ($k_{\mu}$ as a null-like geodesic). The sign of the norm of $v^{\mu}$ is undetermined and may be chosen arbitrarily.

\underline{Case $C$}: the most general case involving first order derivatives of $k_{\mu}$ occurs when the metric is expressed in the form (higher derivatives of $k_{\mu}$ greater than that leads to a metric tensor $\widehat q_{\mu\nu}$ ill-defined.)

$$\widehat n^{\mu\nu} = \eta^{\mu\nu} + \beta \, k^{\mu} \,k^{\nu}+ \delta a^{\mu}a^{\nu}+\lambda a^{(\mu}k^{\nu)}$$
and its inverse is
$$\widehat n_{\mu\nu} = \eta_{\mu\nu} + B \, k_{\mu} \,k_{\nu}+ \Delta a_{\mu}a_{\nu}+ \Lambda  a_{(\mu}k_{\nu)}.$$
The covariant metric coefficients are given by
$$B = - \fracc{\beta( 1 - \delta a^2 )+\lambda a^2}{Z},\quad \Delta = -\fracc{\delta(1+\beta \, N)-\lambda^2N}{Z} \quad \mbox{and} \quad \Lambda = -\fracc{\lambda(2+\dot N \,\lambda)-2 \,\beta \,\delta \, \dot N}{2Z},$$
where $Z=\left[1-\delta \, a^2+ \beta \, N+ \dot N \, \lambda-(\beta \, \delta-\lambda^2)\left(\frac{1}{4}\dot N^2+N \,a^2\right)\right]$.

Once it involves more degrees of freedom, we can regain all outcomes presented before, but with different algebraic relations of course. In particular, the magnitude of the wave vector in $\widehat n_{\mu\nu}$ is set
$$\widehat{N}_{(n)}=( 1+\beta \, N+ \lambda \, \dot N) \, N+\fracc{\delta}{4} \, \dot N^2.$$

Remark that the metric and its inverse have the same number of polynomial terms as required from the beginning. Following this reasoning, one can manipulate the cases $A$ and $C$ in order to describe geometrically the kinematical aspects of a given theory. In particular, it can be done with classical mechanics \cite{nov_bit_drag,nov_bit_gordon} and the electromagnetic case is in progress.

\subsection{Analogue spherical black holes}
\label{Analogue}
In this section we briefly review an example -- from several presented in the literature \cite{hehl,kruglov,visser2011,frauendiener} -- developed recently \cite{bitCGQ} on how it is possible to generate a non-gravitational black-hole for photons. Let us consider a dielectric medium moving with four-velocity $v^{\alpha}=\delta^{\alpha}_0$, subjected to an electric field directed along the radial direction and no magnetic field. For the static spherically symmetric situation we are dealing with, the current four-vector $J^\mu=(\rho,\vec J)$ presents only its time component, the charge density $\rho$. Maxwell equations written in flat spherical coordinates adapted to the dielectric medium, then reduce to
\begin{equation}
\frac{\partial_r(\sqrt{-\gamma}\epsilon E)}{\sqrt{-\gamma}}=\rho
\label{max},
\end{equation}
where $\gamma=-r^4\sin^2\theta$. In this case, the effective geometry generalizes the Gordon metric, yielding
\begin{equation}
g^{\alpha\beta}={\rm diag}\left(\mu(\epsilon+\epsilon'E),\,-\frac{\epsilon+\epsilon'E}{\epsilon},\,-\frac{1}{r^2},\,-\frac{1}{r^2\sin^2\theta}\right)
\label{diag}.
\end{equation}
This form allows one to seek for analogue spherically symmetric static black hole solutions
\begin{equation}
g^{\alpha\beta}={\rm diag}\left(\frac{1}{A},\,-A,\,-\frac{1}{r^2},\,-\frac{1}{r^2\sin^2\theta}\right)
\label{bh},
\end{equation}
where $A=A(r)$ is a given radial function such that $A(r)=1-r_h/r$ describes a Schwarzschild black hole with a Schwarzschild radius $r_h$. Equation (\ref{bh}) includes all spherically symmetric black holes, such as Schwarzschild, Reissner-Nordstr\"om, de-Sitter or combinations thereof, being characterized by the explicit form of the function $A$. The identification of Eqs.\ (\ref{diag}) and (\ref{bh}) gives the two possible solutions
\begin{equation}
\epsilon+\epsilon'E=\pm\sqrt{\epsilon/\mu}
\label{eq_epsilon}.
\end{equation}
In order to integrate this equation, a two-parameter function
$\mu=\mu(\epsilon,E)$ can arbitrarily be chosen. Among all possibilities, we restrict ourselves here to the mathematically convenient form
\begin{equation}
\mu=\frac{\epsilon_0^2}{\epsilon^3}
\label{mu},
\end{equation}
where $\epsilon_0$ is the vacuum permittivity constant (with $\mu_0\epsilon_0=1$). Equation (\ref{eq_epsilon}) then reads $(\epsilon E)'=\pm\epsilon^2/\epsilon_0$, whose integration immediately yields $\epsilon=\epsilon_\pm$, where
\begin{equation}
\epsilon_\pm=\fracc{\epsilon_0}{\frac{E}{E_0}\pm1}
\label{epsilon},
\end{equation}
and $A(r)=\pm\epsilon_\pm/\epsilon_0=1/(1\pm E/E_0)$, where $E_0>0$ is a constant of integration; for $\epsilon=\epsilon_{{}+{}}$ one has $E=E_0$ at $A=1/2$, while the solution $\epsilon=\epsilon_{{}-{}}$ is limited to $E>E_0$ (since $E<E_0$ would correspond to $A>1$ in this case). The usual range $-\infty<A<1$ of an effective black hole can thus be obtained by joining the solution $\epsilon_{{}-{}}$ inside horizon with the solution $\epsilon_{{}+{}}$ outside horizon.
%That is, $\epsilon=\epsilon_0/(E/E_0-1)$ for $r<R$ and $\epsilon=\epsilon_0/(E/E_0+1)$ for $r>R$.

The expressions of the electric field $E$, the electric displacement $D=\epsilon E$, and the charge density $\rho$ in terms of $A$ then give
\begin{equation}
\fracc{E}{E_0}=\pm\fracc{(1-A)}{A},\quad D=\epsilon_0 \, E_0 \, (1-A), \quad \mbox{and} \quad \rho=\epsilon_0 \, E_0 \, \left[\fracc{2(1-A)}{r}-\fracc{{\rm d}A}{{\rm d}r}\right],
\end{equation}
which hold for either $0<A<1$ or $A<0$. In the case of a Schwarzschild analogue black hole, these expressions reduce to $E=\pm E_0r_h/(r-r_h)$ with a quadratic charge density profile $\rho=\epsilon_0E_0r_h/r^2$ and a linear electric displacement $D=\epsilon_0E_0r_h/r$ (note that $\rho/D=1/r$ in this case). Therefore, $E$ diverges at the horizon but remains finite everywhere else, while both $D$ and $\rho$ are finite at the horizon but they both diverge at the center (except for $A-1\sim r^{-2}$, which gives $\rho=0$ everywhere). The inner solution should then be regularized near the center. Our definite proposal to a spherically symmetric black hole analogue with radius $R$ is thus built with a medium whose permittivity is such that
\begin{equation}
\frac{\epsilon}{\epsilon_0}= \frac{E_0}{E+E_0}, \, \mbox{if }r>r_h,
\end{equation}

\begin{equation}
\frac{\epsilon}{\epsilon_0}= \frac{E_0}{E-E_0}, \, \mbox{if }\frac12r_h<r<r_h,
\end{equation}

\begin{equation}
\frac{\epsilon}{\epsilon_0}= 1, \, \mbox{if }r<\frac12r_h.
\label{eps(r)}
\end{equation}

When expressed in terms of the radial coordinate $r$, then Eq.\ (\ref{eps(r)}) gives $\epsilon/\epsilon_0=|A|$, where $-1<A<1$. We can compare this result with previous similar proposals: for example, which rely upon postulating a core absorption coefficient \cite{narimanov}; here, no doping is required, but only a variable volumetric density of the medium. Moreover, as already noted \cite{chen}, that was not a consistent solution of Einstein field equations; this latter instead deals with the cylindric case, and proposed a non-diagonal structure for both $\epsilon$ and $\mu$, while we treat these two parameters both as scalars.

\subsection{Reproducing metrics of GR through moving dielectrics}
Mow let us show that the effective metric can mimic some properties of the geometry discovered by G\"odel, where closed causal paths in spacetime occur.

We start with the background Minkowski geometry written in a cylindrical coordinate system
\begin{equation}
ds^2 = dt^2 - dr^2 - r^2 \,d\varphi^{2} - dz^2.
\protect\label{S1}
\end{equation}
The physical system we analyse consists on a variable magnetic field that induces an electric field such that the component $F^{02}$ does not
vanish. For our purposes here we do not need to specify the field further. We choose the observer that is comoving with the dielectric medium
endowed with a normalized velocity $v^{\mu} =\gamma \left(1, 0, \frac{\dot{\varphi}}{c}, 0 \right),$ where $\gamma \equiv \left(1 - \frac{v^2}{c^2}\right)^{-\frac{1}{2}}$ is the Lorentz factor and  $v \equiv r \dot{\phi}$. It then follows that the non-null electric components $E^{\mu}$ are $E^{0} = -\frac{\,E \gamma r {\dot\varphi}}{c}$ and $E^{2} = -\frac{\,E \gamma}{r}.$ The photons propagate as null geodesics in the effective geometry, whose the non null components of the metric are
\begin{equation}
g^{00}=1+\left(\mu\,\epsilon - 1 + \mu\,\epsilon^{'}\,E \right)\gamma^2 - \frac{\epsilon^{'}\,E}{\epsilon}\gamma^2 v ,
\protect\label{g00}
\end{equation}
\begin{equation}
g^{11}=-1,
\protect\label{g11}
\end{equation}
\begin{equation}
g^{22}= -1+\left(\mu\,\epsilon - 1 + \mu\,\epsilon^{'}\,E \right)(\gamma v)^2 - \frac{\epsilon^{'}\,E}{\epsilon}\gamma^2  ,
\protect\label{g22}
\end{equation}
\begin{equation}
g^{02}= \left(\mu\,\epsilon - 1 + \mu\,\epsilon^{'}\,E \right)(\gamma v)^2 - \frac{\epsilon^{'}\,E}{\epsilon}\gamma^2 v  ,
\protect\label{g02}
\end{equation}
\begin{equation}
g^{33}=-1,
\protect\label{g33}
\end{equation}
Let us consider a curve defined by the equations $ t = constant,$ $r = constant,$ and $z = constant$. The length element of this curve is
\begin{equation}
ds^{2}_{\rm{eff}} = g_{22}\,d\varphi^{2}.
\end{equation}
This curve can be a photon path if the condition $g_{22} = 0$ is satisfied. In other words, if the velocity of the dielectric is such that
\begin{equation}
\frac{v}{c^2} = \sqrt{\epsilon\mu}.
\protect\label{vel1}
\end{equation}
Then, the photons follow along closed time-like geodesics in this effective metric. This velocity is actually possible to be achieved by real materials, when the relative permittivity is less than one, that is, when the electric susceptibility $\chi$, defined as $ \epsilon=\epsilon_0 \left(1+\chi\right)$, is negative. An example of this is found in materials where the dielectric response of induced dipoles is a
resonant phenomenon. In this case, for frequencies above the characteristic frequency of the material $\omega_i$, the electric susceptibility is negative and the equation for the velocity becomes:
\begin{equation}
\frac{v^2}{c^2}= c^2 \mu_0\left( \epsilon_0 - \frac{4\pi N_{\rm{eff}} e^2}{m\omega_i(\omega-\omega_i)} \right),
\protect\label{vel}
\end{equation}
where $(m, e)$ are the mass and charge of a free electron respectively and $N_{\rm{eff}}$ is the oscillator strength times the total number of electrons per unit volume.

\newpage

\section{Nonlinear electromagnetic fields}
As we mentioned before, in a dielectric medium the photon paths can be described as geodesics in an effective metric. Such result can be generalized for nonlinear electrodynamics. Indeed, the electromagnetic force a photon undergoes in a nonlinear regime can be geometrized. This is a rather unexpected result and, at the same time, a beautiful consequence of the analysis of the discontinuities of inhomogeneous nonlinear electromagnetic field. We will show how such geometrization is possible. By the same token, we will show that this property is not restricted to spin-1 fields, but, on the contrary, it is a rather general property of nonlinear field theories.

\subsection{General comments on nonlinear electrodynamics}
Modifications of light propagation in different vacua states have recently been a subject of interest. Such investigation shows that, under distinct non trivial vacua (related to several circumstances such as temperature effects, particular boundary conditions, quantum polarization etc.),
the motion of light can be viewed as  electromagnetic waves propagating through a classical dispersive medium. This medium induces modifications on the equations of motion, which are described in terms of nonlinearities of the field. In order to apply such a {\em medium interpretation} we consider, from the microscopic point of view, modifications of electrodynamics due to virtual pair creation. In this case the effects can be simulated by an effective Lagrangian which depends only on the two gauge invariants $F$ and $G$ of the electromagnetic field \cite{Dittrich,Shore}.

One of the main achievements of such investigation is the understanding that, in such nonlinear framework, photons propagate along geodesics that are no more null in the actual Minkowski spacetime, but in another effective geometry. Although the basic understanding of this fact ---
at least for the specific case of Born-Infeld electrodynamics \cite{born1,born2} --- has been known for a long time \cite{Plebanski}, it has been scarcely noticed in the literature. Moreover, its consequences were not exploited any further. In particular, we emphasize the general application and the corresponding consequences of the method of the effective geometry outlined here.

The exam of the photon propagation beyond Maxwell electrodynamics has a rather diversified history: it has been investigated in curved space-time, as a consequence of non-minimal coupling of electrodynamics with gravity \cite{Drummond,novello,vitPLB} and in nontrivial QED vacua, as an effective modification induced by quantum fluctuations \cite{Latorre,Dittrich,Shore}. As a consequence of this examination some unexpected results appear. Just to point one out, we mention the possibility of {\em faster-than-light} photons.

The general approach of all these theories is based on a gauge invariant effective action, which takes into account modifications of Maxwell electrodynamics induced by different sorts of processes. Such a procedure is intended to deal with the quantum vacuum as if it was a classical medium.  Another important consequence of such point of view is the possibility to interpret all such vacua modifications--with respect to the photon propagation--as an effective change of the spacetime metric properties. This result allows one to appeal to an analogy with the electromagnetic wave propagation in curved spacetime due to gravitational phenomena.

Once the modifications of the vacuum which will be dealt here do not break the gauge invariance of the theory, the general form of the modified Lagrangian for electrodynamics may be written as a functional of the invariants, that is,
$$L = L(F,\,G).$$
We will denote by $L_F$ and $L_G$ the derivatives of the Lagrangian $L$ with respect to $F$ and $G$, respectively; and similarly for the higher order derivatives. We are particularly interested in the derivation of the characteristic surfaces which guide the propagation of the field discontinuities as described in the Hadamard's method before.

\subsection{The method of the effective geometry: one-parameter Lagrangians}
In this section we will investigate the effects of nonlinearities in the equation of evolution of electromagnetic waves.  We consider in the first part to the simple class of gauge invariant Lagrangians defined by $L=L(F)$. From the least action principle, we obtain the field equation
\begin{equation}
\partial_{\mu}\,\left(L_{F}F^{\mu\nu}\right) = 0.
\label{previous}
\end{equation}

Applying the Hadamard conditions for the discontinuity of the field equation (\ref{previous}) through $\Sigma$ we get
\begin{equation}
L_{F}f^{\mu\nu}\, k_{\nu} + 2 L_{FF} \,\xi F^{\mu\nu} k_{\nu} = 0,
\label{gw3}
\end{equation}
where $\xi \doteq  F^{\alpha\beta} \, f_{\alpha\beta}$. The consequence of such discontinuity in the cyclic identity is
\begin{equation}
f_{\mu\nu} k_{\lambda} + f_{\nu\lambda} k_{\mu} + f_{\lambda\mu} k_{\nu} = 0.
\label{gw4}
\end{equation}
In order to obtain a scalar relation we contract this equation with $k_{\alpha}\eta^{\alpha\lambda}F^{\mu\nu}$, resulting in
\begin{equation}
\label{gw33}
\xi k_\nu k_{\mu}\eta^{\mu\nu} + 2F^{\mu\nu} f_\nu{}^\lambda k_\lambda k_\mu = 0.
\end{equation}
Let us consider the case in which $\xi$ does not vanish (cf. the case $\xi = 0$ directly in Lichnerowicz, 1958\cite{Lichne}). Substituting equation (\ref{gw33}) in (\ref{gw3}), we obtain the propagation equation for the field discontinuities
\begin{equation}
\left(L_F\eta^{\mu\nu} - 4L_{FF}F^{\mu\alpha}F_\alpha{}^\nu\right) k_\mu k_\nu = 0.
\label{gww4}
\end{equation}

Expression (\ref{gww4}) suggests that one can interpret the self-interaction of the background field $F^{\mu\nu},$ in what concerns the propagation of electromagnetic discontinuities, as if it had induced a modification on the spacetime metric $\eta_{\mu\nu}$, leading to the effective geometry

\begin{equation}
g^{\mu\nu} = L_{F}\,\eta^{\mu\nu}  - 4\, L_{FF} \,{F^{\mu}}_{\alpha} \,F^{\alpha\nu}.
\label{geffec}
\end{equation}
that is, $g^{\mu\nu}\,k_{\mu}\,k_{\nu} = 0$.

A simple inspection of this equation shows that only in the particular case of linear Maxwell electrodynamics the discontinuity of the  electromagnetic field propagates along null paths in the Minkowski background.

The general expression of the effective geometry can be equivalently written in terms of the vacuum expectation value of the energy-momentum tensor, which is
\begin{equation}
T_{\mu\nu} \equiv \frac{2}{\sqrt{-\gamma}}
\,\frac{\delta\,\Gamma}{\delta\,\gamma^{\mu\nu}},
\end{equation}
where $\Gamma \doteq \int \,d^{4}x \sqrt{-\gamma}\,L$ is the effective action and $\gamma_{\mu\nu}$ the is Minkowski metric written in an arbitrary
coordinate system. In the case of one-parameter Lagrangians, $L=L(F),$ we obtain
\begin{equation}
T_{\mu\nu} = - 4 L_{F}\, {F_{\mu}}^{\alpha} \, F_{\alpha\nu} - L\,\gamma_{\mu\nu},
\end{equation}
In terms of this tensor, the effective geometry (\ref{geffec}) can be rewritten as
\begin{equation}
\label{gT}
g^{\mu\nu}=\left(L_F+\frac{L\,L_{FF}}{L_F}\right)\gamma^{\mu\nu} +\frac{L_{FF}}{L_F}T^{\mu\nu}.
\end{equation}
We remark that once the modified geometry along which the photon propagates depends upon the energy-momentum tensor distribution of the background electromagnetic field, it is tempting to search for an analogy with the corresponding behavior of photons in a gravitational field.

Therefore, the field discontinuities propagate along null geodesics in an effective geometry which depends on the EM energy distribution.
Let us point out that, as it is explicitly shown from the above equation, the stress-energy distribution of the field is the actual responsible for the deviation of the geometry from its Minkowskian form for the photons.

In order to show that the photon path is actually a geodesic curve, it is necessary to know the inverse $g^{\mu\nu}$ of the effective metric $g_{\nu\lambda}$, defined by
\begin{equation}
\label{gT11}
g^{\mu\nu}\,g_{\nu\lambda} = \delta^{\mu}_{\lambda}.
\end{equation}
This calculation is simplified if we take into account the well known properties:
\begin{equation}
F^*_{\mu\nu} \,F^{\nu\lambda} = -\, \frac{1}{4} \,G\, \delta^\lambda_\mu,
\end{equation}
and
\begin{equation}
F^*_{\mu\lambda} \,F^{*\,\lambda\nu} - F_{\mu\lambda} \,F^{\lambda\nu} = \frac{1}{2} \, F \, \delta^{\nu}_{\mu}.
\end{equation}

Thus the covariant form of the metric can be written in the form:
\begin{equation}
\label{gT22}
g_{\mu\nu}= a\,\eta_{\mu\nu} + b\,T_{\mu\nu},
\end{equation}
in which $a$ and $b$ are given in terms of the Lagrangian and its corresponding derivatives by:
\begin{equation}
a = -\,b \,\left(\frac{L_{F}^{2}}{L_{FF}}+L+\frac{1}{2}\,T \right),
\end{equation}
and
\begin{equation}
b  =  16\, \frac{L_{FF}}{L_{F}}\,\left[ \left(F^{2} + G^{2}\right)\, L_{FF}^{2} - 16\,\left(L_{F} + F\,L_{FF}\right)^{2}\,\right]^{-\,1},
\end{equation}
where $T = T^{\alpha}_{\alpha}$ is the trace of the energy-momentum tensor.

\subsection{Two-parameter Lagrangians}
In this section we will go one step further and deal with the general case in which the effective action depends upon both invariants, that is
\begin{equation}
\label{L(FG)}
L=L(F,\,G).
\end{equation}
The equations of motion are
\begin{equation}
\label{field}
\partial_{\nu}\,\left(L_{F}F^{\mu\nu} +
L_{G}F^{*\,\mu\nu}\right) = 0.
\end{equation}
Our aim is to examine the propagation of the discontinuities in such case. Following the same procedure as presented in the previous section one gets
\begin{equation}
[L_F\,f^{\mu\nu}+2A\,F^{\mu\nu} +2B\,F^{*\,\mu\nu}]\,k_\nu = 0,
\label{ref1}
\end{equation}
and contracting this expression with $F^\alpha{}_\mu k_\alpha$ and $F^{*\,\alpha}{}_\mu k_\alpha$, respectively, yields
\begin{equation}
\left[ \xi\,L_{F} + \frac{1}{2}\,B\,G \right]\,\eta^{\mu\nu}\, k_{\mu}\,k_{\nu} - 2 A  {F^{\nu}}_{\alpha}\, F^{\alpha\mu} k_\nu k_\mu = 0
\label{gw331}
\end{equation}
and
\begin{equation}
\left[ \zeta\,L_{F} - B\,F + \frac{1}{2}\, A\,G \right]\,\eta^{\mu\nu}\, k_{\mu}\,k_{\nu} - 2 B  {F^{\nu}}_{\alpha}\, F^{\alpha\mu} k_\nu k_\mu
= 0.
\label{gw332}
\end{equation}
In these expressions we have set $A \doteq 2\,(\xi\,L_{FF} + \zeta\,L_{FG})$, $B \doteq 2\,(\xi\,L_{FG} + \zeta\,L_{GG}),$ and $\zeta \doteq  F^{\alpha\beta} \, f_{\alpha\beta}^{*}$.

In order to simplify our equations it is worth defining the quantity $\Omega \doteq \zeta/\xi$. From equations (\ref{gw331}) and (\ref{gw332}), it follows that
\begin{equation}
\Omega^{2}\, \Omega_{1} + \Omega\,\Omega_{2} + \Omega_{3} = 0,
\label{r22}
\end{equation}
with the quantities $\Omega_i,\,\, i=1,2,3$ given by
\begin{eqnarray}
\Omega_1 & = & - L_FL_{FG}+2FL_{FG} L_{GG} + G(L_{GG}^2 - L_{FG}^2),\label{Omega1}\\
\Omega_2 & = & (L_F+2GL_{FG})(L_{GG} - L_{FF}) +  2F(L_{FF}L_{GG} + L_{FG}^2),\label{Omega2}\\
\Omega_3 & = & L_FL_{FG}+2FL_{FF}L_{FG} + G(L_{FG}^2 - L_{FF}^2).\label{Omega3}
\end{eqnarray}
The quantity $\Omega$ is then given by the algebraic expression
\begin{equation}
\Omega = \frac{-\Omega_2 \,\pm\, \sqrt{\Delta}}{2\Omega_1} ,
\label{Omega}
\end{equation}
where $\Delta \doteq (\Omega_2)^2 - 4\Omega_1\Omega_3.$ Thus, in the general case we are concerned here, the photon path is kinematically described by
\begin{equation}
\label{g2ef}
g^{\mu\nu}\,k_{\mu}\,k_{\nu} = 0,
\end{equation}
where the effective metric $g^{\mu\nu}$ is given by
\begin{equation}
g^{\mu\nu} =L_F\eta^{\mu\nu} - 4\left[\left( L_{FF} + \Omega L_{FG}\right)F^{\mu}\mbox{}_{\lambda} F^{\lambda\nu} + \left(L_{FG} + \Omega L_{GG}\right) F^{\mu}\mbox{}_{\lambda}F^{*\,\lambda\nu}\right].
\label{geral}
\end{equation}
When the Lagrangian does not depend on the invariant $G$, expression (\ref{geral}) reduces to the form (\ref{geffec}).

From the general expression of the energy-momentum tensor for an electromagnetic theory $L=L(F,\,G)$ we have
\begin{equation}
T_{\mu\nu} = - 4 L_{F}\, {F_{\mu}}^{\alpha} \, F_{\alpha\nu} - \left( L - G \, L_{G} \right) \,\eta_{\mu\nu}.
\end{equation}
The scale anomaly is given by the trace
\begin{equation}
\label{scaleA}
T = 4\, \left( -L + F \,L_{F}  + G \,L_{G} \right).
\end{equation}
We can then rewrite the effective geometry in a more appealing form in terms of the energy momentum tensor, that is,
\begin{equation}
g^{\mu\nu} = {\cal M}\,\eta^{\mu\nu} + {\cal N} \,T^{\mu\nu},
\label{AB}
\end{equation}
where the functions $ {\cal M}$ and $ {\cal N}$ are given by
\begin{eqnarray}
\label{A}
{\cal M} & = & L_F + G
\left( L_{FG} + \Omega L_{GG} \right) + \frac{1} { L_{F}} \left( L_{FF} + \Omega L_{FG} \right) \left( L - G L_{G} \right),\\
\label{B}
{\cal N} & = & \frac{ 1 }{ L_{F} } \left( L_{FF} + \Omega L_{FG} \right).
\end{eqnarray}
As a consequence of this, the Minkowskian norm of the propagation vector $k_\mu$ reads
\begin{equation}
\eta^{\mu\nu} k_{\mu}\,k_{\nu} = - \frac{{\cal N}}{{\cal M}} T^{\mu\nu}k_{\mu}k_{\nu}.
\end{equation}

\subsection{The effective null geodesics}
\label{Ageo}

The geometrical relevance of the effective geometry goes beyond its immediate definition. Indeed, as follows it will be shown that the integral curves of the vector $k_\nu$ ({\em i.e.}, the photons trajectories) are in fact geodesics. In order to achieve this result it will be required an underlying Riemannian structure for the manifold associated with the effective geometry. In other words this implies a set of Levi-Civita connection coefficients $\Gamma^\alpha\mbox{}_{\mu\nu}=\Gamma^\alpha\mbox{}_{\nu\mu}$, by means of which there exists a covariant differential operator $\nabla_\lambda$ (the {\em covariant derivative in the effective metric}) such that
\begin{equation}
\label{Riemann}
\nabla_\lambda g^{\mu\nu}\equiv g^{\mu\nu}\mbox{}_{;\,\lambda} \equiv g^{\mu\nu}\mbox{}_{,\,\lambda} + \Gamma^\mu\mbox{}_{\sigma\lambda}g^{\sigma\nu} + \Gamma^\nu\mbox{}_{\sigma\lambda}g^{\sigma\mu}=0.
\end{equation}
From (\ref{Riemann}) it follows that the effective connection coefficients are completely determined from the effective geometry by the usual Christoffel formula.

Contracting (\ref{Riemann}) with $k_\mu k_\nu$ yields
\begin{equation}
\label{N15}
k_\mu k_\nu g^{\mu\nu}\mbox{}_{,\,\lambda} = -2k_\mu k_\nu\Gamma^\mu\mbox{}_{\sigma\lambda}g^{\sigma\nu}.
\end{equation}
Differentiating the dispersion relation we have
\begin{equation}
\label{N16pr}
2k_{\mu,\,\lambda}k_\nu g^{\mu\nu} + k_\mu k_\nu g^{\mu\nu}\mbox{}_{,\,\lambda} = 0.
\end{equation}
Inserting this into the left hand side of (\ref{N15}), we obtain
\begin{equation}
\label{N18}
g^{\mu\nu}k_{\mu;\,\lambda}k_\nu \equiv g^{\mu\nu}\left(k_{\mu,\,\lambda} - \Gamma^\sigma\mbox{}_{\mu\lambda}k_\sigma\right)k_\nu = 0.
\end{equation}
As the propagation vector $k_\mu=\Sigma_{,\,\mu}$ is an exact gradient one can write $k_{\mu;\,\lambda}=k_{\lambda;\,\mu}$. With this identity and defining $k^\mu\doteq g^{\mu\nu}k_\nu$ equation (\ref{N18}) reads
\begin{equation}
\label{geodesic}
k_{\mu;\,\lambda}k^\lambda = 0,
\end{equation}
which states that $k_\mu$ is a null geodesic vector (with respect to the effective geometry $g^{\mu\nu}$), namely, its integral curves are therefore null geodesics.

\subsection{Exceptional Lagrangians}
It seems worth noting that equation (\ref{geral}) contains a remarkable result: the velocities of the photon are, in general, doubled. There are some exceptional cases, however, for  which the uniqueness of the path is guaranteed by the equations of motion\cite{boillat,birula}. Such uniqueness occurs for those dynamics described by Lagrangian $L$ that satisfy the condition
$$\Delta = 0.$$

The most known example of such uniqueness for the photon velocity in a nonlinear theory is the Born-Infeld (BI) electrodynamics, which we will study in details next section. Originally, the BI theory was an attempt to modify Maxwell's electromagnetism that could circumvent the self-energy divergence of the classical point-like charged particle. Hence, one of its main motivations was to establish a consistent classical theory for the electron. Notwithstanding, the BI theory has other interesting features that make it a distinguished theory among non-linear electromagnetic theories: 1) excitations propagate without the shocks, common to generic nonlinear models; 2) single characteristic surface equation, i.e. birefringence is absent 3) fulfills the positive energy density condition and 4) satisfies the duality invariance \cite{novello1}. Nowadays, there has been a renewed interest in BI theories and its non-abelian generalizations in connection with the theory of strings, gauge fields on D-branes, K-essence-like models in cosmology and others. Thus, BI theory is a good prototype of a nonlinear theory with desirable physical properties.

\subsection{The special case of Born-Infeld dynamics}\label{BIeffGeo}
Let us pause for a while in order to analyze the BI theory closely. Then, we start with the BI Lagrangian in the form
\begin{equation}
L = \beta^{2} \, \left( 1 - \sqrt{U} \right),
\end{equation}
where
$$U\equiv1 + \frac{F}{2 \beta^{2}} - \frac{G^{2}}{ 16 \beta^{4}}.$$
Note that the addition of a constant term in the lagrangian has the purpose to make the energy-momentum tensor of the point-charge field goes to zero in the spatial infinity. In a cosmological scenario it can be interpreted as a kind of cosmological constant. However, here it plays no role whatsoever and can be omitted at will.

Born and Infeld had shown that such nonlinear dynamics may  be written in terms of the determinant of an object that has no symmetry $C_{\mu\nu}$ constructed as
$$ C_{\mu\nu} \equiv \eta_{\mu\nu} + F_{\mu\nu}.$$
Indeed they showed that the Lagrangian takes the form $ L \sim \det \, [C^{\mu}{}_{\nu}]$.

We shall see that using the effective metric obtained through the path of photons in this theory it is possible to describe the same Lagrangian using a symmetric tensor that appears naturally in the causal structure of this theory. We have shown that the causal structure represented by the effective metric of a given non-linear theory depends on the dynamics of the field and non the other way round. However we could ask if there exists a particular theory such that its dynamics is given as a functional of its effective geometry.  We shall see that BI is precisely such theory.

In the case of the BI theory all quantities $\Omega_i,\,\, i=1,2,3$ vanish identically. Hence, in this situation we cannot obtain the effective geometry from equation (\ref{geral}). In this very exceptional case we have to return to the original equation (\ref{gw331}). The most direct way to prove this is to to try to solve the following problem. Given an arbitrary Lagrangian of the form
\begin{equation}
L = L ( F, G)
\end{equation}
Its corresponding effective geometry (that controls the path of its discontinuities) in the BI particular case takes the form
\begin{equation}
g^{\mu\nu} = \frac{1}{4 \beta^{2} \, U^{\frac{3}{2}}} \, \left[ (\beta^{2} + \frac{F}{2} ) \, \eta^{\mu\nu} + F^{\mu}{}_{\lambda} \, F^{\lambda\nu} \right]
\end{equation}
Its inverse is
\begin{equation}
g_{\mu\nu} = 4 \, \sqrt{U}  \, \left[  \eta_{\mu\nu} - \frac{1}{\beta^{2}} \, F_{\mu}{}^{\lambda} \, F_{\lambda\nu} \right]
\end{equation}
This correspond to a unique characteristic equation which does not show birefringence--this was obtained by Plebanski \cite{Plebanski} for the first time. From the Cayley-Hamilton formula for $A \equiv \eta^{\mu\alpha}g_{\alpha\nu}=I - (1/\beta^{2}) \, F^{\mu}{}_{\alpha} \, F^{\alpha}{}_{\nu}$, we find the unexpected result that the lagrangian can be written in terms of the determinant of the metric that drives the propagation of the discontinuities:
\begin{equation}
\det A = U^{2}.
\end{equation}
Using this expression we can rewrite BI Lagrangian in the form in which only the determinant of the effective metric appears:
\begin{equation}
L_{BI} = \beta^{2} \left[ 1 - \frac{1}{2} \, ( \det [g_{\mu\nu}])^{\frac{1}{8}} \right]
\end{equation}
We can then state the following equivalent form of BI dynamics that exhibits a very curious phenomenon of self-consistency in action: {\it the BI dynamics of the electromagnetic field is obtained by a variation of the effective metric that extremizes its determinant. This geometry is related to the causal structure of the theory.}

Let us now combine this non-linear electrodynamics to gravity within the description of general relativity. From the expression of the determinant, we can write
\begin{equation}
det \, \left( \delta^{\mu}_{\nu} - \frac{1}{\beta^{2}} \, F^{\mu}{}_{\alpha} \, F^{\alpha}{}_{\nu} \right) = U^{2}
\end{equation}
or equivalently,
\begin{equation}
det \, \left( \delta^{\mu}_{\nu} + \frac{1}{\beta} \, F^{\mu}{}_{\nu} \right) = U.
\end{equation}

In general the energy-momentum tensor for the nonlinear electrodynamics has the form
\begin{equation}
T_{\mu\nu} = - 4 \, L_{F} \,F_{\mu}{}^{\alpha} \, F_{\alpha\nu} + ( G \, L_{G} - L ) \, g^{b}_{\mu\nu},
\end{equation}
where the index $b$ in the metric tensor makes reference to the background geometry. For the BI case we find
\begin{equation}
T_{\mu\nu} = \frac{1}{\sqrt{U}} \, F_{\mu}{}^{\alpha} \, F_{\alpha\nu} + \left( \frac{G^{2}}{16 \beta^{2} \sqrt{U}} - \beta^{2} + \beta^{2} \sqrt{U} \right) \, g^{b}_{\mu\nu}
\end{equation}
We can thus write

\begin{equation}
F_{\mu}{}^{\alpha} \, F_{\alpha\nu}  = \sqrt{U} \, T_{\mu\nu} - \left( \frac{F}{2} + \beta^{2} - \beta^{2} \, \sqrt{U} \right) \, g^{b}_{\mu\nu}.
\end{equation}
Note that the trace is given by
\begin{equation}
T = \frac{- F}{\sqrt{U}} + \frac{G^{2}}{4 \beta^{2} \sqrt{U}} - 4 \beta^{2} + 4 \beta^{2} \sqrt{U}
\end{equation}
consequently the effective metric reduces to
\begin{equation}
g_{\mu\nu} = \frac{4 U}{\beta^{2}} \, \left( - T_{\mu\nu} + \frac{1}{2} \, T \, g^{b}_{\mu\nu} \right)
\end{equation}
from the equations of GR it becomes
\begin{equation}
g_{\mu\nu} = \frac{4 U}{\beta^{2}} R_{\mu\nu}.
\end{equation}
We summarize this result in the following

\begin{theorem}
Let us combine the BI nonlinear theory minimally coupled to the gravitational field obeying the equations of GR. Let $l_{\mu}$ be a null vector for the background metric and $k_{\mu} $ the propagating vector for the discontinuities of the non-linear electrodynamics. We have shown that linear photons follows null geodesics in the background geometry $g^{b}_{\mu\nu}$ and the non-linear photons follow geodesics in the effective geometry $g_{\mu\nu}.$
\end{theorem}

Thus for the linear photons the propagation vector $l^{\nu}$ satisfies the condition $ g^{b}_{\mu\nu} \, l^{\mu} \, l^{\nu} = 0 $ and the BI photons obey $ R^{b}_{\mu\nu} \, l^{\mu} \, l^{\nu} = 0$, where $ R^{b}_{\mu\nu}$ is the Ricci tensor of the background geometry. This expression shows an unexpected relationship between the effective metric and the curvature of the background metric.

\subsection{Special Riemannian geometries}
In this review we deal with many different kinds of geometry that belong to a specific binomial structure. This form of geometry is typical of the effective geometry that controls the propagation of waves in non-linear field theories. However it appears in many other situations and it is much more general as we shall see.

Let $g_{\mu\nu} $ be the Riemannian metric of space-time. We construct an associated geometry through the definition
\begin{equation}
\widehat{q}^{\mu \nu} \equiv a \, g^{\mu \nu}+ b \, \phi^{\mu \nu}.
\label{hum}
\end{equation}
We require that the inverse metric $ \widehat{q}_{\mu \nu}$ defined by $\widehat{q}^{\mu \nu} \, \widehat{q}_{\nu\alpha} = \delta^{\mu}_{\alpha}$ must have the same binomial form, that is

\begin{equation}
\widehat{q}_{\mu \nu} \equiv A \, g_{\mu \nu}+ B \, \phi_{\mu \nu}.
\label{1dois}
\end{equation}
Although parameters $ a $ and $ b $ are completely arbitrary, the associated ones $ A $ and $ B $ are given in terms of $ a$ and $b.$

Thus, the tensor $  \phi^{\mu \nu} $ which is constructed in terms of other fields - scalar, vector, for instance - must satisfy the condition
\begin{equation}
\phi_{\mu \nu} \, \phi^{\nu \lambda} \, = m \,   \delta_{\mu}^{\lambda} + n \,   \phi_{\mu}^{\lambda}
\label{phi}
\end{equation}

We will analyze in these notes some examples like
\begin{itemize}
\item{Scalar field: where $\phi_{\mu\nu}=\partial_{\mu}\varphi \partial_{\nu}\varphi$, $m=0$ and $n=\omega \equiv \partial_{\mu} \varphi \partial^{\mu}\varphi$.}
\item{Electromagnetic field: where $\phi_{\mu \nu} \equiv  F_{\mu}{}^{\alpha}F_{\alpha\nu}$, $m=(1/16)G^{2}$ and $n = - F/2$.}
\item{Spinor fields: $\phi_{\mu \nu}$ is given by a combination between the vector $J^{\mu}$ and axial $I^{\mu}$ currents.}
\end{itemize}

\subsection{Some remarkable consequences: Closed null-like paths }
\label{consequences}
The effective metric method to describe propagation of photons by modification of the underlying metric can be used to mimic unusual solutions of the equations of GR. A particular situation concerns the rotating universe proposed by G\"odel that contains closed time-like paths, as we discussed before. We shall show the possibility of the existence of closed paths for photons in space-time.

The physical system we will analyze consists of a charged wire running through a solenoid. The flat Minkowskian background geometry is written in a
$(t, r, \varphi, z)$ coordinate system as
\begin{equation}
ds^2 = dt^2 - dr^2 - r^2 \,d\varphi^{2} - dz^2.
\protect\label{S12}
\end{equation}
The non-null components of the Maxwell tensor $F^{\mu\nu}$, compatible with the symmetry properties of the system are: $F^{01} = E(r)$ and $F^{12}= B(r)$. In this case the equation of motion of the system reduces to
\begin{eqnarray}
r \, L_{F}\, F^{01} = Q,\protect\label{S2}\\
r \,L_{F}\, F^{12} =\mu,\protect\label{S2,5}
\end{eqnarray}
where $Q$ and $\mu$ are constants that determine the charge density of the wire and the current carried through the solenoid respectively. We are interested in the analysis of the propagation of electromagnetic waves in the region inside the solenoid. Following our previous treatment we can assert that the photons propagate as if the metric structure of space-time was changed into an effective Riemannian geometry. From Eq.\ (\ref{geral}) we obtain the following components of the effective metric
\begin{eqnarray}
g^{00} &=& 1 -\psi\, E^2,\protect\label{S3}\\
g^{11} &=& -1 - \psi\,(B^2 \, r^2 - E^2),\protect\label{S4}\\
g^{02} &=& \psi \, E \, B,\protect\label{S5}\\
g^{22} &=& -\left(\frac{1}{r^2} + \psi\, B^2\right),\protect\label{S6}\\
g^{33} &=& -\,1,\protect\label{S7}
\end{eqnarray}
where $\Psi = 4\,\frac{L_{FF}}{L_{F}}$. The photon paths are null geodesics in this effective geometry. Consider a curve defined by the equations $ t = \mbox{const.},$ $r = \mbox{const.},$ and $z = \mbox{const.}$ The length element of this curve is given by
\begin{equation}
ds^{2}_{\rm{eff}} = g_{22}\,d\varphi^{2}.
\protect\label{zz3}
\end{equation}
This curve can become a photon path if there is a radius
$r = r_{c}$ such that $g_{22}(r_{c}) = 0.$
In terms of the contravariant components of the effective metric listed above this condition results
\begin{equation}
(1-\Psi E^2 )=0.
\protect\label{zzz3}
\end{equation}
 The solution for this equation is
\begin{equation}
r_c=\frac{2Q}{L_F}\left(\frac{L_{FF}}{L_F}\right)^{1/2}
\end{equation}
which implies that
\begin{equation}
\left.\frac{L_{FF}}{L_{F}}\right |_{r_c}>0.
\end{equation}
In order to present a simple example which exhibits closed curves, we work with a BI-like Lagrangian
\begin{equation}
L=\frac{\beta^2}{2}\left(\sqrt{1-\frac{F}{\beta^2}} -1 \right).
\protect\label{zzz}
\end{equation}
Note that although highly speculative, this Lagrangian has the Maxwell limit for weak fields
($F\ll\beta^2$). It also has an interesting property in the situation we are examining. Substituting the Lagrangian (\ref{zzz}) into equation (\ref{zzz3}) we find that the  magnetic field in this case takes the large value $B = \beta/\sqrt{2}$.

Other nonlinear Lagrangians such as the Heisenberg-Euler Lagrangian for QED,  cannot be analyzed with the formalism presented here since they depend on both invariants F and G. The real interest in this phenomenon is that it might be possible to be observed in the laboratory. This possibility rests in the analogy between the propagation of photons described by nonlinear Lagrangians in vacuum and  that of photons described by Maxwell's theory in the presence of a dielectric \cite{novello1}

It has been known from more than half a century that gravitational processes allow the existence of closed paths in space-time. This led to the belief that this strange situation occurs uniquely under the effect of gravity. In the above example we have shown that this is not the case. Indeed, photons can follow closed curves (CC's) in space-time due to electromagnetic forces in a nonlinear regime. In the limiting case in which the nonlinearities are neglected, the presence of CC's is no more possible. Thus we can state that this property depends crucially on the nonlinearity of the electromagnetic processes and it does not exist in Maxwell's theory. In other words, the existence of closed paths in space-time is not an exclusive
property of the gravitational interaction: it appears also in pure electromagnetic processes, depending on the nonlinearities of the background field. The existence of such paths in both gravitational and electromagnetic processes asks for a deep review of the causal structure as displayed by the geodesics of the photons.

\subsection{Analogue model for Kasner Cosmology}
Inside material media electrodynamics are described by nonlinear equations. Indeed, in such situations Maxwell equations must be supplemented with constitutive relations which, in general, are nonlinear and depend on the physical properties of the medium under the action of external fields.
As a consequence, several effects (non usual in the context of linear Maxwell theory) are predicted. Of actual interest is the phenomenon of
artificial birefringence: when an external field is applied in a medium with nonlinear dielectric properties, an artificial optical axis may appear
\cite{Landau,del01,del02,Klippert,bjp,local,Yuri,prl2,del04}.

Analogue models have been proposed in several branches of physics. It deals with acoustics \cite{unr81,mat85,jac93,hoc97,vis98,lib00,gar00,fed03,bar03b}, optics \cite{Plebanski,rez00,leo99,cor99,nov01,shu02,leo03,unr03} among others \cite{gon98,jac98,vol98,vis00,leonetal,bre02,bar03a} with experimental verification including quantum effects \cite{belgiorno,hawking1975,pendry,sheng,unruh2011,yin}, despite of contrary claims \cite{belinski}. These are only representative references; a complete review can be found in Barcel\'o {\it et al} (2011)\cite{visser2011}.

Particularly, nonlinear electrodynamics has been considered as a possible scenario to construct analogue models for GR, either in the context of nonlinear Lagrangian or nonlinear material media. As an example, let us show that homogeneous dielectric media at rest with the dielectric coefficients $\varepsilon^\mu{}_\nu (\vec{E})$  and constant $\mu$, in the limit of geometrical optics, can be used to construct an analogue model for Kasner cosmology. In order to avoid ambiguities with the wave velocity, dispersive effects were neglected by considering only monochromatic waves. It is shown that naturally uniaxial media presenting nonlinear dielectric properties can be operated by external fields in such way to induce anisotropy in the optical metric.

Recently the effective geometry for light rays in local anisotropic dielectric media was obtained \cite{del04}, which can be presented in the symmetrized form:
\begin{equation}
g^{\alpha\beta}_{\pm} = \mu\alpha V^\alpha V^\beta +\frac{1}{2}\left[ C^\nu{}_\nu - \frac{1}{\mu (v^\pm_\varphi)^2}\right]C^{(\alpha\beta)} -
\frac{1}{2}C^{(\alpha}{}_{\nu}C^{\nu\beta)},
\label{um}
\end{equation}
where
\begin{eqnarray}
C^{\alpha}{}_\tau &\doteq&\varepsilon^{\alpha}{}_{\tau} + \frac{\partial \varepsilon^{\alpha}{}_{\beta}}{\partial E^{\tau}}E^\beta +
\frac{1}{\omega}\frac{\partial \varepsilon^{\alpha}{}_{\beta}}{\partial B^{\rho}} \eta^{\rho\lambda\gamma}{}_{\tau}E^{\beta}K_\lambda V_\gamma
\label{dois}
\end{eqnarray}
and the phase velocities $v^\pm_\varphi$ are
\begin{equation}
v^\pm_\varphi = \sqrt{\frac{\beta}{2\alpha}\left(1 \pm \sqrt{1 + \frac{4\alpha \gamma}{\beta^2}}\right)},
\label{3}
\end{equation}
with $\omega \doteq K^\alpha V_\alpha$ identified as the frequency of the electromagnetic wave and the coefficients $\alpha$, $\beta$ and $\gamma$
given by
\begin{eqnarray}
\!\!\!\!\alpha \!&\doteq&\! \frac{1}{6} \left[ (C^\mu{}_\mu)^3 \!- 3 C^\mu{}_\mu C^\alpha{}_\beta C^\beta{}_\alpha \!+2 C^\alpha{}_\beta C^\beta{}_\gamma C^\gamma{}_\alpha \right]\!,\label{4}\\
\!\!\!\!\beta \!&\doteq&\!  \mu^{-1} \left( C^\lambda{}_\alpha C^{\alpha\nu} -C^\alpha{}_\alpha C^{\lambda\nu} \right)\widehat{q}_\lambda \widehat{q}_\nu,
\label{5}\\
\!\!\!\!\gamma \!&\doteq&\! \mu^{-2} C^{\lambda\nu} \widehat{q}_\lambda \widehat{q}_\nu .
\label{6}
\end{eqnarray}
In the last two equations (\ref{5}-\ref{6}) we introduced the 3-dimensional projection of the wave vector $K^\alpha$ as $q^\alpha = h^{\alpha}_\mu K^\mu = K^\alpha - \omega V^\alpha$.

The symmetric tensors $g^{\mu\nu}_{\pm}$ represent the optical metrics (also known as effective geometries) associated with the wave propagation,
and the symbol $\pm$ indicates the possible distinct metrics, one for each polarization mode. Correspondingly to it Eq.\ (\ref{3}) expresses the fact
that, in general, the phase velocity of the electromagnetic waves inside a material medium may get two possible values ($v_\varphi^+, v_\varphi^-$)  which are associated with the two possible polarization modes \cite{Klippert}. For the particular case of  Maxwell linear theory in vacuum, both metrics $g^{\mu\nu}_+$ and $g^{\mu\nu}_-$ reduce to the Minkowski metric $\eta^{\mu\nu}$, as expected.

Now, let us consider a naturally uniaxial medium reacting nonlinearly when subjected to an external electric field as $\varepsilon^\alpha{}_\beta = {\rm diag}[0,\varepsilon_{\parallel} (E),\varepsilon_{\perp} (E),\varepsilon_{\perp} (E)]$. In this case, $\varepsilon^\alpha{}_\beta = \varepsilon^\alpha{}_\beta(E)$ and by setting $\vec{E}$ in the $x$-direction (optical axis) we obtain $C^\alpha{}_\beta={\rm diag}(0,\varepsilon_{\parallel}+ E \varepsilon'_{\parallel}, \varepsilon_{\perp},\varepsilon_{\perp})$, where $\varepsilon'_{\parallel} = d\varepsilon_{\parallel} / dE$. For this particular case $C^{\alpha\beta}$ is a symmetric tensor. The phase velocities reduce to
\begin{eqnarray}
(v_\varphi^+)^2 &=& \frac{1}{\mu\varepsilon_\perp},\label{12}\\
(v_\varphi^-)^2 &=& \frac{1}{\mu\varepsilon_\perp C^1{}_1}\left[\varepsilon_\perp (1-\widehat{q}_1{}^2)+C^1{}_1\widehat{q}_1{}^2\right].\label{13}
\end{eqnarray}
Note that $v_\varphi^-$ depends on the direction of propagation, as it should be expected for the extraordinary ray. The two velocities coincide when either the propagation occurs along the direction of the electric field ($\widehat{q}_1{}^2=1$), or when the no-birefringence condition $\varepsilon_{\parallel} + E \varepsilon'_{\parallel} = 0$ holds \cite{del04}.

Let us also particularize to the model where
\begin{eqnarray}
\varepsilon_\perp = \epsilon_\perp - 3 p E^2, \quad \mbox{and} \quad \varepsilon_\parallel = \epsilon_\parallel - s E^2.
\end{eqnarray}
where $s$ and $p$ are constants. Thus, $C^\alpha{}_\beta={\rm diag}(0,\epsilon_{\parallel} -3 s E^2,\epsilon_{\perp} - 3 p E^2,\epsilon_{\perp} - 3 p E^2)$.

For the ordinary ray the optical metric coefficients are
\begin{eqnarray}
g^{00}_+ &=& \mu \alpha\label{26}\\
g^{ii}_+ &=& -\epsilon_\parallel \epsilon_\perp + 3(s \epsilon_\perp + p \epsilon_\parallel) E^2 - 9spE^4\label{29}
\end{eqnarray}
where
\begin{eqnarray}
\alpha = -27s p^2E^6 +9p(p\epsilon_\parallel+2s\epsilon_\perp)E^4 -3\epsilon_\perp(s\epsilon_\perp+2p\epsilon_\parallel)E^2 + \epsilon_\parallel \epsilon_\perp^2.
\label{22}
\end{eqnarray}
Equations (\ref{26}-\ref{29}) show that for the ordinary ray there will be no anisotropy in the space section.

For the extraordinary ray the optical metric coefficients are
\begin{eqnarray}
g^{00}_- &=& \mu \alpha\label{32}\\
g^{11}_- &=& -(\chi - \epsilon_\parallel + 3sE^2)(\epsilon_\parallel - 3sE^2)\label{33}\\
g^{22}_- &=& g^{33}_- = -(\chi -\epsilon_\perp + 3pE^2)(\epsilon_\perp - 3pE^2)\label{35}
\end{eqnarray}
where $\chi$ depends on the direction on wave propagation as
\begin{eqnarray}
\chi = - \frac{(\epsilon_\parallel - 3sE^2)(\epsilon_\perp - 3pE^2)} {(\epsilon_\perp - 3pE^2)(1-\widehat{q}_1{}^2) + (\epsilon_\parallel - 3sE^2)\widehat{q}_1{}^2} + (\epsilon_\parallel - 3sE^2)+2(\epsilon_\perp - 3pE^2).
\label{30}
\end{eqnarray}
%
%\begin{eqnarray}
%\chi =\frac{(\epsilon_\parallel - 3sE^2)[\epsilon_\parallel+ \epsilon_\perp - 3(s+p)E^2]\widehat{q}_1{}^2
%+2(\epsilon_\perp - 3pE^2)(1-\widehat{q}_1{}^2)}
%{(\epsilon_\perp - 3pE^2)(1-\widehat{q}_1{}^2) + (\epsilon_\parallel - 3sE^2)\widehat{q}_1{}^2}.
%\label{300}
%\end{eqnarray}
%
Equations (\ref{32}-\ref{35}) show that for the extraordinary ray there will be anisotropy in the space section ($g^{11}_- \neq g^{22}_- = g^{33}_-$).

Before closing this section we note that when the propagation occurs in the direction of the optical axes it follows $\chi \left.\right|_{\widehat{q}_1{}^2=1} = (\epsilon_\parallel - 3sE^2)+(\epsilon_\perp - 3pE^2)$ and $g^{11}_- = g^{22}_- = g^{33}_- = -(\epsilon_\parallel - 3sE^2)(\epsilon_\perp - 3pE^2)$. By the other hand, when the propagation occurs perpendicularly to the optical axes it follows $\chi\left.\right|_{\widehat{q}_1{}^2=0} = 2(\epsilon_\perp - 3pE^2)$ and $g^{11}_- \neq g^{22}_- = g^{33}_-$.

Using results presented in the literature describing the propagation of monochromatic electromagnetic waves inside naturally anisotropic material media with nonlinear dielectric properties an analogue model for general relativity presenting anisotropy in the space section has been constructed. Thus light propagation in local anisotropic media can be used as a tool for testing kinematic aspects of Kasner cosmological model in laboratory.

\section{Reproducing metrics of GR through scalar fields}
The phenomenon of induced metric that we have been analyzed for electrodynamics is rather general and may occurs for any nonlinear theory independently of its spin properties. In this section we consider the case of a non linear scalar field. We will show that a class of theories that have been analyzed in the literature, having regular configuration in the Minkowski space-time background is such that the field propagates like free waves in an effective de Sitter geometry. The observation of these waves would led us to infer, erroneously, that we live in a de Sitter universe.

\subsection{The kinematical analogy}
Let us give a simple example. Consider a scalar field $\varphi $ propagating in a flat space-time whose dynamics is provided by the non-linear Lagrangian
$$ L = L(w), $$
where $w \equiv \partial_{\mu} \varphi \, \partial_{\nu} \varphi \, \eta^{\mu\nu} $ is the canonical kinematic term. The equation of
motion for $\varphi$ reads

\begin{equation}
\label{EqMotion}
\partial_\mu \Bigl( L_{w} \, \partial_\nu \varphi \, \eta^{\mu\nu} \Bigr) = 0,
\end{equation}
where $L_w$ denotes the first derivative of $L$ with respect to $w.$ By expanding the left hand side we obtain the explicit form
\begin{equation}\label{eom}
L_{w}\square\varphi+2L_{ww}\partial^{\mu}\varphi\partial^{\nu}\varphi\partial_{\mu}\partial_{\nu}\varphi=0,
\end{equation}
with $\square\equiv\eta^{\mu\nu}\partial_{\mu}\partial_{\nu}$. This constitutes a quasi-linear second order partial differential equation for $\varphi$. We are interested in evaluating the characteristic surfaces of wave propagation in this theory. The most direct and elegant way to pursue this goal is to use the Hadamard formalism discussed in Sec.\ [\ref{had_meth}].

Let $\Sigma$ be a surface of discontinuity of the scalar field $\varphi.$ The field $\varphi$ and its first derivative $\partial_\mu \varphi$ are continuous across $\Sigma$, while the second derivative presents a discontinuity:
\begin{eqnarray}
\label{CondHadamard}
\left[ \varphi \right]_\Sigma = 0,\quad \left[ \partial_\mu \varphi \right]_\Sigma = 0,\quad \mbox{and} \quad \left[ \partial_\mu\partial_\nu\varphi \right]_\Sigma = k_\mu k_\nu \xi(x),
\end{eqnarray}
where $ k_\mu = \partial_\mu \Sigma$ is the propagation vector and $\xi(x)$ the amplitude of the discontinuity. From the above conditions we obtain that both $L_{w}$ and $L_{ww}$ are continuous functions across $\Sigma$. Using these discontinuity properties in the equation of motion (\ref{eom}) it follows that only the second order derivative terms contribute. We obtain the relation
\begin{equation}
L_{w}\eta^{\mu\nu}\left[ \partial_\mu\partial_\nu\varphi \right]_\Sigma+ 2 \, L_{ww}\, \partial^\mu \varphi \partial^\nu\varphi \, \left[
\partial_\mu\partial_\nu\varphi \right]_\Sigma=0.
\end{equation}
Thus, using Hadamard conditions it follows
\begin{equation*}
k_\mu k_\nu \Bigl(L_w \eta^{\mu\nu} + 2 L_{ww} \partial^\mu\varphi\partial^\nu\varphi\Bigr) = 0.
\end{equation*}
This equation suggests the introduction of the effective metric defined by
\begin{equation}
\label{gEff}
\widehat{g}^{\mu\nu} \equiv L_w \eta^{\mu\nu} + 2 L_{ww} \partial^\mu\varphi\partial^\nu\varphi.
\end{equation}
Thus there are two distinct metrics in this framework: the Minkowskian $ \eta^{\mu\nu} $ that enters in the dynamics of the field $\varphi$  and the effective metric $ \widehat{g}^{\mu\nu} $ that controls the propagation of the waves. Note that, once the vector of discontinuity $ k_{\mu}$ is a gradient, discontinuities of the field $\varphi$ propagate through null geodesics in the effective metric $\widehat{g}_{\mu\nu},$ i.e.
\begin{equation}
\widehat{g}^{\mu\nu}k_{\alpha ;\mu}k_{\nu}=0,
\end{equation}
where ``;" stands here for the covariant derivative evaluated with the effective metric. The inverse $\widehat{g}_{\mu\nu}$ of (\ref{gEff})
is obtained through the condition $\widehat{g}^{\mu\alpha}\widehat{g}_{\alpha\nu} = \delta^\mu_\nu:$
\begin{equation}
\label{gEffInv}
\widehat{g}_{\mu\nu} = \frac{1}{L_w} \, \eta_{\mu\nu} - \frac{2 L_{ww}}{L_{w} \,(L_w + 2 w L_{ww})} \partial_\mu\varphi\partial_\nu\varphi.
\end{equation}
We remind that the determinant of a mixed tensor $\textbf{T}=T^{\alpha}_{\phantom a\beta}$ may be expressed in terms of traces of its powers as we exhibited in the first section as an immediate consequence of the Cayley-Hamilton theorem.
Applying this formula to evaluate the determinant of the effective metric (\ref{gEff}) one obtains, after a straightforward calculation
\begin{equation}\label{det}
\sqrt{ -\widehat{g}} = L_{w}^{-2} \, \left(1 + 2\frac{L_{ww}}{L_{w}} \, w\right)^{-1/2},
\end{equation}
Note that, the square-root of the determinant is real only if the condition
\begin{equation}
1+2w \, L_{ww} /L_{w}>0
\end{equation}
is satisfied. This is the same as to guarantee the hyperbolicity of the equations of motion and henceforth the existence of waves. Nevertheless, we remark that the effective metric that controls the propagation of these waves is not unique and is determined up to a conformal factor. However, as it occurs in typical nonlinear theory, the dynamics is not conformal-invariant.

Since the scalar field ``see" the effective geometry one can ask for nonlinear Lagrangians leading to a given effective geometry in a fixed background. To this end one proceeds by choosing a Lagrangian, determining the corresponding effective geometry and solving the Euler-Lagrange equations for $\varphi$. Unfortunately, since the effective metric depends on the field, such an approach is often intractable. A convenient means to
simplify the problem is to choose $L_w + 2 w L_{ww}$, this allows to partly control the interplay between the Lagrangian and the effective geometry (\ref{gEffInv}). Let us examine the simplest case where $L_w + 2 w L_{ww} = 1 = 1$ which we use hereafter. This choice obviously simplify (\ref{gEffInv}). This equation can be straightforwardly integrated to yield
\begin{equation}
L = w + 2 \lambda \sqrt{w} + C,
\label{5Nov2}
\end{equation}
where $\lambda$ is a non-zero c-number and $C$ a constant with respect to $w$, in particular one can set $C = - V(\varphi)$. This case is worth of considering in particular because of the properties it provides for the effective metric, but besides this it can be understood as a perturbation of the standard linear theory. Just to present a toy model that corresponds to a specific ``fake inflation" we will restrict the case in which the potential take the form \cite{branden}
\begin{equation}
V(\varphi) = - \lambda^2 e^{-\frac{2 H}{\lambda} \varphi} \, \left(1 + \frac{e^{-\frac{2 H}{\lambda} \varphi}}{2}\right).
\label{5Nov1}
\end{equation}

\subsection{Effective FRW metric in a Minkowskian background}
We consider the background metric as the Minkowski metric $\eta_{\mu\nu}$ in Cartesian coordinates and a field $\varphi$ depending only on time $\varphi = \varphi(t)$ and satisfying the Lagrangian (\ref{5Nov2}). The effective metric $\widehat{g}^{\mu\nu}$ felt by $\varphi$ is a spatially flat FRW metric:
\begin{equation}
\label{FRW}
ds^2 = dt^2 - a^2(t) \bigl(dr^2 + r^2 d\Omega^2\bigr).
\end{equation}

Since $\varphi$ do not depends on spatial coordinates the Euler-Lagrange equations reduce simply to:
\begin{equation*}
\ddot{\varphi}\bigl(L_w + 2 (\dot{\varphi})^2 L_{ww}\bigr) = - \frac{1}{2} \frac{\delta V}{\delta\varphi},
\end{equation*}
where a dot means a derivative with respect to the time. Now, the effective invariant length element reads
\begin{equation}
\label{ds2}
ds^2 = \widehat{g}_{\mu\nu}dx^\mu dx^\nu =  dt^2 - \frac{1}{L_w}\Bigl(dr^2 + r^2 d\Omega^2\Bigr).
\end{equation}
Note that $2L_{ww} = 1-L_w$ leads to $\widehat{g}_{tt} = 1$. Let us set the expansion factor on the effective FRW geometry to an inflationary form: $a(t) = e^{H t}$, $H$ being a real positive parameter. For that choice the equation $a(t)^{2} = 1/L_w$ leads to
\begin{equation}
\label{EqPhiInfl}
\sqrt{w} = \frac{\lambda}{e^{-2 H t} - 1}.
\end{equation}
Since $\sqrt{w}$ is positive $\lambda$ must be negative.  Assuming
$\dot{\varphi} \leqslant 0$ (calculations for $\dot{\varphi}
\geqslant 0$ are analogous) the above equation can be integrated
to~:
\begin{equation}
\label{PhiIfl}
\varphi = \frac{\lambda}{2 H} \ln (e^{2 H t} - 1) + K,
\end{equation}
where $K$ is a constant, which we set equal to zero. Solving (\ref{PhiIfl}) for $t$ allows to obtain precisely the form exhibited in equation (\ref{5Nov1}) of the potential. In other words, observation of the effective geometry $\widehat{g}_{\mu\nu}$ would led us to believe, erroneously, that we are in a de Sitter geometry. Although it is a toy model \cite{novello-huguet}, a similar situation can occur for other nonlinear theories.

We note that the covariant metric tensor of the effective metric defined by the relation $\widehat g_{\mu\nu}g^{\nu\lambda}=\delta_{\mu}^{\lambda}$, is given by
\begin{equation}
\widehat g_{\mu\nu}=\frac{1}{L_{w}}\,g_{\mu\nu}-\frac{2L_{ww}}{L_{w}\,(L_{w}+2wL_{ww})}\,\nabla_{\mu}\varphi\,\nabla_{\nu}\varphi\,. \label{EL1}%
\end{equation}
A straightforward calculation shows that the evaluation of the determinant of the effective metric yields
\begin{equation}
\det\,\widehat g^{\mu\nu}=L_{w}^{3}\,(L_{w}+2w\,L_{ww})\,.
\label{detg}
\end{equation}
It then follows that the unique theory that can be written in terms of its associated effective metric is the one provided by the BI-like form
\begin{equation}
L_{BI}=-\sqrt{bw+e}\,,
\label{BILag}
\end{equation}
where $b$ and $e$ are constants.

The energy density and pressure of the effective fluid description for the BI dynamics (from Eq.\ \ref{BILag})
\begin{equation}
\rho=e/\sqrt{bW+e}\,,\qquad p=-\sqrt{bW+e}\, \label{rhopChap}%
\end{equation}
are such that the equation of state takes a very simple form
\begin{equation}
p=-\frac{e}{\rho}\,. \label{ChapEOS}%
\end{equation}
A fluid with the above equation of state is known as Chaplygin gas \cite{ugo}.

\section{The method of the Dynamical Bridge}\label{dyn_brid}
Linear Maxwell theory and Born-Infeld electrodynamics are distinct theories. Not only they describe configurations that are not similar, they provide different answers for a same problem. Although such trivial statement seems obvious, it is possible to bypass its constrains and define a framework that exhibits a dynamical equivalence of these two theories. How is it possible? The dynamical bridge consists precisely in a scheme in which such similarity is present. The key point to understand such proposal is related to the distinction of the background space-time structure. We shall see that it is possible to map the dynamical properties of Maxwell theory written in a Minkowski space-time into the Born-Infeld electrodynamics described in a curved space-time, which is defined in terms of the electromagnetic field itself. It is clear that when considered in whatever unique metrical structure these two theories are not the same, they do not describe the same phenomena. However we shall see that by a convenient modification of the metric of the background structure an unexpected equivalence appears that establishes a bridge between these two theories making they represent the same situation. In other words we shall prove that a given dynamics described in a space-time can be mapped in another theory in a modified metric structure. We shall prove this for different kinds of fields, e.g., scalar, vector and spinors.

The main difference between the method of the equivalent metric and the one used in GR concerns the uniqueness of the geometry. In the latter, due to the universality of the metric, the gravitational field should also be described in the same and unique geometry once the gravitational field is identified with this geometry. Let us now describe a distinct situation in the equivalent DM and show how it is possible to represent the dynamics of the scalar field in terms of the associated DM. To clarify our description we compare it with the procedure used in GR.

\subsection{Exceptional dynamics}
In recent years, intense activity on analogue gravity has been developed \cite{nov_viss_volo}.This was concentrated in nonlinear electrodynamics, acoustics, hydrodynamics, optics inside media and various condensed matter systems. This term (analogue gravity) implies the description of distinct physical processes as modifications of the geometrical structure of the background space-time. Until now this analogy had been limited to perturbed aspects restricting it to the propagation of excitations (photons and quasi-particles) through a medium or a background field configuration. However it is possible to go one step ahead and provide an example which goes beyond this limitation describing dynamical features of fields in terms of their respective effective metric \cite{novello2}.

We will analyze this in two steps. In the first moment we will deal with the effective metric that was used to represent the propagation of the discontinuities. This special case is not very general and that is the reason to call the dynamics that are examples of this method as exceptional. In a second stage we will deal with a far reaching structure that concerns the complete equivalence without a stringent limitation of the dynamics. In other words, we would like to stress that in this general case the dynamical aspect has nothing to do with mimicking Einstein's equations through effective metrics as we did before.

The main steps to achieve this result are the following:
\begin{itemize}
\item{Consider a nonlinear field theory described in a flat Minkowski background;}
\item{Note that the propagation of the discontinuities of the field leads to the raising of a second metric $ \widehat{g}^{\mu\nu}$ such that the path of the waves are null geodesics in this effective geometry;}
\item{There exists a special class of theories such that its corresponding dynamics (which we will deserve the name exceptional) can be described alternatively as the gravitational interaction of the field in a given curved geometry;}
\item{The geometry of such gravitational space-time is precisely the effective metric $ \widehat{g}^{\mu\nu}.$ }
\end{itemize}
Thus it is possible to claim that the self-interaction described by exceptional dynamics is described in an equivalent way as the gravitational interaction of the field with its own effective metric. For pedagogical reasons we start our analysis with the simplest case of non linear scalar fields. Generalization to other cases will be described later on.

Let us note that the discontinuities of the field propagate in a curved space-time, although the field $\varphi $ is described by a non-linear theory in Minkowski geometry \cite{novello2}.

In the framework of GR it is the presence of gravity that allows the existence of curvature in the geometry. This has led to the interpretation of the dispersion relation of non-linear fields in terms of an effective metric as nothing but the simplification of its description, that is, a matter of language.

At this point we face the following question: is it possible that among all non-linear theories one can select a special class such that the dynamics of the field itself is described in terms of the effective metric? That is, can the dynamics of the field be unified with the propagation of its waves such that just one metric appears? Let us emphasize that this is not equivalent to the known property --- that occurs in hydrodynamics (and also in field theory) --- that the characteristic propagation of the field coincides with the propagation of its perturbations.

If such quality is possible, then the equation of evolution of $ \varphi $ is equivalent to a sort of gravitational interaction between $ \varphi $ and its effective metric. This means that the equation of motion (\ref{EqMotion}) can be written under the equivalent form
\begin{equation}
\label{1fevereiro2011}
\widehat{\Box}\, \varphi \equiv     \frac{1}{\sqrt{-\widehat{g}}} \,
\partial_\mu\Bigl(\sqrt{-\widehat{g}} \,
  \partial_\nu \varphi \, \widehat{g}^{\mu\nu}\Bigr) = 0.
\end{equation}
In general it is not possible to rewrite the nonlinear equation (\ref{EqMotion}) in the above form for an arbitrary lagrangian. However, we will show that there exist some special situations where this implementation becomes feasible. We first note the following relation
\begin{equation}
\partial_{\nu}\varphi\widehat{g}^{\mu\nu}=(L_{w}+2wL_{ww}) \partial_{\nu}\varphi\eta^{\mu\nu}.
\end{equation}
It then follows that the dynamics described by (\ref{EqMotion}) and (\ref{1fevereiro2011}) will be the same, provided the Lagrangian satisfies the condition
\begin{equation}
L_{w}=\sqrt{-\widehat{g}}(L_{w}+2wL_{ww}).
\end{equation}
Using the expression for the determinant (\ref{det}) the equivalence is provided by the nonlinear differential equation for the lagrangian
\begin{equation}
2 \, w \, L_{ww} + L_{w} - L_{w}^{5} = 0.
\label{1fevereiro2}
\end{equation}
We will call the system described by Lagrangians that satisfies condition (\ref{1fevereiro2}) as Exceptional Dynamics. In other words, non-linear systems described by exceptional dynamics may be alternatively interpreted as fields gravitationally coupled to its own effective geometry. Thus the field and the corresponding waves agree in the interpretation that the geometry of the space-time is given by $\widehat{g}_{\mu\nu}.$

Equation (\ref{1fevereiro2}) is such that the first derivative of the lagrangian with respect to $w$  may be obtained explicitly. Indeed, one obtains that
\begin{equation}\label{first}
L_{w}=\pm\frac{1}{(1-\lambda w^{2})^{\frac{1}{4}}},
\end{equation}
where $\lambda$ is an arbitrary real positive constant. The general solution of equation (\ref{1fevereiro2}) may be obtained as an infinite series given by the hypergeometric function
\begin{equation}
\label{excep}
L(w)=\pm w\,\mbox{hypergeom}\left(\left[\frac{1}{4},\frac{1}{2}\right],\left[\frac{3}{2}\right],\lambda w^{2}\right).
\end{equation}
We note that this dynamics is such that the admissible values of $ w $ are restricted to the domain $w^{2}<\lambda$. Thus, the theory naturally avoids arbitrarily large values of the kinematical term. Second, the lagrangian is a monotonic function of $ w $ as an immediate consequence of equation $(\ref{first})$. Third, it admits the linear theory as a limiting case for small values of $w.$ Lastly, the resulting function is odd, i.e., $L(w)=-L(-w)$.

It is instructive to expand the exceptional lagrangian to obtain a clear idea of its behavior in terms of $\lambda$, obtaining
\begin{equation}
\label{expansion}
L(w)=w+\frac{\lambda}{12}w^{3}+\frac{\lambda^{2}}{32}w^{5}+\frac{15\lambda^{3}}{896}w^{7}+O[x]^{9}
\end{equation}
For small values of the constant ($\lambda\ll1$) the exceptional lagrangian reduces to a cubic expression. We can arrive at this result assuming from the beginning a lagrangian in the vicinity of the linear theory of the form $L(w)=w+\epsilon f(w)$ with $\epsilon^{2}\ll\epsilon$ and solving the simplified equation $wf_{ww}-2f_{w}=0.$ The linear theory is obviously recovered when $\lambda=0$, implying that both the discontinuities and the field dynamics are represented in the same effective metric structure, that is,  $\eta^{\mu\nu}$.\\

Then, we have examined the case in which a self-interacting scalar field generates a geometrical arena for its own propagation. A simple extension can be elaborated in such a way that the original theory, describing the dynamics of the scalar field in a flat Minkowski arena can be alternatively described as the gravitational interaction of the field. Then a non-expected result appeared: the metric field that describes this gravitational effect is nothing but the same effective metric that controls the wave propagation. In other words both the propagation of the field discontinuities and the field dynamics are controlled by the same metric structure. The nonlinearities are such that the equations in Minkowski space-time mimic the dynamics of a ``free-field" embedded in a curved space-time generated by the field itself.

Let us summarize the novelty of such analysis:
\begin{itemize}
\item{For any field theory described on a Minkowski background by a non-linear Lagrangian $ L = L(w)$ the discontinuity of $\varphi$ propagates as null geodesics in an effective metric $\widehat{g}^{\mu\nu};$}
\item{As such, the theory presents a duplicity of metrics: the field is described in flat Minkowski space-time and its corresponding waves propagate as null geodesics in a curved geometry;}
\item{It is possible to unify the description of the dynamics of $ \varphi$ in such a way that only one metric appears. This is possible for those Lagrangians that represent exceptional dynamics;}
\item{In this case, the self-interaction of $ \varphi$ is described equivalently as if it was interacting minimally with its own effective geometry, allowing the interpretation in terms of an emergent gravitational phenomenon.}
 \end{itemize}

The structure of the nonlinear equations of motion suggests that the previous procedure can be adapted to other structures like nonlinear electrodynamics in a moving dielectric, vector and tensor field theories.

\subsection{Dynamical bridge: the case of scalar fields}
In the precedent sections we have displayed the possibility of using the modification of the geometry of space-time to describe paths of accelerated particles, massive or not, in flat or curved backgrounds. These proposals are generically called analogue models of gravity, once
they use the main idea of GR to map accelerated bodies in a given metric structure into geodesics of a curved geometry. In the  next sections we enter in a distinct analysis once we will deal not with forces in a given body but we will consider how different fields obeying a given nonlinear equation of motion can be  described in terms of modifications of the background geometry generated by themselves. We emphasize that this is not a gravitational framework, as we shall see.

We start by showing how a metric $q_{\mu\nu}$ naturally appears in nonlinear scalar field theories. Let us set the following nonlinear Lagrangian in  flat Minkowski space-time:
\begin{equation}
\label{lagr}
L = V(\Phi) \,w,
\end{equation}
where $w\equiv\eta^{\mu\nu} \partial_{\mu}\Phi \, \partial_{\nu} \Phi$. For $V = 1/2$ this is just the standard free massless Klein-Gordon scalar field. In the general case, the usual kinetic term is re-scaled by a field dependent amplitude (potential) $V(\Phi)$. Here we are using $\eta^{\mu\nu}$ but we could have used an arbitrary coordinate system as well, since the theory is generally covariant and there is no privileged reference frame. The field equation is
\begin{equation}
\frac{1}{\sqrt{-\eta}}\partial_{\mu} \, \left(\sqrt{-\eta}  \, \eta^{\mu\nu}\, \partial_{\nu} \Phi\right) + \frac{1}{2} \, \frac{V'}{V} \, w =0,
\label{23julho3}
\end{equation}
where $ V' \equiv dV/d\Phi $ and  $ \eta $ is the determinant of $\eta_{\mu\nu}.$

Now comes a remarkable result: the above field equation (\ref{23julho3}) can be seen as a massless Klein-Gordon field propagating in a curved space-time whose geometry is governed by $\Phi$ itself. In other words, the same dynamics can be written either in a Minkowski background or in another geometry constructed in terms of the scalar field. Following the steps established previously \cite{novello2}, we introduce the contra-variant metric tensor $ q^{\mu\nu}$ by the binomial formula
\begin{equation}
q^{\mu\nu} = \alpha \, \eta^{\mu\nu} + \frac{\beta}{w} \, \partial^{\mu}\Phi \, \partial^{\nu} \Phi,
\label{9junho1}
\end{equation}
where $ \partial^{\mu} \Phi \equiv \eta^{\mu\nu}\,\partial_{\nu} \Phi$ and $\alpha $ and $\beta$ are dimensionless functions of $\Phi$. The corresponding covariant expression, defined as the inverse $q_{\mu\nu} \, q^{\nu\lambda} = \delta^{\lambda}_{\mu}$, is also a binomial expression:
\begin{equation}
q_{\mu\nu} = \frac{1}{\alpha} \, \eta_{\mu\nu} - \frac{\beta}{\alpha \, (\alpha + \beta) \, w} \, \partial_{\mu} \Phi \, \partial_{\nu} \Phi.
\label{9junho11}
\end{equation}

Now we ask whether it is possible to find $\alpha $ and $ \beta,$ in such a way that the dynamics of the field\ (\ref{23julho3}) takes the form
\begin{equation}
\Box \, \Phi = 0,
\label{23julho5}
\end{equation}
where $\Box$ is the Laplace-Beltrami operator relative to the metric $q_{\mu\nu}$, that is
$$ \Box \, \Phi \equiv  \frac{1}{\sqrt{- q}} \partial_{\mu} ( \sqrt{- q} \,q^{\mu\nu} \,\partial_{\nu} \Phi).$$
To answer this question we evaluate the determinant $q$ of the metric $q_{\mu\nu}$, for which a direct calculation yields
\begin{equation}
\sqrt{- \, q} = \frac{\sqrt{- \,{\eta}}}{\alpha \, \sqrt{\alpha \,(\alpha + \beta)}}.
\label{lolo}
\end{equation}
Using the fact that $q^{\mu\nu}\partial_{\nu}\Phi=(\alpha+\beta)\eta^{\mu\nu}\partial_{\nu}\Phi$, the final result is summarized in the following:

\begin{theorem}
\label{lemma_phi}
Given the Lagrangian $L = V(\Phi) w$ with an arbitrary potential $V(\Phi)$, the field theory satisfying Eq.\ (\ref{23julho3}) in Minkowski spacetime is equivalent to a massless Klein-Gordon field $\Box \Phi = 0$ in the metric $q^{\mu\nu}$ provided that the functions $\alpha(\Phi)$ and $ \beta(\Phi) $ satisfy the condition
\begin{equation}
\alpha + \beta = \alpha^{3} \, V.
\label{9julho5}
\end{equation}
\end{theorem}

Remarkably, this equivalence is valid for any dynamics described in the Minkowski background by the Lagrangian $L$. This fact can be extended to other kinds of nonlinear Lagrangian \cite{goulart}.

\subsection{Dynamical bridge: The case of electromagnetic field}
Recently, it was shown that the Born-Infeld dynamics in a curved space-time endowed with an associated metric $ \widehat e_{\mu\nu}$ is dynamically equivalent to the linear Maxwell electrodynamics described in a flat Minkowski background \cite{nov_fal_gou}. Namely, there is a map relating these two dynamics in such a way that they are distinct representations of one and the same physics. It means that \textit{any} solution of the former will also be a solution of the latter, which is given in terms of a preestablished map. This task
is highly nontrivial and  becomes possible only if the space-time metric depends explicitly on the electromagnetic field. Due to the algebraic structure of the electromagnetic tensor $F_{\mu\nu}$ and its dual ${}^*F_{\mu\nu},$ there are some closure relations that allow the existence of this map, generating a connection between those two paradigmatic theories. The apparent disadvantage is to leave aside the simple Minkowski background $\eta_{\mu\nu}$ and to go to a specific curved space-time $\widehat{e}_{\mu\nu}$, which is constructed solely in terms of the background metric and the electromagnetic fields. In principle, one could suspect that this map is useless because it links a very simple electromagnetic theory in flat space to a nonlinear theory in a curved space-time. Notwithstanding, recently \cite{nov_bit} it was shown that physical principles can guide us in the choice of the more appropriate representation to describe the system under consideration.

To describe a physical theory one has to specify not only the lagrangian that contains the dynamics of the fields but also the space-time structure where the theory is defined. To avoid notational cumbersomeness, we shall denote every object in the curved space-time defined by the metric $\widehat{q}_{\mu \nu}$ with an upper hat. Thus, in Maxwell's theory, every tensor is raised and lowered by the Minkowski metric $\gamma_{\mu \nu}$ while in the curved space-time representation, where we will construct the Born-Infeld theory, shall be raised and lowered using the $\widehat{e}_{\mu \nu}$ metric.

Let us begin with Maxwell's theory in Minkowski metric  $\gamma_{\mu \nu}$ is determined by the Lagrangian $L=-\frac{F}{4}$. We set $F^{\mu \nu}\equiv F_{\alpha \beta}\gamma^{\mu \alpha}\gamma^{\nu \beta}$ and $F\equiv F_{\mu \nu}F_{\alpha \beta}\gamma^{\mu \alpha}\gamma^{\nu \beta}$. Born-Infeld theory described in the curved metric $\widehat{e}_{\mu \nu}$ is determined by the Lagrangian $\widehat{L}=\beta^2\left(1-\sqrt{\widehat{U}}\right)$, where we thus have
\begin{equation*}
\widehat{U}=1+\frac{\widehat{F}}{2\beta^2}-\frac{\widehat{G}}{16\beta^4},
\end{equation*}
with
$$\widehat{F}^{\mu \nu}=F_{\alpha \beta} \, \widehat{e}^{\mu \alpha} \, \widehat{e}^{\nu \beta}, \quad \widehat{F}=F_{\mu \nu} \, F_{\alpha \beta} \,\widehat{e}^{\mu \alpha} \, \widehat{e}^{\nu \beta}, \quad \mbox{and} \quad \widehat{G}=\frac{\sqrt{-\gamma}}{\sqrt{-\widehat{e}}} \, \eta^{\mu \nu \alpha \beta} \, F_{\mu \nu} \, F_{\alpha \beta}=\widehat{F}{}^{\stackrel{\ast}{\mu\nu}} \, F_{\mu \nu}.$$
The dynamical equation for the electromagnetic field in the Born-Infeld theory is
\begin{equation}
\label{din_bi}
\partial_{\nu}\left[ \frac{\sqrt{- \, \widehat e }}{\widehat U } \left(  \widehat F^{\mu\nu} - \frac{1}{4 \, \beta^{2}} \, \widehat G \, {}^*\widehat F^{\mu\nu}\right)  \right] = 0,
\end{equation}
where $\widehat{e}$ is the determinant of the electromagnetic metric $\widehat e_{\mu\nu}$, which is defined as
\begin{equation}
\widehat e^{\mu\nu} \equiv a \, \eta^{\mu\nu} + b \, \Phi^{\mu\nu},
\label{17outubro}
\end{equation}
and $ \Phi_{\mu\nu} \equiv F_{\mu\alpha} \, F^{\alpha}{}_{\nu}$. Due to the algebraic relations of the Faraday tensor (\ref{algrel1})-(\ref{algrel4}), there is a unique way to define the electromagnetic metric where $a$ and $b$ are functions of the Lorentz invariants $F$ and $G$ and $\Phi^{\mu \nu}\equiv F^{\mu \alpha}F_{\alpha}{}^{\nu}$.

The term electromagnetic metric is justified by the fact that $\widehat{e}_{\mu \nu}$ depends only on the electromagnetic field. Note the very convenient property that the inverse of the above structure, which in general is given by an infinite series but in the present case is again a binomial quantity, also as a consequence of Eqs.\ (\ref{algrel1})-(\ref{algrel4}), that is
\begin{equation}
\label{qmet2}
\widehat{e}_{\mu \nu}=A \, \eta_{\mu \nu}+ B\, \phi_{\mu \nu},
\end{equation}
where $A=(2-n F)(2aQ)^{-1}$, $B=-n(aQ)^{-1}$ and we are using the definitions $n \equiv b/a$ and $Q \equiv1-(1/2)nF-(1/16)n^2G^2$.

The definition of $\widehat F^{\mu\nu}$ and the choice of the metric $\widehat{e}_{\mu\nu}$ implies that these fields are related through the expression
\begin{equation}
\label{transf_campos}
\left(\begin{array}{c}
\fracc{\widehat F^{\mu\nu}}{a^2}\\[1.5ex]
\fracc{^*\widehat F^{\mu\nu}}{a^2}
\end{array}\right)=
\left(\begin{array}{cc}
p-n \, F q&-\fracc{n \, G \, q}{2}\\[2ex]
-\fracc{n \,  G \, q}{2}&p
\end{array}\right)
\left(\begin{array}{c}
F^{\mu\nu}\\[2ex]
^*F^{\mu\nu}
\end{array}\right),
\end{equation}
where $p=1+\fracc{n^2 \,G^2}{16}$ and $q=1-\fracc{n \, F}{4}$.

We can interpret Eq.\ (\ref{transf_campos}) as nothing but a map from $ (F_{\mu\nu},  F_{\mu\nu}^{*}) $ into $ (\widehat{F}_{\mu\nu}, \widehat{F}_{\mu\nu}^{*}). $ Substituting this equation in Eq.\ (\ref{din_bi}) and making the requirement that Born-Infeld dynamics in the $ \widehat{e}_{\mu\nu} $ corresponds to the Maxwell dynamics in flat Minkowski space-time, we obtain the following equations for the coefficient of the electromagnetic metric \cite{nov_fal_gou}

\begin{equation}
\label{eqs_din_eq}
\left\{\begin{array}{lcl}
p-n Fq+\fracc{\widehat G}{2\beta^2} \, n \, q&=&-\fracc{Q}{4},\\[2ex]
- n \, G \, q+\fracc{\widehat G \, p}{2\beta^2}&=&0.
\end{array}\right.
\end{equation}
The new invariants, for instance, reads
\begin{equation*}
\widehat{F}=a^2\left[F- \frac{n}{2}(F^2+G^2)\left(1-\frac{nF}{2}\right)\right],\quad \mbox{and}\quad \widehat{G}=a^{2}QG.
\end{equation*}
At this stage, the mapping of the two dynamical systems sum up to make the identification term by term in both equations. In other words, choosing appropriately the unknown function of the EM metric (\ref{17outubro}), \textit{Maxwell's equation in Minkowski geometry and Born-Infeld in this
curved space-time $ \widehat{e}_{\mu\nu}$ describes one and the same dynamics for the electromagnetic field}. Our task consists in a direct comparison of the Maxwell equations in the background $\eta_{\mu\nu}$ with Born-Infeld dynamics in the background $\widehat{e}_{\mu\nu}$.

Equating terms proportional to $F^{\mu \nu}$ and to ${}^{*}F^{\mu \nu}$ we find the following conditions to be fulfilled in order to obtain the equality of the dynamics
\begin{eqnarray}
&&\frac{a^2 Q^2}{2 \beta^2}= - nq\label{1}\\
&&p - nqF=\frac{\sqrt{\widehat U}}{Q}\label{2}
\end{eqnarray}
In the above equations we are considering $a$ and $ n $ as the unknown functions and $Q$ and $\widehat{U}$ as functions of the formers. Having two equations and two unknown, one can solve to find the two functions, hence, the metric (\ref{17outubro}). However, using equation (\ref{1}), it follows immediately that equation (\ref{2}) trivialize to $1=1$, i.e. if (\ref{1}) is satisfied then (\ref{2}) is also immediately satisfied. Thus the proof of equivalence of both dynamics (Maxwell in Minkowski space-time and Born-Infeld in $\widehat{e}$ metric) reduces to state that  $a$ and $ n $ satisfy
the constraint (\ref{1}). Note that the Bianchi identities in the electromagnetic metric $\widehat{e}_{\mu\nu}$ is identically satisfied. Thus we have shown that there exist a complete equivalence between the two representations, i.e., Born-Infeld theory represented in the space-time with electromagnetic metric $\widehat{e}_{\mu\nu}$ is equivalent dynamically to Maxwell's theory in Minkowski background. We would like to emphasize that although the two metrical structures are different, we are dealing with two representations of the same and unique dynamical process.

Beside the freedom in one of the functions that define the metric $\widehat{e}_{\mu \nu}$, which seems to be related to the conformal invariance of Maxwell's equations, there are still some conditions that constrict these functions. A straightforward calculation using the Cayley-Hamilton theorem (see the first section) shows that the determinant of the metric (\ref{qmet2}) is $\sqrt{-\widehat{e}}=a^{-2}Q^{-1}\sqrt{-\gamma}$. Therefore, $Q$ that is a function of $ n$ has to be positive definite, i.e. $Q>0$. In addition, we want the map to be for any solution of Maxwell's equations, hence for any value of the two invariants $F$ and $G$. Let us investigate some particular regimes.
\begin{itemize}
\item[i)] \underline{case $F=0$:} In this case, equation (\ref{1}) demands that $ n <0$. Furthermore, by its own definition $n=-\frac{4}{G}\sqrt{1-Q},$ which restrict $Q$ to the domain $0<Q\leq 1$.\\
\item[ii)] \underline{case $G=0$:} the relation between $n $ and $Q$ now gives $n=\frac{2}{F}\left(1-Q\right)$. In addition, from Eq.\ (\ref{1}) we can solve for the function $a$ as
\begin{equation}
a=\frac{\beta}{|Q|}\sqrt{\frac{2}{F}\left(Q^2-1\right)}.
\end{equation}
Since $a$ is a real function, we have
\begin{displaymath}
G=0 \rightarrow \left\{\begin{array}{l}
F>0, \quad \mbox{and}\quad Q>1\\
F<0, \quad \mbox{and}\quad 0<Q<1
\end{array}\right.
\end{displaymath}

\item[iii)] \underline{case $F=G=0$:} the vanishing of both invariants simultaneously, which is the case for plane waves, trivialize all conditions. In fact, we simply have $Q=1$ and $a^2=-2\beta^2 \, n.$ Thus, it suffice to have $n <0$.
\end{itemize}

Having established the connections between Maxwell's electromagnetism and Born-Infeld theory, we can now analyze some examples of metrics generated by this mapping. Maybe the most interesting class of examples are the $G=0$, which includes the pure electrostatic or magneto-static field case. Let us consider $G=0$, i.e.
\begin{eqnarray*}
Q=e^{F/4 \beta^2}\quad &\mbox{and}&\qquad a=\frac{\beta}{Q}\sqrt{\frac{2}{F}\left(Q^2-1\right)}.
\end{eqnarray*}

There are three limiting cases: weak fields $|F|\ll 1$, strong electric field $F\ll -1$ and strong magnetic field $F\gg 1$. When $(i)$ $|F|\ll 1$, we expand $Q$ and $a$ in powers of $F$ and find the electromagnetic metric in first order corrections as
\begin{equation}
\widehat{q}_{\mu \nu}=\eta_{\mu \nu}+\frac{1}{2\beta^2}F_{\mu}{}^{\alpha}F_{\alpha \nu};
\end{equation}
for $F\ll -1$ $(ii)$, we have $Q=e^{F/4 \beta^2}$, $a \approx \beta\sqrt{\frac{2}{|F|}}e^{-F/4 \beta^2}$ and the metric assumes the form
\begin{equation}
\widehat{e}_{\mu \nu}\approx \frac{1}{\beta} \, \sqrt{\frac{2}{|F|}} \, F_{\mu}{}^{\alpha} \, F_{\alpha \nu}.
\end{equation}
Finally, $(iii)$ for $F\gg 1$, it gives a metric of the form
\begin{equation}
\widehat{e}_{\mu \nu}\approx\sqrt{\frac{F}{2\beta^2}} \, \eta_{\mu\nu}
\end{equation}

Consider the regime in which the electromagnetic field is very low in comparison to the critical field, i.e., $F\ll\beta^{2}$ (case $i$). Then, we perform a power series expansion in terms of $F/2\beta^2$ for the objects associated to the bridge and see that, in this regime, the electromagnetic metric reduces to the form
\begin{equation}
\widehat e_{\mu\nu} \approx \eta_{\mu\nu} + \frac{1}{2 \, \beta^{2}} \, \Phi_{\mu\nu}.
\label{1721}
\end{equation}
From now on, we keep this expression for the metric. Remark that it is possible to interpret the presence of the lectromagnetic metric as a small deviation from the flat space-time generated by the electromagnetic field. From the dynamical bridge between the equivalent representations it follows that we can describe pure electromagnetic phenomenon either in terms of Maxwell theory in flat Minkowski space-time or as Born-Infeld dynamics in the electromagnetic curved geometry.

Henceforth we denote Maxwell-Minkowski (MM) representation the case where one chooses to describe electromagnetic processes in the linear Maxwell theory in the flat Minkowski space-time. On the other hand, when one applies the dynamical bridge approach and describes the same processes in the nonlinear Born-Infeld theory in a curved space-time driven by the metric $\widehat{e}_{\mu\nu}$, it will be denoted as $\widehat E-$ representation. Let us emphasize that these representations describe one and the same dynamics.

{\bf The role of the electromagnetic metric.} The presence of a curved structure of the space-time in the realm of electromagnetic fields can be an important theoretical instrument of analysis only if one introduces a prescription of how matter perceives this geometry. Then, the question is: how matter interacts with the electromagnetic metric? This question appears immediately as long as one wants to explore this metric formulation in the presence of matter.

Recently \cite{nov_bit} the idea was proposed to interpret the action of electromagnetic metric in a similar way as in gravitational processes
assuming that through this geometric channel all particles display the same behavior. The statement that defines a Universal Hypothesis of electromagnetic interaction takes the following form: {\it all kind of particles, charged or not charged, interacts with the metric $ \widehat e_{\mu\nu}$ in a universal and unique way.}

The simplest manner to realize this hypothesis is by using the minimal coupling principle that we borrow from gravitational processes as they are described in GR.  Although this hypothesis may appears, at first glance, useless or even no sense for traditionalists, we shall see that it contains a well-posed program with observational consequences that could prove or disprove it. In order to proceed with this idea we examine the case of test spinor fields {in this background and investigate what new effects one can extract from this geometrical scenario.

The extended dynamical bridge concerns the behavior of all kind of matter. In the $\widehat{E}$-representation the assumption of complete democracy---that is, the idea that any kind of matter, charged-or-not, lives in the geometry $\widehat e_{\mu\nu}$---implies an equivalent effect in the MM-representation. In particular, the equation of motion of the neutrino in the electromagnetic geometry, provokes the need to assume a direct interaction between the
neutrino and the electromagnetic field in the MM-representation. Indeed, one is led to accept that in this representation there should exist an extra term in the Lagrangian which is the analogue of the minimal coupling principle between electromagnetic field and the neutrino embedded in the metric $\widehat{e}_{\mu\nu}$. We shall see that, for instance, the role of the parameter $\beta$ that appears explicitly in the $\widehat{E}$-representation, becomes the ratio between the magnetic moment of the particle and its corresponding mass.

{\bf Minimal coupling principle.} Let us describe the electromagnetic effects on the spinor field equation in the $\widehat{E}$-representation. We explore the equivalence displayed by the dynamical bridge and deal with an extended version of the way matter interacts with the electromagnetic field. It is a universal and well-accepted belief that only charged matter is able to couple directly with the electromagnetic field. Nevertheless, from what we have shown above, the analysis in the $\widehat{E}-$representation allows a new possibility of interaction.

The equivalence between the Born-Infeld theory in the $\widehat E$-representation and the Maxwell dynamics in flat space was shown only in the case of free fields. We examine now the extension of such an equivalence by assuming that matter couples universally to the electromagnetic field in the $\widehat e_{\mu\nu}$ framework. Following this approach, let us investigate the consequences of the coupling in the $\widehat{E}$-representation. We start by noticing that it yields two possible forms of interaction of the matter with the electromagnetic field, that is: $(i)$ through the vector potential $A_{\mu}$; $(ii)$ through the metric tensor $\widehat e_{\mu\nu}$.

Once only the geometrical form of interaction was not considered before, we will concentrate our analysis here on this second way. Let us point out that the existence of the electromagnetic geometry makes sense and may have further consequences only if this geometry is perceived for all kind of matter. This means that a test particle can couple to the electromagnetic field without having electric charge or, more precisely, if a particle has an electric charge it couples with the electromagnetic field through the standard channel via the potential $A_{\mu}$. Otherwise, it interacts with the electromagnetic geometry in an unique and the same way, according to the minimal coupling principle stated above. What are the main consequences of this hypothesis? That is the purpose of the next section.

\subsection{A proposal for the origin of the anomalous magnetic moment}
Few years ago, a new form of contribution for the anomalous magnetic moment of all particles was suggested \cite{nov_bit}. This common origin is displayed in the framework of a recent treatment of electrodynamics that is based in the introduction of an electromagnetic metric which has no gravitational character as we commented above. This effective metric constitutes a universal pure electromagnetic process perceived by all bodies, charged or not charged. As a consequence it yields a complementary explanation for the existence of anomalous magnetic moment for charged particles and even for non-charged particles like neutrinos.

In the standard model of particle physics, neutrinos possess the most intriguing and interesting behavior. It is understood that they have only gravitational and weak interaction, although some extensions of the standard model allow that neutrinos could have an effective magnetic moment \cite{sanda,lee,helayel,kau}. For any charged particle the classical magnetic moment ($\mu$) is inversely proportional to the mass $\mu=e\hbar/2m$, where $\hbar$ is the reduced Planck constant, $e$ is the elementary charge and $m$ is the mass of the particle. In the case of electrons, for example, $\mu\equiv\mu_B = e\hbar/2m_e$ is the Bohr magneton. However, in the case of neutrinos $\nu$, it has been suggested that $\mu_{\nu}$ depends linearly on the neutrino mass. As a consequence, this type of coupling between the spinor field and the electromagnetic field presents some difficulties concerning the variational principle and the re-normalization process.

The idea proposed is that the magnetic moment of any particle is composed of two very distinct parts: the standard one that depends
on the charge of the particle---from dimensional analysis, it is inversely proportional to the corresponding mass---and another part (called the geometrical magnetic moment) that depends linearly on the mass. This second case, which is many orders of magnitude lower than the standard one, is common for all particles and does not depend on their charge. In other words, we shall see that part of the anomalous magnetic moment of charged particles is a consequence of such geometrical magnetic moment and for neutral particles, like neutrinos, such proposal provides a nonzero magnetic moment. We shall compare the effects of both terms from experimental data.

Of course, this idea is based on the equivalence between the electromagnetic Born-Infeld theory in a specific curved geometry and Maxwell's theory in the Minkowski space-time, as we have revisited previously. This means that any solution of the former is also a solution of the latter, which is given in terms of a prescribed map. This is possible because the associated curved space-time depends only on the electromagnetic fields. Indeed, due to the algebraic structure of the electromagnetic two-form $F_{\mu\nu}$ and its dual, there exist a kind of closure relation that allows the existence of this mapping, thus generating a dynamical bridge between these two paradigmatic theories.

The application of the dynamical bridge to specific situations suggests a new manner to understand the origin of the anomalous magnetic moment, which
depends linearly on the mass, as a direct consequence of the electromagnetic geometry as we shall analyze next.

{\bf The $\widehat{E}-$metrical origin of the neutrino magnetic moment.} In the case of neutrinos, which are uncharged, there is only the geometrical channel to couple with the electromagnetic field. We will apply the minimal coupling principle in the $\widehat E$-representation and analyze the consequences of this in the Maxwell-Minkowski representation.

We start by defining the Dirac matrices $ \widehat\gamma^\alpha $ associated to the curved space by the relation
\begin{equation}
\{\widehat \gamma^{\mu},\widehat \gamma^{\nu}\}= 2 \, \widehat e^{\mu\nu} \, \textbf{1},
\end{equation}
where $ \textbf{1} $ is the identity matrix of the Clifford algebra and the curly brackets means anti-commutation. Analogously, in the MM-representation, we have
\begin{equation}
\{\gamma^{\mu},\gamma^{\nu}\}= 2 \, \eta^{\mu\nu} \, \textbf{1}.
\end{equation}

It then follows that we can set
\begin{equation}
\label{gamma_hat}
\widehat\gamma^{\mu}=\gamma^{\mu}- \frac{1}{4\, \beta^{2}} \Phi^{\mu}{}_{\alpha}\gamma^{\alpha}
\end{equation}
and the closure relations above are automatically satisfied. Therefore, the dynamical equation for the spinor field, in the metric $\widehat e_{\mu\nu}$, becomes
\begin{equation}
i \hbar c\, \widehat \gamma^{\mu} \widehat \nabla_{\mu} \, \Psi -  m c^{2}\, \Psi = 0. \label{746}
\end{equation}
The covariant derivative $\widehat \nabla_{\mu} \equiv
\partial_{\mu}-\widehat \Gamma^{FI}_{\mu}-\widehat V_{\mu}$ is given by the Fock-Ivanenko coefficients $\widehat \Gamma^{FI}_{\mu}$ plus an arbitrary element of the algebra \cite{novello_73}, which yields
\begin{equation}
\widehat \nabla_{\mu} \, \widehat \gamma^{\nu} = [ \widehat V_{\mu},  \widehat \gamma^{\nu} ], \label{1235}
\end{equation}
still holding the metricity condition $ \widehat \nabla_{\alpha} \,\widehat q^{\mu\nu} = 0$. The presence of $V_{\mu}$ suggests that the interaction between the electromagnetic field and $\Psi$ occurs through the internal space. In the absence of any kind of matter, we are free to assume that the commutator on the right-hand side of Eq.\ (\ref{1235}) vanishes. However, when matter (of any kind) exists, $\widehat V_{\mu}$ depends simultaneously on the electromagnetic field and on the properties of the matter field and, in the case of spinors, we set
\begin{equation}
\label{choice_v} \widehat V_{\mu} = i\frac{m \, c}{\hbar \,\beta} \,
F_{\mu\nu} \, \widehat \gamma^{\nu}\gamma_5,
\end{equation}
where $m$ is the neutrino mass in the minimal extended version of the standard model. In the general case, it was demonstrated that $\widehat V^{\mu}$ must transform like the Fock-Ivanenko connection in order to verify the conservation laws \cite{formiga}. Thus, according to the choice of $\widehat V^{\mu}$ given by\ (\ref{choice_v}), that contains the anti-symmetric tensor $F_{\mu\nu}$, guarantees the conservation law. In order to guarantee that we are dealing only with uncharged particles, we assume that the Fock-Ivanenko connection does not have any term proportional to the identity ${\bf 1}$. It should be remarked that even if we consider massless neutrinos, we can rewrite $\widehat V_{\mu}$ is terms of the neutrino energy and then the coupling to the electromagnetic field is still present. This replacement of $m$ by $E$ suggests that such interaction should be investigated in different scenarios where the right-handed neutrinos could appear \cite{shapos,geng}.

Using Eqs.\ (\ref{1721}) and\ (\ref{choice_v}) into the equation of motion (\ref{746}), we get

\begin{equation}
i \, \hbar c \, \gamma^{\mu} \partial_{\mu} \, \Psi  + \frac{\,m c^{2}}{\beta} \,
F_{\mu\nu} \, \sigma^{\mu\nu}\gamma_5 \, \Psi - m \, c^{2} \, \Psi = 0,
\label{1730}
\end{equation}
which, according to the regime under consideration, corresponds to the first order approximation in $F/2\beta^2$ where $\sigma^{\mu\nu} = (\gamma^{\mu}\gamma^{\nu} -\gamma^{\nu}\gamma^{\mu})/2$.

The effect of the universal minimal coupling between matter of any kind and the electromagnetic field using the metric $\widehat e_{\mu\nu}$ provides an effective magnetic moment that depends on the mass of the spinor field and on the parameter $\beta$, that is
\begin{equation}
\mu_{G} \doteq \frac{m \, c^{2}}{\beta}.
\label{1112011}
\end{equation}

Remarkably, the hypothesis of universality of $ \widehat e_{\mu\nu} $ reveals the existence of an alternative origin for the magnetic moment of all particles (charged or not) from first principles. If we had applied the minimal coupling principle to the MM representation, we have never obtained such term. In this representation, it corresponds to a non-minimal coupling, leading to some difficulties to be interpreted inside the standard approach, as pointed in the literature. The compatibility of this result induced by the $\widehat E$-representation in the MM-representation implies that the Lagrangian describing the interaction between the neutrino and the electromagnetic field contains an extra term in the MM-representation which is usually introduced by hand and without further justification. In other words, the presence of a magnetic moment for the neutrino in the standard MM-representation should not be viewed as an exotic surprise but instead should be understood---on the light of the
Dynamical Bridge method---as a consequence of the universality of the geometry $\widehat e_{\mu\nu}$ in the $\widehat E$-representation. Next section, we shall introduce this map for charged particles and then compare with experiments.

{\bf The geometrical magnetic moment for charged particles.} It is important to emphasize that the value of the magnetic moment obtained in the previous section contains only the general contribution for any particle. Charged particles have an extra source for $ \mu $ that is related to their charge, for instance, the Bohr magneton $\mu_{B} = e \, \hbar /2 m_{e}$ for the electron. Thus, the total value of the electron magnetic moment ($\mu_e$) should be read as
$$\mu_{e} = \mu_{B} + \frac{m_{e} \, c^{2}}{\beta}+\mbox{quantum corrections}.$$
The first part corresponds to the standard magnetic moment, the second one corresponds to the $\widehat{E}-$metrical contribution and the last one comes from loop-quantum corrections. A comparison with theoretical predictions and experiments \cite{balant,roberts,marciano,gabrielse,alex,lych,passera} is presented in the original paper \cite{nov_bit}.

\subsection{Dynamical bridge: the case of spinor fields}
\protect\label{algebraic}

In the precedent sections we have shown how fundamental fields can satisfy equivalent dynamical equations in different background geometries. Namely, a given nonlinearity of the field in the flat space can be hidden in the curvature of an effective curved space. This was shown to be true for spin $0$ and $1$ fields \cite{novello2,goulart,JCAP,nov_fal_gou,nov_bit} and can also be applied in the kinematical context as we described in this review \cite{nov_bit_gordon,nov_bit_drag}. In this section we complete our task and demonstrate that such procedure also holds for spinor fields. In particular, we shall see that the self-interacting term of the Heisenberg-Nambu-Jona-Lasinio (HNJL) model \cite{heisenberg,njl,hatsuda}--can be derived from the minimal coupling of a Dirac field with an effective curved geometry. This Dynamical Bridge goes together with the chiral symmetry breaking for massive and massless spinor fields if the coupling constant is sufficiently large. This approach was suggested recently \cite{novello-edu-faci} as providing a new geometrical explanation for the non observation of the right-handed neutrinos.

Let $\Psi$  be a massless spinor field satisfying the Dirac equation
$$i \widehat\gamma^\mu \widehat\nabla_{\mu} \Psi = 0 $$
in a curved background (with no gravitational character) and assume that such curved geometry represents the modification of the flat space caused by the field $\Psi$ itself through the specific form of the metric:
\begin{equation}
\label{first-eq}
\widehat{g}^{\mu\nu}  =  \eta^{\mu\nu} + 2\,\alpha\, H^{\mu} H^{\nu}.
\end{equation}
The arbitrary function $\alpha$ depends on the scalars $A\equiv\bar\Psi \Psi$ and $B\equiv i \bar\Psi \gamma_5 \Psi$. The vector $H^{\mu}$ is given in terms of the vector $J^{\mu}$ and axial $I^{\mu}$ currents. It was proved that the Dirac equation in such curved space-time is dynamically equivalent to the dynamics of Nambu Jona-Lasinio equation in the flat Minkowski space-time. Afterwards we rewrite this dynamics in Minkowski space (equipped with the metric $\eta_{\mu\nu}$). This leads to the two following equations
\begin{eqnarray}
& i\gamma^\mu \partial_{\mu}\Psi_L=0,\\
& i\gamma^\mu \partial_{\mu}\Psi_R + s (A+iB\gamma_5) \, \Psi_L=0, \label{zzz22}
\end{eqnarray}
where $s$ depends on $\alpha$ and its derivatives, $\Psi_L\equiv(1/2)({\bf 1}-\gamma_5)\Psi$ and $\Psi_R\equiv(1/2)({\bf 1}+\gamma_5)\Psi$ are respectively the left-handed and right-handed chirality components of $\Psi$, with $\Psi= \Psi_L+\Psi_R$. The left-handed component still obey the Dirac dynamics in Minkowski space while the right-handed component verifies the equation (\ref{zzz22}), which is a generalization of the HNJL dynamics. The curvature of the effective space goes to the nonlinearity of HNJL dynamics in Minkowski space, but only for the right-handed component of the spinor field. This introduces a new answer to the experimentally observed chiral symmetry breaking of neutrinos and stands as an alternative for the Standard Model interpretation which assumes that right-handed neutrinos do not exist, i.e., they do not interact through the standard model forces.

{\bf Overview on the mathematical tools for the DB.} In order to properly construct the Dynamical Bridge for spinor fields let us describe some mathematical ingredients. It is well known that any metric tensor $\widehat{g}^{\mu\nu}$ can always be decomposed into
\begin{equation}
\label{met}
\widehat{g}^{\mu\nu} = g^{\mu\nu} + \Sigma^{\mu\nu},
\end{equation}
where $g^{\mu\nu}$ is some background metric and $\Sigma^{\mu\nu}$ is a rank two tensor field. As we discussed before, this metric tensor admits an inverse with the same binomial form if we impose the condition $\Sigma^{\mu\nu} \, \Sigma_{\nu\lambda} = p \, \delta^{\mu}_{\lambda} + q \, \Sigma^{\mu}{}_{\lambda}$, where $p$ and $q$ are arbitrary functions of the coordinates. For the sake of simplicity, we set the background metric $g_{\mu\nu}$ to be the Minkowski one $\eta_{\mu\nu}$ in arbitrary coordinate systems. For eventual generalizations concerning the background, we refer to the Dami\~ao-Soares lectures \cite{spin_curv}.

With a generic spinor field $\Psi$ we construct two scalars $A\equiv\bar\Psi \Psi$ and $B\equiv i \bar\Psi \gamma_5 \Psi$ and two currents defined as $J^{\mu} \equiv \bar\Psi_{} \gamma^{\mu} \Psi_{}$ and $I^{\mu} \equiv \bar\Psi_{} \gamma^{\mu} \gamma_5 \Psi_{}$. The $\gamma^{\mu}$'s are the Dirac matrices which satisfy the closure relation of the Clifford algebra.
Using the Pauli-Kofink identity for an arbitrary element $Q$ of the Clifford algebra
\begin{equation}
\label{pkofink}
(\bar\Psi Q\gamma_{\lambda}\Psi)\gamma^{\lambda}\Psi=(\bar\Psi Q\Psi)\Psi - (\bar\Psi Q\gamma_5\Psi)\gamma_5\Psi,
\end{equation}
the following relations are easily derived $J^2 = - I^2 = A^2+B^2$ and $J_{\mu}I^{\mu} = 0$, where $X^2\equiv \eta_{\mu\nu} \,  X^{\mu}X^{\nu}$ for the vectorial objects.

Here the method consists in writing $\Sigma^{\mu\nu}$ of the curved space-time metric (\ref{met}) in terms of the dynamical field, which is given by the spinor $\Psi.$ The simplest way to do this is to set
\begin{equation}
\label{Sigma2}
\Sigma^{\mu\nu} \doteq 2\,\alpha\, H^{\mu} H^{\nu},
\end{equation}
where $H^{\mu} \doteq J^{\mu} + \epsilon I^{\mu}$ is a linear combination of the currents, $\epsilon$ is an arbitrary constant and $\alpha$ is an arbitrary function of $A$ and $B$. Eq.\ (\ref{pkofink}) leads to the following identities
\begin{equation}
\label{gamma.H}
H_{\mu}\gamma^{\mu}\,\Psi = ({\bf 1}+\epsilon\gamma_5) (A+iB\gamma_{5})\Psi, \qquad \mbox{and} \qquad H^2=(1-\epsilon^2)J^2.
\end{equation}
Finally, using the relation $\widehat g^{\mu\nu} \widehat g_{\nu\lambda} = \delta^\mu_{\ \lambda}$, the metric tensor and its inverse can be written as
\begin{equation}\label{hatg}
\begin{array}{l}
\widehat g^{\mu\nu} = \eta^{\mu\nu} + 2\, \alpha \, H^{\mu} H^{\nu},\\[2ex]
\widehat g_{\mu\nu} = \eta_{\mu\nu} -  \fracc{2\, \alpha}{1+ 2\, \alpha H^2} \, H_{\mu} H_{\nu}.
\end{array}
\end{equation}

Introducing the Weyl-Cartan formalism \cite{tetrad}, we define two tetrad bases $e^{\mu}{}_A$ and $\widehat{e}^{\mu}{}_A$ which relate the tangent space provided with the metric $\eta_{AB}$ to the physical spaces endowed with the two metrics $\eta_{\mu\nu}$ and $\widehat g_{\mu\nu}$. The two bases satisfy the following relations
\begin{equation}
\label{tetrad}
\widehat{g}^{\mu\nu} = \eta^{AB} \, \widehat e^{\mu}{}_{A} \, \widehat e^{\nu}{}_{B},\quad \mbox{and} \quad \eta^{\mu\nu} = \eta^{AB} \, e^{\mu}{}_{A} \, e^{\nu}{}_{B},
\end{equation}
where once more Greek indices refer to the physical spaces and are manipulated with $\eta_{\mu\nu}$ or $\widehat g_{\mu\nu}$, while Latin indices refer to the tangent space and are operated by $\eta_{AB}=\mbox{diag}(1,-1,-1,-1)$. The inverse tetrad bases $e_{\mu}{}^{A}$ and $\widehat{e}_{\mu}{}^{A}$ should satisfy $ e_{\mu}{}^{A} \, e^{\nu}{}_{A} = \widehat e_{\mu}{}^{A} \,\widehat e^{\nu}{}_{A} = \delta_{\mu}^\nu$ and $e_{\mu}{}^{A} \, e^{\mu}{}_{B}= \widehat e_{\mu}{}^{A} \,\widehat e^{\mu}{}_{B}= \delta^{A}_{B}$. Furthermore, any vector $X^\mu$ (or $\widehat X^\mu$) in the space-time $\eta_{\mu\nu}$ (or $\widehat g_{\mu\nu}$) has a counterpart $X^A\equiv e^A_\mu  X^\mu$ (or $\widehat X^A\equiv\widehat e^A_\mu \widehat X^\mu$) in the internal space. In particular, for the Dirac matrices we assume that
\begin{equation}
\gamma^A = \widehat{e}_{\mu}{}^{A} \, \widehat\gamma^\mu= e_{\mu}{}^{A} \, \gamma^\mu,
\end{equation}
where $\gamma^A$'s are the constant Dirac matrices. Note that the $\widehat\gamma^\mu$'s also verify the algebra related to the curved space $\widehat g_{\mu\nu}$.

For consistency with Eqs.\ (\ref{hatg}), the two tetrad bases must obey a condition like
\begin{equation}
\label{mapa_tetr}
\widehat e^{\mu}{}_{A}= e^{\mu}{}_{A} + \beta \, H_{A} H^\mu,
\end{equation}
with the constraint $\alpha =  2\beta/(2+\beta H^2)$ in this case. Therefore, the inverse tetrad bases are such that
\begin{equation}
\label{inv_mapa_tetr}
\widehat e_{\mu}{}^{A}= e_{\mu}{}^{A} - \fracc{\beta}{1+\beta H^2} \, H^{A} H_\mu.
\end{equation}
Then we end the overview on the tetrad formalism we need to deal with the bridge between the spinorial dynamics (a generalization of this approach can be found in \cite{blg}).

{\bf The hidden breaking of chiral symmetry.} Now we describe how the Dynamical Bridge works for spinor fields. As mentioned before, we start with the linear Dirac equation in an effective curved background then we expand the formulas to reach a new and different, though physically equivalent, dynamics in the Minkowski background. The resulting equation is a generalization of the HNJL dynamics.

In the effective curved geometry given by (\ref{hatg}), Dirac equation for the spinor field $\Psi$ reads
\begin{equation}
\label{Dirac}
i\widehat\gamma^\mu\widehat\nabla_{\mu}\Psi=0,
\end{equation}
where $\widehat\nabla\equiv\partial_{\mu} - \widehat\Gamma_{\mu}$ and the Fock-Ivanenko connection. Introducing the tetrads allows to rewrite $\widehat\Gamma_{\mu}$ as
\begin{equation}
\widehat\Gamma_{A}=\widehat e^\mu_A\,\widehat\Gamma_{\mu}=-\frac{1}{8}\,\widehat\gamma_{BCA}[\gamma^{B},\gamma^{C}],
\end{equation}
which is called spin connection and is defined by
\begin{equation}
\label{C_ABC}
\widehat\gamma_{ABC} = \frac{1}{2} (\widehat C_{ABC} - \widehat C_{BCA} - \widehat C_{CAB}), \quad \mbox{and} \quad
\widehat C_{ABC} = - \widehat e_{\nu A} \widehat e^{\mu}{}_{[B} \widehat e^{\nu}{}_{C], \mu}.
\end{equation}
Note that  $\widehat C_{ABC} = - \widehat C_{ACB}$ and, consequently, $\widehat\gamma _{ABC} = - \widehat\gamma_{BAC}$.

Setting $\epsilon=-1$ the spinor field equation (\ref{Dirac}) takes the following form in the flat space
\begin{equation}
\label{dirac_simpl}
i\left(\gamma^{\mu} \partial_{\mu} +\beta\gamma^{\mu} H_{\mu} H^{\nu} \partial_{\nu} +\frac{\beta}{4}\dot H_{\mu}\gamma^{\mu} + \frac{\dot\beta}{2} H_{\mu}\gamma^{\mu}\right) \, \Psi =0,
\end{equation}
where $X_\mu\equiv X_{\mu,\nu}H^{\nu}$. This nonlinear dynamical equation for $\Psi$ originates from the curved space connection interpreted as self-interacting terms in the Minkowski space (for nonlinear spinor equations see \cite{fushchych} and references therein). In particular, there is a set of solutions of Eq.\ (\ref{dirac_simpl}) provided by the Inomata \cite{Inomata} condition $\Psi_{, \mu} = -(1/2)\dot \beta H_\mu \Psi$, will lead to the chiral symmetry breaking without any assumption on $\beta$.

In the standard model the right-handed neutrinos are not present since weak interactions couple only with the left-handed neutrinos. However, since 1998 various independent and different experiments have detected their family oscillation \cite{pdg}, indicating that neutrinos are massive and break the chiral symmetry. A natural question appears: where are the right-handed neutrinos? There are two possible answers: either neutrinos are of Majorana type or they do not interact weakly. Several experiments have unsuccessfully tried direct or indirect detection of such neutrinos \cite{drewes,Canetti}.

Notwithstanding, a third possible explanation is provided according to the above geometrical arguments. Indeed, in the regime $\dot\beta\gg\beta$, when the nonlinear propagating terms of Eq.\ (\ref{dirac_simpl}) can be neglected, it reduces to
\begin{equation}
\label{dirac_simpl_ext}
i\gamma^{\mu} \partial_{\mu}\Psi + i\frac{\dot \beta}{2}({\bf 1}-\gamma_5)(A+iB\gamma_5) \, \Psi =0,
\end{equation}
where we used Eq.\ (\ref{gamma.H}). This equation is similar to the HNJL equation in Minkowski space-time. This means that the nonlinear self-interacting term can be seen as a modification of the space-time structure. Up to this point, they are just two equivalent equations written in two different spaces. However, let us decompose $\Psi$ into its chiral components, that is
$$\Psi=\Psi_L+\Psi_R=\frac{1}{2}({\bf 1}-\gamma_5)\Psi+\frac{1}{2}({\bf 1}+\gamma_5)\Psi,$$
where $\Psi_L$ and $\Psi_R$ represent the left and right-handed chiralities. Then  Eq (\ref{dirac_simpl_ext}) splits into two distinct parts
\begin{eqnarray}
& i\gamma^\mu \partial_{\mu}\Psi_L=0,\\
& i\gamma^\mu \partial_{\mu}\Psi_R + i\dot\beta (A+iB\gamma_5) \, \Psi_L=0.
\end{eqnarray}

From these equations, it follows the remarkable result: each chiral component $\Psi_L$ and $\Psi_R$ satisfies a different dynamical equation in the Minkowski space. The left-handed component propagates as a free Dirac field when the right-handed component is trapped by the self-interacting term. If the coupling parameter $\dot \beta$ is sufficiently large, the right-handed component needs very high energies to be detected. This leads to a new and geometrical explanation for the non observation of the right-handed neutrinos \cite{novello-edu-faci} .

\section{The theory of the Geometric Scalar Gravity}\label{gsg}
Although the possibility to eliminate accelerations of arbitrary bodies can be described equivalently by particular changes on the metric of space-time, one for each body, there is interest in follow the main ideas of GR and unify all these geometries into a single one. The path followed by Einstein was to describe the gravitational phenomenon in terms of a unique universal metric. The prize to pay is to impose a dynamics for the geometry by its own. Few years ago \cite{JCAP} an alternative was proposed that received the name of Geometric Scalar Gravity (GSG) that do not follow such procedure. We shall make an overview of this theory and its main achievements, highlighting how it bypasses the main drawbacks of the previous proposals \cite{gibbons,giulini,finn,nord,coll-eins,shapiro}, describing well the local tests of gravity \cite{adler,eddington,schiff,will} and can be constructed in the lines of a field theory formulation \cite{deser,feynman,gpp,gupta}. Let us start by enumerating the main properties of GSG:

\begin{itemize}
\item{The gravitational interaction is described by a scalar field $ \Phi$;}
\item{The field $\Phi$ satisfies a nonlinear dynamics;}
\item{The theory satisfies the principle of general covariance. In other words, this is not a theory restricted to the realm of flat spaces;}
\item{All kind of matter and energy interact with $\Phi$ only through the pseudo-Riemannian metric
\begin{equation}
q^{\mu\nu} = \alpha \, \eta^{\mu\nu} + \beta \, \frac{\partial^{\mu} \, \Phi \,\partial^{\nu} \Phi}{\eta^{\gamma\delta}\partial_{\gamma} \Phi \partial_{\delta} \Phi};
\label{112}
\end{equation}}
\item{Test particles follow geodesics relative to the gravitational metric $ q_{\mu\nu};$}
\item{ $\Phi$ is related in a nontrivial way with the Newtonian potential $\Phi_N$;}
\end{itemize}

The quantities $ \alpha $ and $ \beta$ are functionals of $\Phi$ which were specified by fixing the dynamics of the scalar field. The auxiliary (Minkowski) metric $ \eta^{\mu\nu}$ is unobservable because the gravitational field couples to matter only through $ q^{\mu\nu}$ an hypothesis borrowed from GR where a unique geometrical entity interacts with all forms of matter and energy and the geometry underlying all events is
controlled by the gravitational phenomena.

It is worthwhile to point out that the scalar field is not the (special) relativistic generalization of the Newtonian potential. Indeed, following the scheme of GR \cite{einstein} and assuming that the test particles follow geodesics relative to the geometry $q_{\mu\nu}$, we have that
\begin{equation}
\frac{d^{2} x^{i}}{dt^{2}} = - \, \Gamma^{i}_{00} = - \, \partial^{i} \, \Phi_{N},
\label{22julho1}
\end{equation}
where we are assuming static weak field configuration and low velocities for test particles.

From Eq. (\ref{9junho11}), we have
$$  \Gamma^{i}_{00}\approx - \, \frac{1}{2} \partial^{i} \, \ln \alpha. $$
It follows that the Newtonian potential $ \Phi_{N} $ is
approximately given by
$$ \Phi_{N} \approx -\, \frac{1}{2} \, \ln \alpha,$$
which yields the relation between the metric and the Newtonian potential $\Phi_N$ as
$$ q_{00} = \frac{1}{\alpha} \approx 1 + 2 \,\Phi_{N}.$$

Using equation (\ref{23julho5}) one obtains the right (vacuum) Newtonian limit:
$$\nabla^2\Phi_N=0.$$
This was the starting point of Einstein\rq s path in building his tensorial theory of gravity. The geometric scalar gravity follows another path that we describe next. From now on we will explore the consequences of extrapolating from the above approximation the general expression
\begin{equation}
\label{alpha_phi}
\alpha = e^{- 2 \,\Phi}.
\end{equation}

The next task is to determine the functional dependence of $\beta$ on $\Phi$ or either the form of the potential $ V(\Phi)$ as we have presented in the Lemma \ref{lemma_phi} concerning the DB for scalar fields, that is
\begin{equation}
\beta = \alpha \, ( \alpha^{2} \, V -1).
\label{18novembro}
\end{equation}

{\bf The dynamics of scalar gravity.} In (GSG) the dynamics is controlled by the variational principle
$$ \delta S_{1}= \delta \, \int\sqrt{-\eta}\,d^4x\,V(\Phi) w, $$
we get:
\begin{eqnarray}
\delta \, S_{1} &=& - \, \int \, \sqrt{- \eta} \, d^4x\, \left( V' \, w + 2 \, V \, \Box_{M} \Phi \right) \, \delta\Phi
\end{eqnarray}
where $\Box_{M}$ is the d\rq Alembert operator in flat space. Again, using the Lemma \ref{lemma_phi}, we rewrite the equation above as
\begin{equation}
\delta \, S_{1} = - \, 2 \, \int   \sqrt{- q} \, d^4x \, \sqrt{V} \,\Box \Phi  \, \delta \Phi.
\end{equation}
In presence of matter we add a corresponding term $L_{m}$ to the total action:
\begin{equation}
S_{m} = \int \, \sqrt{- q} \, d^4x\, L_{m}.
\label{24julho15}
\end{equation}
The first variation of this term as usual yields
\begin{equation}
\delta S_{m} = - \, \frac{1}{2} \, \int \, \sqrt{- q} \, d^4x\, T^{\mu\nu} \, \delta \, q_{\mu\nu},
\label{24julho17}
\end{equation}
where the energy-momentum tensor is given by its standard form
$$ T_{\mu\nu} \equiv  \, \frac{2}{\sqrt{- q}} \, \frac{\delta( \sqrt{- q} \, L_{m})}{\delta q^{\mu\nu}}.$$
General covariance leads to conservation of the energy-momentum tensor $T^{\mu\nu}{}_{;\nu}=0.$ The equation of motion is obtained by the action principle $\delta S_{1} + \delta S_{m}= 0$. Up to this point we are following the paths of GR. Here, however, in the GSG, the metric $ q_{\mu\nu} $ is not the fundamental quantity. We have to write the variation $ \delta q_{\mu\nu}$ as function of $ \delta \Phi$. While we are using the unobservable background Minkowski metric for simplicity, at the end all expressions should be written in terms of the gravitational metric $q_{\mu\nu}$. Finally, the equation of motion for the gravitational field $ \Phi$ takes the form:
\begin{equation}
\sqrt{V} \, \Box\Phi= \kappa \, \chi,
\label{12out1}
\end{equation}
where
$$ \chi = \fracc{1}{2} \, \left( \frac{\alpha'}{2\alpha} \, (T-E) + \frac{Z'}{2Z}E - \nabla_{\lambda} \, C^{\lambda}\right),$$
with $Z\equiv\alpha+\beta$, $T \equiv T^{\mu\nu} \, q_{\mu\nu}$, $E \equiv \frac{T^{\mu\nu} \, \partial_{\mu}\Phi \, \partial_{\nu}\Phi}{Z \,w}$, $ C^{\lambda}\equiv\frac{\beta}{\alpha \, Z \,w} \, \left( T^{\lambda\mu} - E \, q^{\lambda\mu} \right) \, \partial_{\mu}\Phi$ and prime means derivative w.r.t. $\Phi$.

Substituting $ \alpha = e^{- 2 \Phi}$ and using the expression for the potential derived in the original paper \cite{JCAP}
$$V = \frac{(\alpha - 3)^{2}}{4 \, \alpha^{3}} $$
we rewrite the source term under the form
$$ \chi = \fracc{1}{2} \, \left( \frac{3 \, e^{2 \Phi} + 1}{3 \, e^{2\Phi}- 1} \, E - T - \nabla_{\lambda} \, C^{\lambda}\right).$$

This equation describes the dynamics of GSG in the presence of matter. The quantity $\chi$ involves a non-trivial coupling between the gradient of the scalar field  $\nabla_{\mu} \Phi$ and the complete energy-momentum tensor of the matter field $T_{\mu\nu}$  and not uniquely its trace. This property allows the electromagnetic field to interact with the gravitational field. The Newtonian limit gives the identification
$$\kappa\equiv\fracc{8\pi G}{c^4}.$$

{\bf Natural decomposition.} The form of the metric, containing the derivative of $ \Phi $ suggests a simplification in the description of the matter terms which is useful for exploring the cosmological consequences of GSG. Suppose that $\partial_{\mu} \Phi$ is time-like, that is $ \Omega
> 0$. We then define the normalized vector

\begin{equation}
I_{\mu} = \frac{\partial_\mu \Phi}{\sqrt{\Omega}}.
\end{equation}

This vector can be used to decompose the energy-momentum tensor of a perfect fluid in the "co-moving" representation by setting

\begin{equation}
T^{\mu\nu} = ( \varrho + p) \, I^{\mu}  \, I^{\nu}  - p \, q^{\mu\nu},
\end{equation}
it then follows
$$  T^{\mu\nu} \, \partial_{\mu}\Phi = \sqrt{\Omega} \, \varrho \, I^{\mu}, $$
and
$$  T^{\mu\nu} \, \partial_{\mu}\Phi \, \partial_{\nu}\Phi = \Omega \, \varrho.  $$

Thus, in this frame the quantities $ E $ and $ T $ reduces to
\begin{equation}
E = \varrho, \hspace{1cm} T = \varrho - 3 p.
\end{equation}
Using these results it follows that $ C^\mu = 0.$ In the natural frame associated to the gradient of the gravitational field $ \Phi $ the equation of motion for the scalar gravity, reduces to the form
\begin{equation}
\sqrt{V} \, \Box \, \Phi =  - \frac{\kappa}{2} \,  \left( \frac{2 \, \alpha}{\alpha-3} \, \varrho - 3 p \right).
\end{equation}
This is the form of the dynamics of $ \Phi$ when the source is a perfect fluid. In the next section we provide a simple example of GSG in the analysis of the global properties of the universe.

\section{A Geometrical Description of Quantum Mechanics Q-WIS}
In the literature, there are many attempts of a quantum theory formulation for the space-time from which some are very successful, but none of them is complete. On the other hand, the geometrical approaches for the quantum theory have also their remarkable results. The most common is based on a K\"ahler structure of a complex projective Hilbert space, where people have tried to derive a ``natural" metric in the realm of the so-called geometric quantum mechanics \cite{anandan,anandan2,ashtekar,brody,minic,santamato1,santamato}. There are also other geometrization procedures where quantum mechanics appears as a sort of an emergent theory \cite{abraham,koch,koch2,koch3,others1,others3}.

In this vein, one of us with collaborators have shown that Quantum Mechanics (QM) can be indeed interpreted as a modification of the Euclidean nature of 3-dimensional space into a particular Weyl affine space (called Q-WIS) by using the de Broglie-Bohm causal formulation of QM \cite{nsf}. In the Q-WIS geometry, the length of extended objects changes from point to point, the deformation of the standard rulers used to measure physical distances are in the core of quantum effects allowing a geometrical formulation of the uncertainty principle. How is it possible? Several works have advocated a possible connection between non-Euclidean geometry with quantum effects \cite{canuto,cufaro,gueret,hkv,kyprianidis,israelit,israelit2,israelit3,israelit4,london,wheeler} (see also \cite{caroll,carroll,carroll-book} and references therein) and, in this section, we briefly present one of them in the realm of this review. The relativistic generalization in the case of the Klein-Gordon equation was done in \cite{falciano}.

\subsection{Weyl Geometry}
In the early years of twenty century H. Weyl \cite{Weyl1,Weyl2} suggested a modification of the Euclidean geometry that was an extension of the proposal made by Riemann. The main motivation was related to the Unified Program that followed the geometrization of gravity displayed by GR. Weyl proposed to identify the electromagnetic interaction as a modification of the geometry of space-time along similar lines as it was made for gravity. For various reasons this proposal did not succeeded. Neverthless, the result of this research achieved a well-defined and consistent generalization of Riemannian geometry. In fact, a Riemannian geometry can be understood as a special case of a Weyl geometry. In the first section of this review, we have described in more details the properties of Weyl geometry, but it is worth to mention here its main difference that is related to the notion of a standard ruler.

In a Weyl geometry the length of an extended object changes from point to point. This means that a ruler of length $l$ will change by an amount
\[
\delta \, l = l \, f_{a} \, \dd x^{a} \quad .
\]
This effect may become an obstacle to the notion of a local ruler and thus to local measurement of distance \cite{perlick}. However, there is a special sub-class of Weyl geometries known as Weyl Integrable Space (WIS) that is free of such difficulty. This is provided by the condition that the vector $f_{i}$ is a gradient of a function, i.e., $f_a=f_{,\, a}.$  The geometry WIS is distinguished precisely by the fact that the length of the ruler transported along a closed curve does not change. Hence, if the change of the ruler's length is $\dd l$, for a closed path in WIS we have
\[
\oint \dd l=0 \qquad ,
\]
which guarantees the uniqueness of any local measurement. The allowance of an intrinsic modification of the standard rulers is the main geometrical hypothesis that allows to associate this geometrical modification to the origin of quantum effects. We restrict this review only to the case of an isolated point-like particle possibly subjected to an external potential.

\subsection{Quantum Mechanics}
In the end of last century, the development of quantum cosmological scenarios brought to light some difficulties intrinsic to the Copenhagen interpretation \cite{copenhagen1,copenhagen3,bdb1,bdb2,holland,holland2,others2,others52,others53}. More specifically, the measurement process in a quantum closed universe seems inconsistent \cite{nelson1,nelson2,nelson3}. Fortunately, there are some alternative interpretations that can be applied simultaneously to cosmology and to the micro-world, for instance the Many-Worlds interpretation \cite{others41,others42}, the Consistent Histories formulation \cite{others51,others54} and, in particular, the so-called causal interpretation or also Bohm-de Broglie interpretation \cite{debroglie1,debroglie3,debroglie6,broglie,madelung}, since it is amongst the well defined interpretation that can be applied to any kind of system, including the universe as a whole, and up to date it is completely equivalent to the Copenhagen interpretation when applied to the micro-world.

QM has been seen as a modification of the classical laws of physics to incorporate the uncontrolled disturbance caused by the macroscopic apparatus necessary to realize any kind of measurement. This statement, known as Bohr's complementary principle, contains the main idea of the Copenhagen interpretation of QM. The quantization program continues with the correspondence principle promoting the classical variables into operators and the Poisson brackets into commutation relations.

In this non-relativistic scenario, the Schr\"odinger equation establishes the dynamics for the wave function describing the system. Note that as in Newtonian mechanics time is only a external parameter and the 3-d space is assumed to be endowed with the Euclidean geometry.

Using the polar form for the wave function, $\Psi=A\, e^{iS/\hbar}$, the Schr\"odinger equation is decomposed in two equations for the real functions $A\left(x\right)$ and $S\left(x\right)$
\begin{eqnarray}
&&\frac{\partial S}{\partial t}+\frac{1}{2m}\nabla S\cdot\nabla S +V -\frac{\hbar^{2}}{2m}\frac{\nabla^{2}A}{A}=0 \quad ,\quad \label{HJ}\\[2ex]
&&\frac{\partial A^2}{\partial t}+\nabla \left(A^2\frac{\nabla S}{m}\right)=0 \qquad .\label{cont}
\end{eqnarray}

Solving these two equations is completely analogous to solving the Schr\"odinger equation. The probabilistic interpretation of QM associate $\| \Psi \|^2=A^2$ with the probability distribution function on configuration space. Hence, Eq.\ (\ref{cont}) has exactly the form of a continuity equation with $A^2\, \nabla S/m$ playing the role of current density.\\

\subsection{Short synthesis of Bohm-de Broglie interpretation}
The causal interpretation of QM, propose that the wave function does not contain all the information about the system. An isolated system describing a free particle (or a particle subjected to a potential) is defined simultaneously by a wave function and a point-like particle. In this case, the wave function still satisfies the Schr\"odinger equation but it should also work as a guiding wave modifying the particle trajectory.

Note that Eq.\ (\ref{HJ}) is a Hamilton-Jacobi-like equation with an extra term that is often called quantum potential
\begin{equation}
\label{Q}
Q\equiv-\frac{\hbar^{2}}{2m}\frac{\nabla^{2}A}{A}\quad ,
\end{equation}
while, as already mentioned, Eq.\ (\ref{cont}) is a continuity-like equation. The Bohm-de Broglie interpretation takes these analogies seriously and postulates an extra equation associating the velocity of the point-like particle with the gradient of the phase of the wave function, namely,
\begin{equation}
\label{guide}
\dot{x}=\frac{1}{m}\nabla S\,.
\end{equation}
Integrating this equation yields the quantum Bohmian trajectories. The unknown or hidden variables are the initial positions necessary to fix the constant of integration of the above equation.

The quantum potential is the sole responsible for all novelties of quantum effects such as non-locality or tunneling processes. As a matter of fact, the Bohm-de Broglie interpretation has the theoretical advantage of having a well formulated classical limit. Classical behavior is obtained as soon as the quantum potential, which has dimensions of energy, becomes negligible compared to other energy scales of the system.

In what follows, we will show that it is possible to reinterpret QM as a manifestation of non-Euclidean structure of the 3-dimensional space, leading to a geometrical interpretation of the quantum effects.

{\bf DM for the Q-potential.} The quantum force that a particle suffers has two important properties:
\begin{itemize}
\item{It is universal;}
\item{It admits a potential.}
\end{itemize}

These properties are quite similar to Newtonian gravity. This means that it is possible to follow a similar path as proposed in GR and in the formulation of DM (cf.\ previous sections) to eliminate the Bohmian forces in terms of a Riemannian  geometry. Instead of this, we will follow another path by a different change on the geometry of the 3-dimensional space.

%\subsection{Non-Euclidean geometry}\label{NEG}
%Since ancient times, Euclidean geometry was considered as the most adequate mathematical formulation to describe the physical space. However, its validity can only be established a posteriori as long as its construction yields useful notions to connect physical quantities such as the Euclidean distance between two given points.

%Special Relativity modified the notion of 3-dimensional Euclidean space to incorporate time in a four-dimensional continuum (Minkowski spacetime). Later on, GR generalized the absolute Minkowski spacetime to describe gravitational phenomena, considering the spacetime manifold as a dynamical field that can be deformed and stretched but in such a way that it always preserves its Riemannian structure. It is worth noting that both the Euclidean and Minkowskian spaces are nothing more than special cases of Riemaniann spaces.

%Nonetheless, Riemannian manifold are not the most general type of geometrical spaces. In the same way as above, Riemannian geometries can be understood as a special subclass of a more general structure known as Weyl spaces. As to the matter of which geometry is actually realized in Nature, it has to be determined by physical experiments.

Instead of imposing a priori that QM has to be constructed over an Euclidean background as it is traditionally done,  quantum effects can be interpreted as a manifestation of a non-Euclidean structure derived from a variational principle. The validity of the specific geometrical
structure proposed can be checked a posteriori comparing it to the usual non-relativistic QM.

Thus, consider a point-like particle with velocity $v=\nabla S/m$ and subjected to a potential $V$. Following Einstein's idea to derive the geometrical structure of space from a variational principle by considering the connection as an independent variable, we start with
\begin{equation}
\label{I}
I=\int{ \dd t\dd^3x\sqrt{g}} \, \Omega^2 \left(\lambda^2\mathcal{R}-\frac{\partial S}{\partial t}-\mathcal{H}_m\right)
\end{equation}
and consider the connection of the 3-d space $\Gamma^i_{j k}$, the Hamilton's principal function $S$ and the scalar function $\Omega$ as independent variables. Note that indices $ i, j, ...\ $ run from $1$ to $3$.

Considering the line element in Cartesian coordinates
\[
\dd s^2=g_{ij}\dd x^i \dd x^j=\dd x^2+\dd y^2+\dd z^2,
\]
with $g={\rm det}\, g_{ij}$, the Ricci curvature tensor is defined in terms of the connection through
\[
\mathcal{R}_{ij}=\Gamma^m_{ mi\, ,j}-\Gamma^m_{ij\, ,m}+\Gamma^l_{mi}\Gamma^m_{jl}-\Gamma^l_{ij}\Gamma^m_{lm}
\]
and its trace defines the scalar curvature $\mathcal{R} \equiv g^{ij}\mathcal{R}_{ij}$ which has dimensions of inverse length squared, $[\mathcal{R}]=L^{-2}$. The constant $\lambda^2$ has dimension of energy times length squared, $[\mathcal{\lambda}^2]=E\,L^{2}$, and the $\frac{\partial S}{\partial t}$ term is related to the particle's energy. In the case of point-like particle, the Hamiltonian is
\[
\mathcal{H}_m=\frac{1}{2m}\nabla S\cdot\nabla S +V \,.
\]

Variations of the action $I$ with respect to $\Gamma^i_{j k}$ yield
\begin{eqnarray}
g_{ij;k}=-4\left(\ln \Omega\right)_{,k}\, g_{ij} \, ,
\label{weq}
\end{eqnarray}
where ``$;$'' denotes covariant derivative and ``$,$'' simple derivative. Eq.\ (\ref{weq}) characterizes the affine properties of the physical space. Hence, the variational principle naturally defines a WIS; variation with respect to $\Omega$ gives
\begin{eqnarray}
\lambda^2\mathcal{R}=\frac{\partial S}{\partial t}+\frac{1}{2m}\nabla S\cdot\nabla S+V \,.
\label{HJ3}
\end{eqnarray}

The right-hand side of this equation has dimension of energy while the curvature scalar has dimension of $[\mathcal{R}]=L^{-2}$. Furthermore, apart from the particle's energy, the only extra parameter of the system is the particle's mass $m$. Thus, there is only one way to combine the unknown constant $\lambda^2$, which has dimension of $[\lambda^2]=E\,L^{2}$, with the particle's mass such as to form a physical quantity. Multiplying them, we find a quantity that has dimension of angular momentum squared $[m\,\lambda^2]=\hbar^2$.

In terms of the scalar function $\Omega$, the curvature scalar is given by
\begin{equation}
\label{R}
\mathcal{R}=8\frac{\nabla^2 \Omega}{\Omega}.
\end{equation}
Hence, setting $\lambda^2=\hbar^2/16 m$, equation (\ref{HJ3}) becomes
\begin{equation}
\frac{\partial S}{\partial t}+\frac{1}{2m}\nabla S\cdot\nabla S+V -\frac{\hbar^2}{2m}\frac{\nabla^{2}\Omega}{\Omega}=0 \, ,
\label{HJ4}
\end{equation}
Finally, varying the Hamilton's principal function $S$ we find
\begin{eqnarray}
\frac{\partial \Omega ^2}{\partial t}+\nabla \left(\Omega^2\frac{\nabla S}{m}\right)=0 .
\label{cont3}
\end{eqnarray}

Equations (\ref{HJ4}) and (\ref{cont3}) coincide with (\ref{HJ}) and (\ref{cont}) if we identify $\Omega=A$. Thus, the ``action'' of a point-like particle non-minimally coupled to geometry given by
\begin{equation}
I=\int{ \dd t \dd^3x\sqrt{g}} \, \Omega^2 \left[\frac{\hbar^2}{16\, m}\mathcal{R}-\left(\frac{\partial S}{\partial t}+\mathcal{H}_m\right)\right],
\end{equation}
exactly reproduces the Schr\"odinger equation and thus the quantum behavior.

The straightest way to compare this geometrical approach to the quantum theories is through the Bohm-de Broglie interpretation. Note that this formulation has the advantage of giving a physical explanation for the appearance of the quantum potential (\ref{Q}). In Q-WIS (Q for quantum), this term is simply the curvature scalar of the WIS. The inverse square root of the curvature scalar defines a typical length $L_w$ (Weyl length) that can be used to evaluate the strength of quantum effects
\[
L_w \equiv \frac{1}{\sqrt{\mathcal{R}_w}} \,.
\]

As we have already mentioned, the classical limit of Bohm-de Broglie interpretation is achieved when the quantum potential is negligible compared to other energy scales of the system. In the scope of this geometrical approach, the classical behavior is recovered when the length defined by the Weyl curvature scalar is small compared to the typical length scale of the system. Once the Weyl curvature becomes non-negligible the system goes into a quantum regime.

{\bf Geometrical uncertainty principle.} As long as we accept that QM is a manifestation of a non-Euclidean geometry, we are faced with the need of reinterpreting geometrically all theoretical issues related to quantum effects. As a first step, we derive the uncertainty principle as a break down of the classical notion of a standard ruler.

It is well known amongst relativistic physicists that there is no absolute notion of spatial distance in curved spacetime. However, this is no longer true when there is an absolute Newtonian time and only the spatial manifold is allowed to be curved. In this case, it is possible to define distance as the smallest length between two given points calculated along geodesics in 3-d space. This is a consistent definition since the 3-d space has a true metric in the mathematical sense that its eigenvalues are all positives. However, this definition does not encompass the classical definition of a standard ruler.

Hence, we are unable to perform a classical measurement to distances smaller than the Weyl curvature length. In other words, the size of a measurement has to be bigger than the Weyl length
\begin{equation}
\label{princ1}
\Delta L \geq L_w= \frac{1}{\sqrt{\mathcal{R}_w}}.
\end{equation}

The quantum regime is extreme when the Weyl curvature term dominates. Thus, from equations (\ref{R}) and (\ref{HJ4}) we have
\begin{equation}
\label{princ2}
\mathcal{R}_w =2\left(\frac{2\Delta p}{\hbar}\right)^2-\frac{16m}{\hbar^2}\left(E-V\right) \leq 2\left(\frac{2\Delta p}{\hbar}\right)^2
\end{equation}
and finally combining equations (\ref{princ1}) and (\ref{princ2}) we obtain
\[
\Delta L \,\Delta p\geq \frac{\hbar}{2\sqrt{2}}.
\]

We should emphasize that now the Heisenberg's uncertainty relation has a pure geometrical meaning. This argument resembles Bohr's complementary principle inasmuch as the impossibility of applying the classical definitions of measurements.

Bohr's complementary principle is based on the uncontrolled interference of a classical apparatus of measurement. On the other hand, the notion of a classical standard ruler breaks down because its meaning is intrinsically dependent on the validity of Euclidian geometry. Once it becomes necessary to include the Weyl curvature, we are no longer able to perform a classical measurement of distance.

There is another way to interpret the uncertainty principle. For a given particle of mass $m$ and energy $E$ there is only one combination with the free parameter of the theory $(\hbar)$ that furnishes a quantity with dimensions of length. We take this value as a definition of the classical size of the particle, namely
\begin{equation}
\label{size}
l_{part} \equiv \sqrt{\frac{\hbar^2}{E \, m}}.
\end{equation}

Note that this definition coincides with the Compton's wavelength of the particle which is related to the limits of validity of non-relativistic QM.

Considering a free stationary particle, from equation (\ref{HJ3}) we have
\[
E=\frac{\hbar ^2}{16m}\mathcal{R}_W  \; \Rightarrow \; l_{part} =\frac{4}{\sqrt{\mathcal{R}_W}} \, ,
\]
and from equation (\ref{princ2})
\begin{equation}
l_{part}\, \Delta p \geq \sqrt{2}\, \hbar \,.
\end{equation}

From this point of view, the uncertainty principle indicates that it is impossible to perform a measurement smaller than the classical size of the particle defined by equation (\ref{size}). In other words, it is impossible to perform a classical measurement inside the particle.

\subsection{Additional Comments}
It is well known that as soon as we consider high velocities or high energies one has to abandon the Euclidean geometry as a good description of the physical space. These brought two completely different modifications where the physical space loses its absolute and universal character. In fact, these are the core of classical relativistic physical theories, namely, Special relativity and GR.

In a similar way, one should be allowed to consider that the difficulties that appears while going from classical to QM comes from an inappropriate extrapolation of the Euclidean geometry to the micro-world. Hence, the unquestioned hypothesis of the validity of the 3-d Euclidean geometry to all length scales might be intrinsically related to quantum effects.

We overview here the close connection between the Bohm-de Broglie interpretation of QM and the modification of the geometry of space by passing from Euclidean 3-d to a 3-d Q-WIS. The Bohmian quantum potential can be identified with the curvature scalar of the Q-WIS and through a variational principle it is possible to reproduce the Bohmian dynamical equations which are equivalent to Schr\"odinger's QM.

The Palatini procedure, in which the connection acts as an independent variable while varying the action, naturally endows the space with the appropriate Q-WIS structure. Thus, this geometry enters into the theory less arbitrarily than the implicit ad hoc Euclidean hypothesis of QM. The identification of the curvature scalar as the ultimate origin of quantum effects leads to a geometrical version of the uncertainty principle. This geometrical description considers the uncertainty principle as a break down of the classical notion of standard rulers.
Thus, it arises an identification of quantum effects to the length variation of the standard rulers.

%Finally, we have used a variational principle to arrive at the Q-WIS structure. Let us provide a quick proof of this. Start with the action
%\begin{equation}
%I = \int \dd t\, \dd ^{3}x \sqrt{g} \; \, \Omega^{2} \, \mathcal{R}
%\end{equation}
%then, variation of the connection yields
%\begin{equation}
%\delta I = \int \dd t\, \dd ^{3}x \sqrt{g} \; \, \Omega^{2}\, g^{ab} \, \delta R_{ab}= \int \dd t \, \dd^{3}x\; Z^{a b}_{m}\, \delta \Gamma^{m}_{a b}
%\end{equation}
%with $ Z^{a b}_{m} \equiv (\sqrt{g} \, g^{ab} \, \Omega^{2})_{; \, m} - \frac{1}{2} \, (\sqrt{g} \, g^{a k} \, \Omega^{2})_{; \, k } \, \delta^{b}_{m} - \frac{1}{2} \, (\sqrt{g} \, g^{b k} \, \Omega^{2})_{; \, k } \, \delta^{a}_{m}$. Taking the trace of this expression yields $\left(\sqrt{g} \, g^{a k} \, \Omega^{2}\right)_{; \, k } =0$ and substituting it back, we obtain the condition for a Weyl integrable geometry
%\begin{equation}
%g_{a b \,; \, k} = - \, 4 \, \frac{\Omega_{, \, k}}{\Omega} \, g_{a b} .
%\end{equation}

\newpage
\section{Comments and conclusions}
In the realm of Special Relativity, the Newtonian idea of an absolute clock for all inertial observer is replaced by the concept of a proper time for each observer, namely, the particular measure of the physical time is different depending on the relative motion of a given inertial observer with respect to a rest frame. Then, due to its universality, the gravitational interaction was interpreted as a modification of the geometry of space-time \cite{mercati}. In the electrodynamics of moving dielectrics, a similar result was made possible by Gordon's analysis of light rays inside a material media and its propagation was interpreted in terms of an associated geometry.

Recently, the D'Alembert principle of Classical Mechanics --which transforms any dynamical problem into a statical one - was generalized by adopting as fundamental principle that the acceleration induced by any kind of forces is eliminated through a modification of the background metric. We can understand this procedure as an extension of the Special Relativity idea of \textit{relative time} to the geometry of space-time, once it follows that for each body acted upon by arbitrary forces one can attribute a particular geometry such that it is interpreted as following a free-motion. This property that allows the elimination of any force by a change on the metric properties on the space-time is called \textit{Metric Relativity}.

The second step presented here concerns the Dynamical Bridge (DB). Maxwell and Born-Infeld's theories in the same background correspond to different descriptions of the electromagnetic interaction. However, we have seen that the DB leads to an equivalence between them, that is, they have the same dynamical properties when each one of these theories is written in a distinct space-time geometry. The crucial point is related to the dependence on the electromagnetic field of the associated  metrics. This indicates that both theories can be understood from a more fundamental point of view depending on which aspects of electromagnetism is to be emphasized. This general structure of maps between explicit distinct theories allows connections which were not considered before and the physical phenomena can thus be comprehended more deeply. In other words, distinct theories are seen only as different languages representing the same phenomenon. This property of the DB can be enlarged through other processes.  In particular, we can point out the recent geometric scalar gravity (GSG), that is, a theory initially written in flat Minkowski space is interpreted as a nonlinear theory in an associated curved space-time. Finally, we described the geometrization of Quantum Mechanics (QM), through the use of Weyl geometry, which structure is the responsible for the origin of the quantum potential needed to properly describe QM in the de Broglie-Bohm framework.

In practice, we elaborate in this report a thorough review on the following issues:
\begin{itemize}
\item{From a kinematical point of view, we have shown that it is possible to display a geometrical description of any kind of force assuming that the space-time metric is not unique and not given a priori. In particular, for any accelerated vector field $v^{\mu}$ in a background metric $g_{\mu\nu}$, it is always possible to find another metric tensor $\widehat{g}_{\mu\nu}$ in which $v^{\mu}$ follows a geodesic motion. As we have seen the procedure described in Sec.\ [\ref{geo_acc_path}] can be applied to any vector field lying in any space-time metric, for instance, we selected the Schwarzschild, G\"odel and Kerr solutions of GR equations to exemplify the method. Furthermore, the generalization of Gordon's results to arbitrary material media and nonlinear EM theories in Sec.\ [\ref{accel_mov_diel}] can also be considered as a consequence of it.}
\item{In Sec.\ [\ref{dyn_brid}], we introduced the concept of DB and we applied this to the most relevant cases involving scalar, vector or spinor fields, trying to explain with some details how the DB can provide alternative explanations to current open problems in physics. Still in the realm of the DB, we then present in Sec.\ [\ref{gsg}] a recent attempt to describe the gravitational interaction only in terms of a scalar field.}
\item{Finally, we summarized the geometrical formulation of quantum mechanics according to the de Broglie-Bohm interpretation, emphasizing that the quantum potential can be seen as a consequence of a possible non-metricity of the Euclidean space. This non-metricity is represented by a particular choice of the Weyl affine space called Q-WIS. The main result is that the deformation of the standard rulers used to measure physical distances allows a geometrical formulation of the uncertainty principle.}
\end{itemize}
We conclude this review stating that the most adequate mathematical formulation to describe the geometry of the physical world should not be established a priori \cite{poincare}; the physical experiments indicate which space-time geometry (or class of) is actually realized in Nature and in which scale (macrophysics or microphysics) it is relevant to develop a {\bf measurable/testable} and useful theory.

\newpage
\appendix

\section{Lanczos tensor}
The Weyl tensor $W_{\alpha\beta\mu\nu}$ can be expressed in terms of the 3-index Fierz-Lanczos potential tensor \cite{Fierz,Lanczos} that we will denote by $L_{\alpha\beta\mu}.$  Let us summarize here some definitions and properties of $L_{\alpha\beta\mu}$, since the literature has very few papers on this matter. This tensor was introduced in the 30's to provide, in a similar way as the symmetric tensor $\varphi_{\mu\nu}$ does---in a more used approach---an alternative description of spin-2 field in Minkowski background. In the 60's Lanczos rediscovered it---without recognizing he was dealing with the same object---as a Lagrange multiplier in order to obtain the Bianchi identities in the context of GR. However, a complete analysis of Fierz-Lanczos object was undertaken and it was discovered that its generic (Fierz) version describes not only one but two spin-2 fields \cite{NovelloNelson}. The restriction to just a single spin-2 field is usually called the Lanczos tensor. We will limit all our considerations here to this restricted quantity.

\underline{{\bf Basic Properties:}} In any 4-dimensional Riemannian geometry there is a 3-index tensor $L_{\alpha\beta\mu}$ which has the two following symmetries:

$$L_{\alpha\beta\mu} + L_{\beta\alpha\mu} = 0, \quad \mbox{and} \quad L_{\alpha\beta\mu} + L_{\beta\mu\alpha} + L_{\mu\alpha\beta} = 0$$
and with such $L_{\alpha\beta\mu}$ we write the Weyl tensor in the form of a homogeneous expression that is
\begin{equation}
W_{\alpha\beta\mu\nu}  =  L_{\alpha\beta [\mu ;\nu ]} + L_{\mu\nu[\alpha ;\beta ]} + \frac{1}{2} [ L_{(\alpha\nu)} g_{\beta\mu} + L_{(\beta\mu)}
g_{\alpha\nu} - L_{(\alpha\mu)} g_{\beta\nu} - L_{(\beta\nu)} g_{\alpha \mu}] + \frac{2}{3}\, {L^{\sigma\lambda}}_{  \sigma ;\lambda} \, g_{\alpha\beta\mu\nu},
\protect\label{b3}
\end{equation}
where $ L_{\alpha\mu} \equiv {{L_{\alpha}}^{\sigma}}_{\mu ;\sigma} - L_{\alpha ;\mu}$ and $L_{\alpha} \equiv {{L_{\alpha}}^{ \sigma}}_{\sigma}$.

Let us point out that, due to the above symmetry properties, Lanczos tensor has 20 degrees of freedom. Since Weyl tensor has only 10 independent components, it follows that there is a gauge symmetry involved. This gauge symmetry can be separated into two classes:

$$ \Delta^{(1)} L_{\alpha\beta\mu} = M_{\alpha} \hspace{0.1cm}g_{\beta\mu} - M_{\beta} \hspace{0.1cm}g_{\alpha\mu}, $$

\begin{equation}
\Delta^{(2)} L_{\alpha\beta\mu} = W_{\alpha\beta ;\mu} - \frac{1}{2} W_{\mu\alpha ;\beta} + \frac{1}{2} W_{\mu\beta ;\alpha} + \frac{1}{2} g_{\mu\alpha} \, {{W_{\beta}}^{\lambda}}_{;\lambda} - \frac{1}{2} g_{\mu\beta}\,{{W_{\alpha}}^{\lambda}}_{;\lambda},
\protect\label{b5}
\end{equation}
in which the vector $M_{\alpha}$ and the antisymmetric tensor $W_{\alpha\beta}$ are arbitrary quantities. Then, we can associate Lanczos tensor to the parameters of a congruence of curve for different geometries. This can be more clearly seen by stating certain Lemmas \cite{Novello_Velloso} that we exhibit here without proofs. The interested reader can find them in the quoted article.

\begin{theorem}
If in a given Riemannian geometry there is a congruence of observers $ v^{\mu}$ that is shear-free and irrotational, then the magnetic part of Weyl tensor vanishes for these observers.
\end{theorem}

\begin{theorem}
If in a given Riemannian geometry there is a congruence of observers $ v^{\mu}$ that is shear-free and irrotational, then the Lanczos potential is given by (up to gauges)
$$ L_{\alpha\beta\mu} = a_{\alpha} \, v_{\beta} \, v_{\mu}  -  a_{\beta} \, v_{\alpha} \, v_{\mu} $$
\end{theorem}

\begin{theorem}
If in a given Riemannian geometry there is a congruence of observers $ v^{\mu}$ that is geodesic and irrotational, and such that $ H_{\mu\nu} $ vanishes, then the Lanczos potential is given by (up to gauges)
$$ L_{\alpha\beta\mu} = \sigma_{\mu\alpha} \, v_{\beta}  -  \sigma_{\mu\beta}  \, v_{\alpha}. $$
\end{theorem}

An example is provided by Kasner metric
$$ ds^{2} = dt^{2} - t^{2 \, p_{1}} \, dx^{2} - t^{2 \, p_{2}} \, dy^{2} - t^{2 \, p_{3}} \, dz^{2},$$
where $ p_{1} + p_{2} + p_{3} = p_{1}^{2} + p_{2}^{2} + p_{3}^{2} = 1.$ Consider the velocity field $ v^{\mu} = \delta^{\mu}_{0}, $  which is irrotational but has a non-null shear and the Lanczos potential is
$$ L_{0ij} = \frac{(p_{i} - 1/3)}{3} \, t^{2 \, p_{i} - 1} \, \delta_{ij}.$$
Note that there is no sum on the 3-d indices.

\begin{theorem}
If in a given Riemannian geometry there is a congruence of observers $ v^{\mu}$ that is geodesic, non-expanding and
shear-free and such that the vorticity vector is constant then the Lanczos potential is given by
$$ L_{\alpha\beta\mu} = \omega_{\alpha\beta} \, v_{\mu} + \frac{1}{2} \, \omega_{\alpha\mu} \, v_{\beta} - \frac{1}{2} \, \omega_{\beta\mu} \, v_{\alpha}. $$
\end{theorem}

An example is provided by G\"odel geometry which can be written as
$$ ds^{2} = dt^{2} - dx^{2} + 2 \, e^{ax} \, dt \, dy + \frac{e^{2ax}}{2} \, dy^{2} - dz^{2},$$
where $a$ is a constant. Set the velocity field as $  v^{\mu} = \delta^{\mu}_{0}, $ to obtain for the vorticity $ 2\omega_{12} = - a\, e^{2ax}$. It is straightforward to show that  this vector has no shear and the vorticity is such that $ \omega_{\mu \, ; \nu} = 0. $ The Lanczos potential is given by the formula expressed in this Lemma.

\section{Fluid description of the electromagnetic field: the Dirac gauge}
Some years ago Dirac \cite{dirac} suggested to choose a specific gauge such that the potential $A_{\mu}$ may be identified to the normalized velocity $ v_{\mu}$ of a fluid. This means to set
$$ A_{\mu} \, A_{\nu} \, \eta^{\mu\nu} = 1.$$
We are interested here in the description of the energy-momentum distribution that one obtains when using such Dirac choice for the gauge. Using the
identification of the electromagnetic potential as the four-velocity of a fluid we set for the electromagnetic tensor $ F_{\mu\nu} = A_{\mu \, , \nu} - A_{\nu \, , \mu} = v_{\mu \, , \nu} - v_{\nu \, , \mu}. $ From the standard decomposition of the velocity of a congruence of curves we  decompose the derivative of the velocity field in terms of its irreducible components (see the first section). We set
$$ v_{\mu \, , \nu} = \frac{\Theta}{3} \, h_{\mu\nu} + \sigma_{\mu\nu} + \omega_{\mu\nu} + a_{\mu} \, v_{\nu}.$$
Thus it follows that the electromagnetic field, in the Dirac gauge, assumes the form
$$  F_{\mu\nu} = 2 \, \omega_{\mu\nu} + a_{\mu} \, v_{\nu} - a_{\nu} \, v_{\mu}, $$
where the acceleration $ a_{\mu} $ and the vorticity tensor $\omega_{\mu\nu}$ are defined in the first section in the standard way using the velocity field as the potential vector $ A_{\mu}.$

The electric $ E_{\mu}$ and the magnetic $ H_{\mu}$ vectors are defined as usual
$$ E_{\mu} = F_{\mu\nu} \, v^{\nu} $$
$$ H_{\mu} = F^{*}_{\mu\nu} \, v^{\nu}.$$
In the Dirac gauge and under the identification of $ A_{\mu}$ with  the velocity of the fluid it follows that the electric vector is given by the acceleration $ a_{\mu}$ and the magnetic field is the vorticity $\omega_{\mu}$, that is $E_{\mu} = a_{\mu}$ and $ H_{\mu} = - \, 2 \, w_{\mu}$, where the vorticity vector $ w_{\mu}$ is related to the tensor $\omega_{\mu\nu}$ by the formula
$$ w_{\mu} = \frac{1}{2} \, \eta^{\alpha\beta\rho}{}_{\mu} \, \omega_{\alpha\beta} \, v_{\rho}.$$

The energy-momentum tensor
$$ T^{\mu\nu} = F^{\mu}{}_{\alpha} \, F^{\alpha\nu} + \frac{F}{4} \, \eta^{\mu\nu}, $$
where $ F \equiv F_{\mu\nu} \, F^{\mu\nu},$ decomposed in its irreducible components yields respectively: the density of energy
$ \rho = - \, \frac{1}{2} \, ( a_{\mu} \, a^{\mu} + 4 \, \omega_{\mu} \, \omega^{\mu} );$ the pressure $ p = \frac{1}{3} \, \rho;$ the heat flux (Poynting vector) $ q_{\lambda} = \eta_{\lambda}{}^{\mu\rho\sigma} \, a_{\mu} \, v^{\sigma} \, \omega_{\rho}$ and the anisotropic pressure
$$ \pi_{\mu\nu} = - \, a_{\mu} \, a_{\nu} + \frac{1}{3} \, a_{\tau} \, a^{\tau} \, h_{\mu\nu} - 4 \, \omega_{\mu} \, \omega_{\nu} +
\frac{4}{3} \, \omega_{\lambda} \, \omega^{\lambda} \, h_{\mu\nu}.$$
We note that the shear $ \sigma_{\mu\nu} $ and the expansion coefficient $\theta$ are absent from these formulas, only the acceleration and the vorticity appear.

Once in the Dirac gauge the electric field is identified to the acceleration in this fluid description, from what we have learned in the previous sections it follows that it is possible to introduce an auxiliary DM to annihilate the acceleration, that is, to eliminate the electric field. Let us restrict the analysis here to the case the electric field is a gradient. We set $ E_{\mu} = \partial_{\mu} \Psi$ and once the acceleration is orthogonal to the velocity $ a_{\mu} \, v^{\mu} = E_{\mu} \, A^{\mu} = 0$, following the Lemma \ref{drag_lemma}, the condition that the path $v^{\mu}$ be a geodesics in a DM is provided by the standard condition
\begin{equation}
1 + b = \exp(-2 \, \Psi).
\label{23agosto2014}
\end{equation}
The generalization for an arbitrary electric field goes along the same lines.

\section{Fluid description of Heisenberg-Nambu-Jona-Lasinio field}
From the standard definition of the energy-momentum tensor we obtain the expression of the Heisenberg dynamics in the fundamental solution as
$$ T_{\mu\nu} = - \, a_{1} \, J_{\mu} \, J_{\nu} - b_{1} \, I_{\mu} \, I_{\nu} - s \, J^{2} \, \eta_{\mu\nu}. $$
Let us define the four-velocity field as the normalized current $ v_{\mu} = J_{\mu} / \sqrt{J^{2}} $ and let us use this velocity field to decompose the energy-momentum tensor in its irreducible parts and set

\begin{equation}
T_{\mu \nu}= \rho v_{\mu}v_{\nu}-ph_{\mu \nu}+q_{\mu (}\ v_{\nu)}+ \pi_{\mu \nu}
\end{equation}
to obtain that there is no heat flux, that is  $ q_{\mu} $ vanishes identically and the remaining quantities are given by
$$ \rho = - \, (a_{1} + s) \, J^{2},\quad \mbox{and} \quad p = \lambda \, J^{2}$$
where $ \lambda = (s - a_{1}) \, J^{2}/3$ and
$$ \pi_{\mu\nu} = \frac{b_{1}}{3} \,(J_{\mu}\, J_{\nu} - 3 \,I_{\mu} \, I_{\nu} - J^{2} \, \eta_{\mu\nu}) $$

Before taking into account the conservation laws, let us examine the consequences of the choice for the velocity fluid in terms of its irreducible quantities. From the definition of $v_{\mu}$ in terms of the current it follows
\begin{equation}
v_{\alpha \, , \beta} = 2 \, a_{0} \, \, \frac{I_{\alpha} \, I_{\beta}}{\sqrt{J^{2}}}
\label{8setembro1}
\end{equation}
where the comma represents partial derivative. It is straightforward to prove that $v^{\mu}$ has no acceleration. Therefore, although the field has a self-interaction term, the fluid does not acquires a self-acceleration. Using the expression of the current, the shear in this case is
$$ \sigma_{\mu\nu}  = \frac{- \,2 \, a_{0}}{3 \, \sqrt{J^{2}}} \, \left( J_{\mu} \, J_{\nu} - 3 \, I_{\mu} \, I_{\nu} - J^{2} \, \eta_{\mu\nu} \right).$$

It then follows that the anisotropic pressure and the shear are proportional: $ \pi_{\mu\nu} = \xi \, \sigma_{\mu\nu}$, where $ \xi = - b_{1} \, \sqrt{J^{2}} / 2 a_{0}.$ From the equation of evolution of the current we get $ \dot{\xi} = - \, b_{1} \, J^{2}$. In the case parameters $a$ and $ b$ are such that $ b_{1} + 3 \, a_{1} = 0$, then the fluid behaves as an {\it anisotropic vacuum}, with $ \rho + p = 0$ and $ \pi_{\mu\nu} = \xi \, \sigma_{\mu\nu}$.

\section{Fluid description of the scalar field}
For an arbitrary non-linear dynamics driven by a Lagrangian $L$ that depends on the invariant $ W \equiv \partial_{\mu} \varphi \, \partial^{\mu} \varphi$, the energy-momentum tensor is always given by
\begin{equation}
T_{\mu\nu}=-L\,g_{\mu\nu}+2L_{w}\,\partial_{\mu}\varphi\,\partial_{\nu}\varphi\,,
\label{PF3}%
\end{equation}
where $L_{w}\equiv\partial L/\partial W.$ In the comoving frame to the gradient of the field, defined by the normalized vector
\[
v_{\mu}\equiv\frac{\partial_{\mu}\varphi}{\sqrt{W}}
\]
the energy-momentum tensor (\ref{PF3}) become equivalent to a perfect fluid, since the heat flux $q_{\alpha}$ and the anisotropic pressure $\pi_{\mu\nu}$ vanish identically. In this case, the non identically zero quantities are only the energy density and pressure, given by:
\begin{equation}
\rho=-L+2L_{w} W\,,\qquad p=L\,. \label{rhopL}%
\end{equation}
Moreover, since both $p$ and $\rho$ are given as functions of $W$ only, we can write $p=p\left( \rho\right) $. This shows that the dynamics given by any purely kinetic Lagrangian is equivalent to a perfect fluid with a specific equation of state that depends on the dynamics of the field.

\begin{acknowledgments}
This work was partially supported by {\em Conselho Nacional de Desenvolvimento Cient\'{\i}fico e Tecnol\'ogico} (CNPq) and {\em Funda\c{c}\~ao de Amparo \`a Pesquisa do Estado de Rio de Janeiro} (FAPERJ) of Brazil and the CAPES-ICRANet program (BEX 13956/13-2).

\end{acknowledgments}


\begin{thebibliography}{300}
%A
\bibitem{abraham}
S. Abraham, P. Fernandez de Cordoba, J.M. Isidro and J.L.G. Santander, {\em Int.\ J.\ Geom.\ Meth.\ Mod.\ Phys.} {\bf 6} 759 (2009).
\bibitem{adler-bazin}
R. Adler, M. Bazin, M. Schiffer, \textit{Introduction to General Relativity}, McGraw-Hill, New York, (1975).
\bibitem{adler}
R.J. Adler and A.S. Silbergleit, \textit{Nonlinear gravitodynamics: the Lense-Thirring effect}, Ed.\ R. Ruffini and C. Sigismondi, World Scientific Publishing, Singapore (2003).
\bibitem{anandan}
J. Anandan and Y. Aharonov, {\em Phys.\ Rev.\ Lett.} {\bf 65} 1697 (1990).
\bibitem{anandan2}
J. Anandan, {\em Found.\ Phys.} {\bf 21} 1265 (1991).
\bibitem{anderson}
J.L. Anderson, \textit{Principles of Relativity Physics}, Academic Press, London (1967).
\bibitem{ashtekar}
A. Ashtekar and T.A. Schilling, %\textit{Geometrical Formulation of Quantum Mechanics},
\textit{On Einstein's Path: Essays in Honor of Engelbert Schucking}, Ed.\ A. Harvey, Springer-Verlag, New York (1999)

%B
\bibitem{balant}
A.B. Balantekin, {\em Proceedings of Origin of Matter and the Evolution of Galaxies (OMEG05)}, Tokyo, Japan (2005).
\bibitem{bar03a}
C. Barcel\'o, S. Liberati, and M. Visser, {\em Int.\ J.\ Mod.\ Phys.\ D} {\bf 12} 1641 (2003).
\bibitem{bar03b}
C. Barcel\'o, S. Liberati, and M. Visser; {\em Phys.\ Rev.\ A} {\bf 68} 053613 (2003).
\bibitem{visser2011}
C. Barcel\'o, S. Liberati and M. Visser, {\em Liv.\ Rev.\ Rel.} {\bf 8} 12 (2005). ibidem {\bf 14}, 3 (2011).
\bibitem{belgiorno}
F. Belgiorno et al., {\em Phys.\ Rev.\ Lett.} {\bf 105} 203901 (2010).
\bibitem{helayel}
H. Belich, L.P. Colatto, T. Costa-Soares, J.A. Helay\"el-Neto and M.T.D. Orlando, {\em Eur.\ Phys.\ J.\ C} {\bf 62} 425 (2009).
\bibitem{belinski}
V.A. Belinski, {\em Phys.\ Lett.\ A} {\bf 209} 13 (1995).
\bibitem{pdg}
J. Beringer et al., \textit{Particle Data Group}, {\em Phys.\ Rev.\ D} {\bf 86} 010001 (2012).
\bibitem{kau}
K. Bhattacharya and P.B. Pal, {\em Proc.\ Indian\ Natn\ Sci\ Acad} {\bf 70, A} 145 (2004).
\bibitem{birula}
Z. Bialynicka-Birula and I. Bialynicki-Birula, {\em Phys.\ Rev.\ D} {\bf 2} 2341 (1970).
\bibitem{bitCGQ}
E. Bittencourt, V.A. De Lorenci, R. Klippert, M. Novello and J. M.Salim {\em Class.\ Quantum\ Grav.} {\bf 31} 145007 (2014).
\bibitem{novello-edu-faci}
E. Bittencourt, S. Faci and M. Novello, {\em Int.\ J.\ of Mod.\ Phys.\ A} {\bf 29} 1450145 (2014).
\bibitem{blg}
E. Bittencourt, I.P. Lobo and G.G. Carvalho, (accepted in {\em Class.\ Quantum Grav.}), arXiv:1505.03415 [gr-qc].

\bibitem{bdb1}
D. Bohm, {\em Phys.\ Rev.} {\bf 85} 166 (1952); ibidem {\bf 85} 180 (1952).
\bibitem{bdb2}
D Bohm and B.J. Hiley, \textit{The undivided universe}, Routledge, London (1993).
\bibitem{boillat}
G. Boillat, {\em J.\ Math.\ Phys.} {\bf 11} 941 (1970).
\bibitem{born1}
M. Born, {\em Nature (London)} {\bf 132} 282 (1933); {\em Proc.\ R.\ Soc.\ A} {\bf 143} 410 (1934);
\bibitem{born2}
M. Born and L. Infeld, {\em Proc.\ R.\ Soc.\ A} {bf 144} 425 (1934).
\bibitem{bre02}
I. Brevik and G. Halnes, {\em Phys.\ Rev.\ D} {\bf 65} 024005 (2002).
\bibitem{brody}
D.C. Brody and L.P. Hughston, {\em J.\ Geom.\ and Phys.} {\bf 38} 19 (2001).

%C
\bibitem{Canetti}
L. Canetti, M. Drewes and M. Shaposhnikov, {\em Phys.\ Rev.\ Lett.} {\bf 110} 061801 (2013).
\bibitem{canuto}
V. Canuto, P.J. Adams, S.H. Hsieh, and E. Tsiang, {\em Phys.\ Rev.\ D} {\bf 16} 1643 (1977).
\bibitem{caroll}
R. Carroll,  arXiv:0406004v2 [gr-qc].
\bibitem{carroll}
R. Carroll, arXiv:0705.3921 [gr-qc].
\bibitem{carroll-book}
R. Carroll, \textit{On the emergence theme of Physics}, World Scientific Publishing Co., Singapore (2010).

\bibitem{tetrad}
E. Cartan {\em Ann.\ Sci.\ Ec.\ Norm.\ Sup.} {\bf 40} 325 (1923); generalized by H. Weyl, {\em Zeit.\ Phys.} {\bf 56} 330 (1929).
\bibitem{chen}
H.-Y. Chen, R.-X. Miao, and M. Li, {\em Opt.\ Express} {\bf 18} 15183 (2010).
\bibitem{roberts}
I.C. Cloet, C.D. Roberts and A.W. Thomas, {\em Phys.\ Rev.\ Lett.} {\bf 111} 101803 (2013).
\bibitem{copenhagen3}
C. Cohen-Tannoudji, B. Diu, and F. Lalo\"e, \textit{Quantum mechanics}, Wiley \& Sons, France (1977).
\bibitem{cor99}
S. Corley and T. Jacobson, {\em Phys.\ Rev.\ D} {\bf 59} 124011 (1999).
\bibitem{cufaro}
N. Cufaro-Petroni, C. Dewdney, P. Holland, T. Kyprianidis, J. P. Vigier, {\em Phys.\ Lett.\ A} {\bf 106} 368 (1984).
\bibitem{marciano}
A. Czarneck and W.J. Marciano, {\em Phys.\ Rev.\ D} {\bf 64} 013014 (2001).

%D
\bibitem{debroglie1}
L. De Broglie, {\em C.\ R.\ Acad.\ Sci.\ Paris} {\bf 183} 24 (1926); ibidem {\bf 183} 447 (1926); {\bf 184} 273 (1927); {\bf 185} 380 (1927).
\bibitem{debroglie3}
L. De Broglie, {\em Nature} {\bf 118} 441 (1926).
\bibitem{debroglie6}
L. De Broglie, {\em J.\ de Phys.} {\bf 8} 225 (1927).
\bibitem{broglie}
L. De Broglie \textit{Non-Linear Wave Mechanics: a causal interpretation}, Elsevier, Amsterdam (1960).
\bibitem{felice1971}
F. De Felice, {\em Gen.\ Rel.\ Grav.} {\bf 2} 347 (1971).
\bibitem{del02}
V.A. De Lorenci, {\em Phys.\ Rev.\ E} {\bf 65} 026612 (2002).
\bibitem{Klippert}
V.A. De Lorenci and R. Klippert, {\em Phys.\ Rev.\ D} {\bf 65} 064027 (2002).
\bibitem{bjp}
V.A. De Lorenci and R. Klippert, {\em Braz.\ J. Phys.\ } {\bf 34} 1367 (2004).
\bibitem{local}
V.A. De Lorenci and R. Klippert, {\em Phys.\ Lett.\ A} {\bf 357} 61 (2006).
\bibitem{vitPLB}
V.A. De Lorenci, R. Klippert, M. Novello and J.M. Salim, {\em Phys.\ Lett.\ B} {\bf 482} 134 (2000).
\bibitem{Yuri}
V.A. De Lorenci, R. Klippert and Yu.\ N. Obukhov, {\em Phys.\ Rev.\ D} {\bf 68} 061508R (2003).
\bibitem{del04}
V.A. De Lorenci, R. Klippert and D.H. Teodoro. {\em Phys.\ Rev.\ D} {\bf 70} 124035 (2004).
\bibitem{del01}
V.A. De Lorenci and M.A. Souza, {\em Phys.\ Lett.\ B} {\bf 512} 417 (2001);
\bibitem{deser}
S. Deser, {\em Class.\ Quantum\ Grav.} {\bf 4} L99 (1987).
\bibitem{nelson2}
B.S. DeWitt, {\em Physics Today} {\bf 23} 30 (1970).
\bibitem{nelson1}
B.S. DeWitt, N. Graham, \textit{The Many-Worlds Interpretation of Quantum Mechanics}, Princeton Univ. Press, UK  (1973).
\bibitem{copenhagen1}
P.A.M. Dirac, \textit{The principles of quantum mechanics}, Oxford Univ. Press, UK (1958).
\bibitem{dirac}
P.A.M. Dirac, {\em Proc.\ Roy.\ Soc.\ London} {\bf A 333} 403 (1973).
\bibitem{Dittrich}
W. Dittrich and H. Gies, {\em Phys.\ Rev.\ D} {\bf 58} 025004 (1998); ibid, {\em Phys.\ Lett.\ B} {\bf 431} 420 (1998).
\bibitem{drewes}
M. Drewes, arXiv:1303.6912 [hep-ph].
\bibitem{Drummond}
I.T. Drummond and S.J. Hathrell, {\em Phys.\ Rev.\ D} 22 343 (1980).

%E
\bibitem{eddington}
A.S. Eddington, \textit{The mathematical theory of relativity}, Cambridge University Press, New York (1988).
\bibitem{einstein}
A. Einstein, \textit{The meaning of relativity}, Princeton University Press, (1950).
\bibitem{copenhagen2}
E. Elbaz, \textit{Quantum: the quantum theory of particles, fields, and cosmology}, Springer-Verlag, Berlin (1998).
\bibitem{others41}
H. Everett, {\em Rev.\ Mod.\ Phys.} {\bf 29} 454 (1957).

%F
\bibitem{falciano}
F.T. Falciano, M. Novello and J.M. Salim, {\em Found.\ Phys.} {\bf 40} 1885 (2010).
\bibitem{fed03}
P.O. Fedichev and U.R. Fischer, {\em Phys.\ Rev.\ Lett.} {\bf 91} 240407 (2003).
%\bibitem{feshbach}
%H. Feshbach, F. Villars, {\em Rev.\ Mod.\ Phys.} {\bf 30} 24 (1958).
\bibitem{others1}
I. Feyn\`es, {\em Z.\ Phys.} {\bf 132} 81 (1952).
\bibitem{feynman}
R.P. Feynman, F.B. Morinigo and W.G. Wagner, \textit{Feynman\ lectures on gravitation}, Addison Wesley Pub. Company, Massachusetts, (1995).
\bibitem{Fierz}
M. Fierz, {\em Helv.\ Phys.\ Acta} {\bf 12} 379 (1939)
\bibitem{branden}
F. Finelli and R. Branderberger, arXiv:0112249 [hep-th].
\bibitem{formiga}
J.B. Formiga, arXiv:1210.0759 [hep-th].
\bibitem{fushchych}
W.I. Fushchych, R.Z. Zhdanov, {\em Phys.\ Rep.} {\bf 172} 4 123 (1989).

%G
\bibitem{gar00}
L.J. Garay, J. R. Anglin, J. I. Cirac, and P. Zoller, {\em Phys.\ Rev.\ Lett.} {\bf 85} 4643 (2000); {\em Phys.\ Rev.\ A} {\bf 63} 023611 (2001).
\bibitem{others53}
M. Gell-Mann, J.B. Harle, \textit{Complexity, entropy and the physics of information}, Ed. W. Zurek, Addison Wesley (1990).
\bibitem{geng}
C.Q. Geng and R. Takahashi, {\em Phys.\ Lett.\ B} {\bf 710} 324 (2012).
\bibitem{gibbons}
G. Gibbons and C. Will, {\em Studies in History and Philosophy of Modern Physics} {\bf 39} 41 (2008).
\bibitem{giulini}
D. Giulini, {\em Studies in History and Philosophy of Modern Physics} {\bf 39} 154 (2008).
%\bibitem{goenner}
%H.F.M. Goenner, {\em Liv.\ Rev.\ Rel.} {\bf 7} 2 (2004).
\bibitem{godel}
K. G\"odel, {\em Rev.\ Mod.\ Phys.} {\bf 21} 447 (1949).
\bibitem{gon98}
P.F. Gonz\'alez-Diaz and C.L. Singuenza, {\em Phys. Lett. A} {\bf 248} 412 (1998).
\bibitem{gordon}
W. Gordon {\em Ann.\ Phys.\ (Leipzig)} {\bf 72} 421  (1923).
\bibitem{goulart}
E. Goulart, M. Novello, F. T. Falciano and J. D. Toniato, {\em Class.\ Quantum\ Grav.} {\bf 28} 245008 (2011).
\bibitem{others52}
R.B. Griffiths, {\em J.\ St.\ Phys.} {\bf 36} 219 (1984).
\bibitem{gpp}
J. P. Grishchuk, A. N. Petrov and A. D. Popova, {\em Commun.\ Math.\ Phys.} {\bf 94} 379 (1984).
\bibitem{gueret}
Ph. Gueret, P. Holland, A. Kyprianidis, J.P. Vigier, {\em Phys.\ Lett.\ A} {\bf 107} 379 (1985).
\bibitem{gupta}
S. N. Gupta, {\em Phys.\ Rev.} {\bf 96} 1683 (1954) .

%H
\bibitem{hadamard}
J. Hadamard, \textit{Le\c cons sur la propagation des ondes et les \'equations de hydrodynamique} (Hermann, Paris, 1903);
V. D. Zakharov, \textit{Gravitational waves in Einstein's theory} (John Wiley \& Sons, Inc., New York, 1973).
%\bibitem{halbwachs}
%F. Halbwachs \emph{Th\'eorie Relativiste des Fluides a Spin}, Gauthier-Villars, Pais (1960).
\bibitem{gabrielse}
D. Hanneke, S. Fogwell, and G. Gabrielse, {\em Phys.\ Rev.\ Lett.} {\bf 100} 120801 (2008).
\bibitem{others54}
J.B. Hartle, $13^{th}$ \emph{International conference on General Relativity and Gravitation}, eds. R.J. Gleiser, C.N. Kozameh and O.M. Moreschi, Institute of Physics Publishing, (1993).
\bibitem{hatsuda}
T. Hatsuda and T. Kunihiro, {\em Phys.\ Rep.} {\bf 247} 221 (1994).
\bibitem{hawking1975}
S. W. Hawking, {\em Commun.\ Math.\ Phys.\ } {\bf 43} 199 (1975).
\bibitem{hehl}
F. W. Hehl and Y. N. Obukhov, \textit{Foundations of classical electrodynamics: charge, flux, and metric}, Progress in Mathematical Physics, v.\ 33 (Birkh\"auser, Boston, 2003).
\bibitem{heisenberg}
W. Heisenberg, {\em Rev.\ Mod.\ Phys.} {\bf 29} 3 (1957).
\bibitem{hoc97}
D. Hochberg and J. P\'erez-Mercader, {\em Phys.\ Rev.\ D} {\bf 55} 4880 (1997).
\bibitem{holland}
P.R. Holland, {\em Found.\ Phys.} {\bf 17} 345 (1987).
\bibitem{holland2}
P.R. Holland, \textit{The quantum theory of motion}, Cambridge Univ. Press, Cambridge (1993).
\bibitem{hkv}
P. Holland, A. Kyprianidis, J.P. Vigier, {\em Phys.\ Lett.\ A} {\bf 107} 376 (1985).

%I
\bibitem{israelit}
Israelit, M., {\em Found.\ Phys.} {\bf 28} 205 (1998).
\bibitem{israelit2}
Israelit, M., {\em Found.\ Phys.} {\bf 29} 1303 (1999).
\bibitem{israelit3}
Israelit, M., {\em Found.\ Phys.} {\bf 32} 295 (2002).
\bibitem{israelit4}
Israelit, M., {\em Found.\ Phys.} {\bf 32} 945 (2002).
\bibitem{Inomata}
A. Inomata and W.A. McKinley, {\em Phys.\ Rev.} {\bf 140} 1467 (1965).

%J
\bibitem{jac93}
T. Jacobson, {\em Phys.\ Rev.\ D} {\bf 48} 728 (1993).
\bibitem{jac98}
T. Jacobson, {\em Phys.\ Rev.\ D} {\bf 58} 064021 (1998).
%\bibitem{jac99}
%T. Jacobson, {\em Phys.\ Rev.\ D} {\bf 44} 1731 (1999).

%K
\bibitem{ugo}
A.Yu.\ Kamenshchik, U. Moschella and V. Pasquier, {\em Phys.\ Lett.\ B} {\bf 511} 265 (2001).
\bibitem{others2}
D. Kershaw, {\em Phys.\ Rev.\ B} {\bf 136} 1850 (1964).
\bibitem{koch}
B. Koch, arxiv:0810.2786 [hep-th].
\bibitem{koch2}
B. Koch, arxiv:0901.4106 [gr-qc].
\bibitem{koch3}
B. Koch, {\em AIP Conf.\ Proc.} {\bf 1232} 313 (2010).
\bibitem{kruglov}
S.I. Kruglov, {\em Int.\ J.\ Geom.\ Methods Mod.\ Phys.} {\bf 12} 1550073 (2015).
\bibitem{alex}
A.V. Kuznetsov and N.V. Mikheev {\em JCAP} {\bf 11} 031 (2007).
\bibitem{kyprianidis}
A. Kyprianidis, {\em Phys.\ Lett.\ A} {\bf 111} 111 (1985).

%L
\bibitem{Lanczos}
C. Lanczos, {\em Rev.\ Mod.\ Phys.} {\bf 34} 379 (1962).
\bibitem{lanczos}
C. Lanczos, \textit{The Variational Principles of Mechanics}, Ed. 4, Dover, New York (1970);
\bibitem{Landau}
L.D. Landau and E.M. Lifshitz, \textit{Electrodynamique des milieux Continus}, Ed. Mir, Moscow (1969);
\bibitem{landau1971}
L.D. Landau and E.M. Lifshitz, \textit{The classical theory of fields}, Pergamon Press Ltd., Oxford, (1971).
\bibitem{Latorre}
J.I. Latorre, P. Pascual and R. Tarrach, {\em Nuclear\ Phys.\ B} {\bf 437} 60 (1995).
\bibitem{lee}
B.W. Lee and R.E. Shrock, {\em Phys.\ Rev.\ D} {\bf 16} 1444 (1977).
\bibitem{leo03}
U. Leonhardt and P. \"Ohberg, {\em Phys.\ Rev.\ A} {\bf 67} 053616 (2003).
\bibitem{leo99}
U. Leonhardt and P. Piwnicki, {\em Phys.\ Rev.\ A} {\bf 60} 4301 (1999).
\bibitem{leonetal}
U. Leonhardt and P. Piwnicki, {\em Phys.\ Rev.\ Lett.} {\bf 85} 5253 (2000).
\bibitem{lib00}
S. Liberati, S. Sonego and M. Visser, {\em Class.\ Quantum\ Grav.} {\bf 17} 2903 (2000).
\bibitem{Lichne}
A. Lichnerowicz, \textit{Geometrie des groupes de transformations}, Dunod, Paris, (1958).
\bibitem{london}
F. London, {\em Zeitschrift f\"ur Physik} {\bf 42} 375 (1927).
\bibitem{lych}
O. Lychkovskiy and S. Blinnikov, {\em Phys.\ Atom.\ Nucl.} {\bf 73} 614 (2010).

%M
%\bibitem{mach}
%E. Mach, \textit{The Science of Mechanics}, ed.\ 6, The Open Court Publishing CO., Illinois (1960).
%\bibitem{maqueijo}
%J. Magueijo, {\em Phys.\ Rev.\ D} {\bf 79} 043525 (2009).
\bibitem{madelung}
E. Madelung, {\em Z.\ Phys.} {\bf 40} 332 (1926).
\bibitem{sanda}
W.J. Marciano and A.I. Sanda, {\em Phys.\ Lett.\ B} {\bf 67} 303 (1977).
\bibitem{mercati}
These Newtonian concepts have been revisited in other contexts, leading to other theories of gravity. See, for instance, F. Mercati, \textit{A Shape Dynamics Tutorial} arXiv:1409.0105 [gr-qc].
\bibitem{mat85}
S. Matarrese, {\em Proc.\ R.\ Soc.\ Lond.\ A} {\bf 401} 53 (1985).
\bibitem{Maupertuis}
M. de Maupertuis, %\textit{Accord de différentes loix de la nature qui avoient jusqu’ici paru incompatibles}
Histoire de l'Acad\'emie Royale des Sciences de Paris, {\bf 1744} 417 (1748).
\bibitem{minic}
D. Minic and C.-H. Tze {\em Phys.\ Lett.\ B} {\bf 536} 305 (2002).

%N
\bibitem{njl}
Y. Nambu and G. Jona-Lasinio, {\em Phys.\ Rev.} {\bf 122} 345 (1961); Y. Nambu and G. Jona-Lasinio, {\em Phys.\ Rev.} {\bf 124} 246 (1961).
\bibitem{narimanov}
E.E. Narimanov and A.V. Kildishev, {\em Appl.\ Phys.\ Lett.\ } {\bf 95} 041106 (2009).
\bibitem{others3}
E. Nelson, {\em Phys.\ Rev.} {\bf 150} 1079 (1966).
\bibitem{novello_73}
M. Novello, {\em Phys.\ Rev.\ D} {\bf 8} 2398 (1973).
%\bibitem{novdirac}
%M. Novello, {\em Europhys.\ Lett.} {\bf 80} 41001 (2007).
\bibitem{nov_bit_gordon}
M. Novello and E. Bittencourt, {\em Phys.\ Rev.\ D} {\bf 86} 124024 (2012).
\bibitem{nov_bit_drag}
M. Novello and E. Bittencourt, {\em Gen.\ Rel.\ Grav.} {\bf 45} 1005 (2013).
\bibitem{nov_bit}
M. Novello and E. Bittencourt, {\em Int.\ J.\ Mod.\ Phys.\ A} {\bf 29} 1450075 (2014).
\bibitem{JCAP}
M. Novello, E. Bittencourt, U. Mosquella, E. Goulart, J.M. Salim and J.D. Toniato, {\em JCAP} {\bf 06} 014 (2013).
\bibitem{nov_bit_sal}
M. Novello, E. Bittencourt and J. M. Salim, {\em Braz.\ J.\ Phys.} {\bf 44} 832 (2014).
\bibitem{novellosoarestiomno}
M. Novello, I. Dami\~ao Soares and J. Tiomno, {\em Phys.\ Rev.\ D} {\bf 27} 4 779 (1983).
\bibitem{novello1}
M. Novello, V.A. De Lorenci, J.M. Salim and R. Klippert {\em Phys.\ Rev.\ D} {\bf 61} 045001 (2000).
\bibitem{nov_fal_gou}
M. Novello, F.T. Falciano, E. Goulart, arXiv:1111.2631 [gr-qc].
\bibitem{novello2}
M. Novello and E. Goulart {\em Class.\ Quantum\ Grav.} {\bf 28} 145022 (2011).
\bibitem{novello-huguet}
M. Novello, E. Huguet and J. Queva, {\em Phys.\ Rev.\ D} {\bf 73} 123531 (2006).
\bibitem{novello}
M. Novello, L.A.R. Oliveira, J.M. Salim, E. Elbaz, {\em Int.\ J.\ Mod.\ Phys.} {\bf D 1} 641-677 (1992).
%\bibitem{novello3}
%M. Novello, S.E. Perez-Bergliaffa and J.M. Salim, {\em Phys.\ Rev.\ D} {\bf 69} 127301 (2004).
\bibitem{novello4}
M. Novello, S.E. Perez-Bergliaffa, J.M. Salim, V.A. De Lorenci, and R. Klippert, {\em Class.\ Quantum Grav.} {\bf 20} 859 (2003).
\bibitem{NovelloNelson}
M. Novello and N. Pinto-Neto, {\em Fortschrift der Physik} {\bf 40} 173 (1992); ibidem {\bf 40} 195 (1992).
\bibitem{nov01}
M. Novello and J. M. Salim, {\em Phys.\ Rev.\ D} {\bf 63} 083511 (2001).
\bibitem{nsf}
M. Novello, J.M. Salim and F.T. Falciano, {\em Int.\ J.\ Geom.\ Meth.\ Mod.\ Phys.} {\bf 8} 87 (2011).
\bibitem{novellonamiemilia}
M. Novello, N.F. Svaiter and M. E. X. Guimarães {\em Mod.\ Phys.\ Lett.\ A} {\bf 7} 5 381 (1992);
\bibitem{Novello_Velloso}
M. Novello and A.L. Velloso {\em Gen.\ Rel.\ and Grav.} {\bf 19} 1251 (1987).
\bibitem{nov_viss_volo}
M. Novello, M. Visser and G. Volovik (Eds.), \textit{``Artificial Black Holes", Proceedings of the workshop: Analog Models of General Relativity}, World Scientific (2002).
\bibitem{nord}
G. Nordstr\"om, {\em Phys.\ Zeit.} {\bf 13} 1126 (1912).

%O
\bibitem{others51}
R. Omn\`es, \textit{The interpretation of quantum mechanics}, Princeton Univ. Press, UK (1994).

%P
\bibitem{passera}
M. Passera, {\em Nucl.\ Phys.\ Proc.\ Suppl.} {\bf 169} 213 (2007).
\bibitem{pendry}
J.B. Pendry et al., {\em Science} {\bf 312} 1780 (2006).
\bibitem{perlick}
V. Perlick, {\em Class.\ Quant.\ Grav.} {\bf 8} 1369 (1991).
\bibitem{nelson3}
N. Pinto-Neto, \textit{Cosmology and Gravitation II} , ed.\ M. Novello, Editions Fronti\`eres, France, (1996).
\bibitem{Plebanski}
J. Plebanski, {\em Phys.\ Rev.} {\bf 118} 1396 (1960).
\bibitem{poincare}
H. Poincar\'e, \textit{Science and Hypothesis}, Ed. Scott, Michigan University (1905).
%\bibitem{poplawski}
%N.J. Poplawski, arXiv:0710.3982 [gr-qc].

%R
\bibitem{finn}
F. Ravndal, \textit{Scalar gravitation and extra dimensions}, arXiv:0405030 [gr-qc].
\bibitem{rez00}
B. Reznik, {\em Phys.\ Rev.\ D} {\bf 62} 044044 (2000).
\bibitem{riemann}
B. Riemann, \textit{On the hypothesis which lie at the foundations of geometry} in D.E. Smith, \emph{A Source Book in Mathematics} - vol.\ 2, ed. Dover, New York (1959).
\bibitem{rindler}
W. Rindler, \textit{Relativity: Special, General, and Cosmological}, Oxford University Press (2001).
%\bibitem{rosen}
%N. Rosen, {\em Found.\ of Phys.} {\bf 12} 213 (1982).
%\bibitem{rosen2}
%N. Rosen, {\em Found.\ of Phys.} {\bf 13} 363 (1983).
\bibitem{prl2}
T. Roth and G.L.J.A. Rikken, {\em Phys.\ Rev.\ Lett.} {\bf 85} 4478 (2000); {\it ibidem} {\bf 88} 063001 (2002).

%S
\bibitem{santamato1}
E. Santamato, {\em J.\ Math.\ Phys.} {\bf 25} 2477 (1984).
\bibitem{santamato}
E. Santamato, {\em Phys.\ Rev.} {\bf D 29} 216 (1984).
%\bibitem{schenberg1971}
%M. Schenberg, {\em Revista Brasileira de Fisica}, {\bf 1} 91 (1971).
\bibitem{schiff}
L.I. Schiff, {\em Phys.\ Rev.\ Lett.} {\bf 4} 215 (1960).
\bibitem{shu02}
R. Sch\"utzhold, G. Plunien, and G. Soff, {\em Phys.\ Rev.\ Lett.} {\bf 88}, 061101 (2002).
\bibitem{schuring}
D. Scuring et al., {\em Science} {\bf 314} 977 (2006).
\bibitem{shapiro}
S.L. Shapiro and S.A. Teukolsky, {\em Phys.\ Rev.\ D} {\bf 47} 1529 (1993).
\bibitem{shapos}
M. Shaposhnikov, {\em Proceedings of the 11th Marcel Grossmann Meeting on General Relativity } 1006, Berlin, Germany (2006).
\bibitem{sheng}
C. Sheng, H. Liu, Y. Wang, S.N. Zhou and D.A. Genov, {\em Nature Photonics} {\bf 7} 902 (2013).
\bibitem{Shore}
G.M. Shore, {\em Nuclear\ Phys.\ B} {\bf 460} 379 (1996); R.D. Daniels and G.M. Shore, {\em Nuclear\ Phys.\ B} {\bf 425} 634 (1994).
\bibitem{smith}
D.R. Smith et al., {\em Science} {\bf 305} 788 (2009).
\bibitem{spin_curv}
I.D. Soares, {\em Proc.\ II Brazilian School of Cosmology and Gravitation}, Ed. M. Novello, impressed by J. Sasson \& Cia. Ltd., Rio de Janeiro (1980).
\bibitem{coll-eins}
J. Stachel et al., editors,\textit{The collected papers of Albert Einstein}, vol. 1-9, Princeton University Press, Princeton, New Jersey (1987-2005).
\bibitem{synge}
J.L. Synge, \textit{Relativity: the special theory}, North-Holland, Amsterdam (1958).

%U
\bibitem{unr81}
W.G. Unruh, {\em Phys.\ Rev.\ Lett.} {\bf 46} 1351 (1981); {\em Phys.\ Rev.\ D} {\bf 51} 2827 (1995).
\bibitem{unr03}
W.G. Unruh and R. Sch\"utzhold, {\em Phys.\ Rev.\ D} {\bf 68} 024008 (2003).

%V
%\bibitem{veltman}
%M. Veltman, (lectures) in CERN 97-05 for a historical review.
\bibitem{vis98}
M. Visser, {\em Class.\ Quantum\ Grav.} {\bf 15} 1767 (1998).
\bibitem{vis00}
M. Visser, {\em Phys.\ Rev.\ Lett.} {\bf 85} 5252 (2000).
\bibitem{vol98}
G.E. Volovik, arXiv:9806010 [cond-mat].

%T
\bibitem{frauendiener}
R.T. Thompson and J. Frauendiener, {\em Phys.\ Rev.\ D} {\bf 82} 124021 (2010).
\bibitem{others42}
F. Tipler, {\em Phys.\ Rept.} {\bf 137} 231 (1986).
\bibitem{tolman}
R.C. Tolman, \textit{Relativity Thermodynamics and Cosmology}, Oxford Univ. Press, Oxford (1962).

%W
\bibitem{unruh2011}
S. Weinfurtner, E. W. Tedford, M.C.J. Penrice, W.G. Unruh, and G.A. Lawrence, {\em Phys.\ Rev.\ Lett.\ } {\bf 106}, 021302 (2011).
\bibitem{Weyl1}
H. Weyl,{\em Sitz.\ Preuss.\ Akad.\ Wiss.} {\bf 26} 465 (1918).
\bibitem{Weyl2}
H. Weyl, \emph{Space, Time, Matter}, Dover Publications, UK (1922).
\bibitem{wheeler}
J.T. Wheeler, {\em Phys.\ Rev.} {\bf D 41} 431 (1990).
\bibitem{will}
C. Will, {\em Liv.\ Rev.\ Rel.} {\bf 9} 3 (2006).

%Y
\bibitem{yin}
M. Yin, X. Y. Tian, L. L. Wu, and D. C. Li, {\em Opt.\ Exp.} {\bf 21} 19082 (2013).
\end{thebibliography}
\end{document}